# The NUMEN project: NUclear Matrix Elements for Neutrinoless double beta decay

F. Cappuzzello[1,2], C. Agodi[2], M. Cavallaro[2], D. Carbone[2], S. Tudisco[2], D. Lo Presti[1,3], J. R. B. Oliveira[4], P. Finocchiaro[2], M. Colonna[2], D. Rifuggiato[2], L. Calabretta[2], D. Calvo[5], L. Pandola[2], L. Acosta[6], N. Auerbach[7], J. Bellone[1,2], R. Bijker[8], D. Bonanno[3], D. Bongiovanni[2], T. Borello-Lewin[4], I. Boztosun[9], O. Brunasso[5], S. Burrello[2,10], S. Calabrese[1,2], A. Calanna[2], E.R. Chávez Lomelí[6], G. D'Agostino[1,2], P.N. De Faria[11], G. De Geronimo[12], F. Delaunay[5,13], N. Deshmukh[2], J.L. Ferreira[11], M. Fisichella[5], A. Foti[3], G. Gallo[1,2], H. Garcia[14,25], V. Greco[1,2], A. Hacisalihoglu[2,15], F. Iazzi[5,16], R. Introzzi[5,16], G. Lanzalone[2,17], J.A. Lay[2,10], F. La Via[18], H. Lenske[19], R. Linares[11], G. Litrico[2], F. Longhitano[3], J. Lubian[11], N. H. Medina[4], D.R. Mendes[11], M. Moralles[20], A. Muoio[2], A. Pakou[21], H. Petrascu[22], F. Pinna[5,16], S. Reito[3], A. D. Russo[2], G. Russo[1,3], G. Santagati[2], E. Santopinto[14], R.B.B. Santos[23], O. Sgouros[2,21], M. A. G. da Silveira[23], S.O. Solakci[9], G. Souliotis[24], V. Soukeras[2,21], A. Spatafora[1,2], D. Torresi[2], R. Magana Vsevolodovna[14,25], V.A.B. Zagatto[11], A. Yildirin[9]

1. Dipartimento di Fisica e Astronomia, Università di Catania, Italy
2. Istituto Nazionale di Fisica Nucleare, Laboratori Nazionali del Sud, Catania, Italy
3. Istituto Nazionale di Fisica Nucleare, Sezione di Catania, Italy
4. Departamento de Fisica Nuclear, Universidade de Sao Paulo, Sao Paulo, Brazil
5. Istituto Nazionale di Fisica Nucleare, Sezione di Torino, Italy
6. Instituto de Física, Universidad Nacional Autónoma de México, Mexico
7. School of Physics and Astronomy Tel Aviv University, Israel
8. Instituto de Ciencias Nucleares, Universidad Nacional Autónoma de México, Mexico
9. Akdeniz University, Antalya, Turkey
10. Departamento de FAMN, Universidad de Sevilla, Sevilla, Spain
11. Universidade Federal Fluminense, Niteroi, Brazil
12. Stony Brook University, NY, USA
13. LPC Caen, Normandie Université, ENSICAEN, UNICAEN, CNRS/IN2P3, Caen, France
14. Istituto Nazionale di Fisica Nucleare, Sezione di Genova, Italy
15. Institute of Natural Science, Karadeniz Teknik Universitesi, Trabzon, Turkey
16. DISAT-Politecnico di Torino, Italy
17. Università degli Studi di Enna "Kore", Enna, Italy
18. CNR-IMM, Sezione di Catania, Italy
19. University of Giessen, Germany
20. Instituto de Pesquisas Energeticas e Nucleares IPEN/CNEN, Sao Paulo, SP, Brazil
21. Department of Physics and HINP, The University of Ioannina, Ioannina, Greece
22. IFIN-HH, Bucarest, Romania
23. Centro Universitario FEI, Sao Bernardo do Campo, Brazil
24. Laboratory of Physical Chemistry, Department of Chemistry, National and Kapodistrian University of Athens, Athens, Greece
25. Dipartimento di Fisica dell'Università di Genova, Genova, Italy

**Abstract.** The article describes the main achievements of the NUMEN project together with an updated and detailed overview of the related R&D activities and theoretical developments. NUMEN proposes an innovative technique to access the nuclear matrix elements entering the expression of the lifetime of the double beta decay by cross section measurements of heavy-ion induced Double Charge Exchange (DCE) reactions. Despite the two processes, namely neutrinoless double beta decay and DCE reactions, are triggered by the weak and strong interaction respectively, important analogies are suggested. The basic point is the coincidence of the initial and final state many-body wave-functions in the two types of processes and the formal similarity of the transition operators. First experimental results obtained at the INFN-LNS laboratory for the $^{40}$Ca($^{18}$O,$^{18}$Ne)$^{40}$Ar reaction at 270 MeV, give encouraging indication on the capability of the proposed technique to access relevant quantitative information.

The two major aspects for this project are the K800 Superconducting Cyclotron and MAGNEX spectrometer. The former is used for the acceleration of the required high resolution and low emittance heavy ion beams and the latter is the large acceptance magnetic spectrometer for the detection of the ejectiles. The use of the high-order trajectory reconstruction technique, implemented in MAGNEX, allows to reach the experimental resolution and sensitivity required for the accurate measurement of the DCE cross sections at forward angles. However, the tiny values of such cross sections and the resolution requirements demand beam intensities much larger than manageable with the present facility. The on-going upgrade of the INFN-LNS facilities in this perspective is part of the NUMEN project and will be discussed in the article.

# SUMMARY

# 1 Introduction

Neutrinoless double beta decay ($0\nu\beta\beta$) is potentially the best resource to probe the Majorana or Dirac nature of neutrino and to extract its effective mass. Moreover, if observed, $0\nu\beta\beta$ decay will signal that the total lepton number is not conserved. Presently, this physics case is among the most important research "beyond the Standard Model" and might guide the way toward a Grand Unified Theory of fundamental interactions and to unveil the source of matter-antimatter asymmetry observed in the Universe.

Since the $\beta\beta$ decay process involves transitions in atomic nuclei, nuclear structure issues must be also accounted for, in order to describe it. In particular, the $0\nu\beta\beta$ decay rate $[T_{1/2}]^{-1}$ can be factorized as a phase-space factor $G_{0\nu}$, the Nuclear Matrix Element (NME) $M_{0\nu}$ and a term $f(m_i, U_{ei}, \xi_i)$ containing a combination of the masses $m_i$, the mixing coefficients $U_{ei}$ of the neutrino species and the Majorana phases $\xi_i$:

$$[T_{1/2}]^{-1} = G_{0\nu} |M_{0\nu}|^2 |f(m_i, U_{ei}, \xi_i)|^2 \qquad (1.1)$$

where $M_{0\nu}$ is the transition amplitude from the initial $\varphi_i$ to the final $\varphi_f$ nuclear state of the $\beta\beta$ process through the $0\nu\beta\beta$ decay operator:

$$|M_{0\nu}|^2 = |\langle \varphi_f | \hat{O}^{0\nu\beta\beta} | \varphi_i \rangle|^2 \qquad (1.2)$$

Thus, if the NMEs are established with sufficient precision, the $f(m_i, U_{ei}, \xi_i)$ function, containing physics beyond the standard model, can be extracted from $0\nu\beta\beta$ decay rate measurements or bounds.

The evaluation of the NMEs is presently based on state-of-the-art model calculations from different methods, e.g. Quasi-particle Random Phase Approximation (QRPA), large scale shell-model, Interacting Boson Model (IBM), Energy Density Functional (EDF), ab-initio [1], [2], [3], [4], [5]. All of these approaches propose different truncation schemes of the still unsolved full nuclear many-body problem into a solvable one, limited to a model space. The purpose is to include, as much as possible, the relevant degrees of freedom which allow a complete description of the problem. However, this condition cannot be easily checked without a comparison with experimental data. Indirect hints of the reliability of model calculations could come from their relative convergence to common values, even if this condition would not exclude that common unverified assumptions are still present in all models. High precision experimental information from Single Charge Exchange (SCE), transfer reactions and Electron Capture (EC) are also used to constrain the calculations [6], [7], [8], [9], [10]. However, the ambiguities in the models are still too large and the constraints too loose to provide accurate values of the NMEs. Discrepancy factors larger than two are presently reported in literature [11]. In addition, some assumptions, common to the different competing calculations, could cause unknown overall systematic uncertainties [12]. A pertinent example is about the use of quenched coupling constants within the nuclei, especially for the axial-vector weak interaction, which is strongly debated nowadays [13], [14], [15], [16], [17], [18], [19], [20].

In this scenario, the experimental study of other nuclear transitions where the nuclear charge is changed by two units leaving the mass number unvaried, in analogy to the *ββ*-decay, could provide important information. Past attempts to use pion-induced double charge exchange reactions [21], [22], [23], [24] to probe *ββ*-decay NMEs were abandoned due to the large differences in the structure of the operators [12]. Early studies of heavy-ion induced Double Charge Exchange (DCE) reactions were also not conclusive. The reason was the lack of zero-degree data and the poor yields in the measured energy spectra and angular distributions, due to the very low cross sections observed, ranging from about 5-40 nb/sr [25], [26] to 10 µb/sr [27]. Actually, this wide range of observed cross sections has not yet been deeply discussed. An additional complication in the interpretation of the data was due to possible contributions of multi-nucleon transfer reactions leading to the same final states [28], [29], [30].

Recently, the use of modern high resolution and large acceptance spectrometers has been proven to be effective in order to face the main experimental challenges and to extract quantitative information from DCE reactions. The measurement of DCE high-resolution energy spectra and accurate absolute cross sections at very forward angles is crucial to identify the transitions of interest [31] [32]. The concurrent measurement of the other relevant reaction channels allows to isolate the direct DCE mechanism from the competing multi-nucleon transfer processes. These are at least of 4[th] order in the nucleus-nucleus interaction and can be effectively minimized by the choice of the proper projectile-target system and incident energy [33].

Based on these results, the NUMEN (*NUclear Matrix Elements for Neutrinoless double beta decay*) project was recently proposed, with the aim to investigate the nuclear response to DCE reactions for all the isotopes explored by present and future studies of *0νββ* decay [34] [35]. Several aspects of the project require the development of innovative techniques, for both experiment set-up and theoretical analysis of the collected data. Consequently, NUMEN represents a challenging research opportunity for nuclear physics, besides its main objective.

Here we present an updated and detailed overview of the NUMEN project, along with selected newly achieved results and prominent scientific perspectives. In Section 2 the DCE reactions are presented as tools to explore the nuclear response to isospin and spin-isospin operators. The main features and objectives of the NUMEN project are then described in Section 3, while the recently achieved results are discussed in Section 4. Particular attention is given to the limits of the present experimental set up and to the required upgrades of both experimental and theoretical aspects. In Section 5, the guidelines for the development of a state of art microscopic theory for DCE reactions are given, which will be suitable, to extract the relevant information for $0\nu\beta\beta$ from the measured DCE cross sections. The proposed solutions for the upgrade of the accelerator and detector equipment are discussed in Section 6. Conclusions and perspectives are summarized in Section 7.

# 2 Nuclear response to Charge Exchange reactions

## 2.1 *A View of Heavy Ion Single Charge Exchange Reactions*

Single Charge Exchange (SCE) reactions are well established tools for spectroscopic studies of nuclear states. In a SCE reaction induced by a projectile *a* on a target *A*, a proton (neutron) of the target is converted into a neutron (proton), $\Delta Z_A = \pm 1$, $\Delta N_A = \mp 1$, keeping the mass number *A* unchanged, with opposite transition simultaneously occurring in the projectile, $\Delta Z_a = \mp 1$, $\Delta N_a = \pm 1$. In the isospin representation, SCE reactions probe the isovector excitations generated, at two-body level, by $\tau_{a\pm}\tau_{A\mp}$ combination of the isospin rising and lowering operators acting on a nucleon in the projectile *a* and the target *A*, respectively. After the first pioneering explorations [36], [37], the study of SCE reactions was soon extended to transitions associated to spin degrees of freedom [38], [39]. In particular the monopole component $\Delta L = 0$ has attracted special interest, since the associated $\sigma\tau$ operator is analogous to the Gamow-Teller (GT) one acting in the spin transferring *β*-decay. Important results have been obtained for example at TRIUMF, IUCF, LAMPF and other laboratories [40], [41], [42], [43], [44], [45]. In addition, a similar operator drives magnetic dipole (M1) transitions in γ-decay and (e,e') inelastic scattering. In the years, a wealth of studies of SCE reactions has been reported. Excellent reviews of the early activities can be found in key articles by F. Osterfeld [46] for the theoretical aspects and by W. P. Alford and B.M. Spicer [47] for a survey of the experimental explorations. Also important is the paper by T. N. Taddeucci et al. [48], which proposed a useful factorization of the CE cross section. More recently updated reviews of the field are found in Refs. [49], [50], [51]. Here we focus the attention on heavy-ion induced SCE, not much discussed previously.

SCE reactions are induced by the strong interaction, mediated by the exchange of isovector mesons, the lightest of which are the pions *π*. For momentum transfer sensibly smaller than the *π* mass, the meson form factors do not influence appreciably the SCE dynamics and a simpler description in terms of smoothly energy dependent coupling factors is possible. This is similar to the weak interaction where constant coupling factors $g_v$ and $g_a$ scale the spin-isospin operators. In this way, the analogy between $\tau$ (Fermi) and $\sigma\tau$ (GT) operators of the strong and weak interactions becomes closer. As a consequence, SCE reactions offer the opportunity to complement *β*-decay studies of the nuclear response to isovector probes. The best example is the study of isovector monopole nuclear response, ($\Delta J^\pi=0^+$, $\Delta L=0$; $\Delta\sigma=0$; $\Delta\tau=1$) for the Fermi and ($\Delta J^\pi=1^+$, $\Delta L=0$; $\Delta\sigma=1$; $\Delta\tau=1$) for the GT, which is intrinsically limited to a reduced energy window accessible by β-decay, but not for SCE reactions. Due to the isospin symmetry, the Fermi response is concentrated in a unique transition, named Isobaric Analogue State (IAS) [52], which practically exhausts the model-independent sum rule for the Fermi operator, making the study of this excitation mode not very distinctive for nuclear structure purposes. On the other hand, since the $\sigma\tau$ is not a symmetry for nuclear systems, the associated GT strength is spread over many broadly distributed states as a function of the excitation energy around the Gamow-Teller Resonance (GTR) [53], [54]. The exact GT distribution is a unique property of each nucleus, reflecting in a detailed way its peculiar many-body aspects. For that reason, the exploration of GT strength has soon gained a central relevance in the development of nuclear physics. A relevant finding is that only part of the strength (from about 50% to 70%) predicted by the model-independent sum rule for GT [55], [46] is found in the experiments [56], at least in the region of the GTR or even up to about 50 MeV excitation energy. Beyond this limit, it is hard to extract the monopole strength from the experiments with the necessary accuracy. In addition, the GT strengths extracted from measured cross sections of isolated transitions are typically smaller than shell model predictions and a quenching factor of about 0.7 is needed to reproduce the data. The problems of the missing overall GT strength and of the quenching for individual GT states have been deeply investigated in the past (see Ref. [47] for a detailed discussion). The off-shell excitation of nucleons to the $\Delta$-resonance ($M_\Delta$=1232 MeV) is considered a possible mechanism that pushes the GT strength at excitation energies higher than those experimentally accessed in typical analyses of SCE reaction data. In addition, the coupling of the one-particle-one-hole (1p-1h) GT modes with 2p-2h and higher order correlations is another mechanism proposed for the observed behavior of GT strength, even if in this case such an effect should be observed for Fermi transitions and for other multipolarities.

    An important aspect of SCE reactions, when used for investigation of GT modes in nuclei, is that the momentum transfer should be kept as small as possible in order to filter out $\Delta L \neq 0$ components in the collision or easily distinguished in the data analysis. This also ensures that the tensor components of the isovector nucleon-nucleon interaction ($\Delta J^\pi=1^+$, $\Delta L=2$; $\Delta\sigma=1$; $\Delta\tau=1$) have a small impact on the observed $\Delta J^\pi=1^+$ strength. Such condition is best matched when the incident

energy is typically above 100 MeV/u and the scattering angle is close to zero degree. Following this strategy the measured cross sections for (n,p) and (p,n) reactions at energies above 100 MeV were found proportional to known $\beta^+$ and $\beta^-$ strengths, respectively, even if the achieved experimental resolution does not allow to separate all the GT states in the energy spectra, somewhat reducing the sensitivity of these experimental tools. Complementary results have been achieved by SCE reactions induced by heavier projectiles, such as the (d,$^2$He), (t,$^3$He), ($^7$Li,$^7$Be) ($^{12}$C,$^{12}$N) ($^{18}$O,$^{18}$F) for the $\beta^+$-like target transitions, or the ($^3$He,t), ($^{12}$C,$^{12}$B) for the $\beta^-$-like class. In general, the F and GT modes are not separated in the projectile transition, unless the projectile is spin-less ($J_a^\pi = 0^+$) as in the case of ($^{12}$C,$^{12}$B), ($^{12}$C,$^{12}$N) or ($^{18}$O,$^{18}$F). For the (n,p), (p,n) or ($^3$He,t) reactions both F and GT modes are possible and can be separated only if proper selection rules hold for the target transitions, as in the case of transition induced in $J_A^\pi = 0^+$ even-even targets. Alternatively, the F component in the SCE transitions should be minimized. This condition again is better matched at energies of about 100-200 MeV/u, where the volume integral of the free στ nucleon-nucleon interaction is sensibly larger than the τ component. Since the average GT and F contribution to the SCE cross section scale approximately with the square of the volume integrals one finds that at these energies GT studies are accurate enough even for projectiles with $J_a^\pi \neq 0^+$.

From the experimental side, state-of-art results have been obtained by the ($^3$He,t) reaction performed at 140 MeV/u with the Grand Raiden magnetic spectrometer of RCNP in Osaka [57], [58], [59]. The zero degree mode available for the spectrometer and the high energy resolution (FWHM typically ~ 25 keV), achieved thanks to the implemented energy and angle dispersion matching technique [57], [60] are the key peculiarities of this facility. A remarkable proportionality (better than 5%) between measured cross sections and known $\beta^-$ strengths have been reported as a general finding, at least for not suppressed transitions, for a large number of states in many targets. As a consequence, the RCNP facility has become the ideal place for high resolution GT studies. For the $\beta^+$ transitions remarkable results have been obtained by the (d,$^2$He) reactions at KVI and RIKEN laboratories [61], [62], [63], [64], [65], [51]. Experimentally, the high efficient detection of the two protons decaying from $^2$He has allowed to get an overall energy resolution of about 100 keV in the missing mass spectra. About 100 MeV/u bombarding energy was chosen, as discussed above, and the center-of-mass detection angle for the $^2$He system was around zero degrees. Again a close proportionality between nuclear matrix elements extracted from SCE cross sections and those extracted from $\beta^+$ and EC studies was found.

An interesting application of high-resolution ($^3$He,t) and (d,$^2$He) reactions is to map the GT response of specific nuclei, which are intermediate systems in known two-neutrino double beta decays (*2νββ*). The GT response in the intermediate system is separately explored from the parent and the daughter side. Among the many $1^+$ states populated in the two reactions, it is possible to infer what states give relevant contribution to the *2νββ*, as those which are significantly populated in both SCE processes. A drawback of this technique is that only the transition probabilities to individual $1^+$ states are extracted from the experiments for each step, while in the *2νββ* the amplitudes are needed with the proper phase, since they add coherently. A simple case is obtained when a single $1^+$ state is found to be dominant in the intermediate state, since in this case no coherent sum is needed. Approximate schemes have also been proposed for $1^+$ transitions close to the Fermi level [66].

When the study is extended using the heavier projectiles, a typical problem is their complex many body nature, for the SCE cross section analyses. The projectile-target potential needs to be described with high accuracy both in the entrance (Initial State Interaction, ISI) and the outgoing (Final State Interaction, FSI) channel. In this case, the quasi-elastic SCE reactions are localized in the nuclear surfaces of the colliding systems, due to the strong absorption of the incoming waves in the inner part of heavy nuclei. This aspect of the heavy-ion reaction mechanism is crucial, since it allows to convert the full many-body reaction problem into a much simpler one, where direct reactions as SCE can be treated as small perturbations of the direct elastic scattering, which is described by an appropriate nucleus-nucleus average optical potential. Modern techniques to build ISI and FSI potentials by double folding integrals of the nucleon-nucleon interaction with the densities of the colliding systems have proven to be accurate enough for this purpose [67] [68] [69] [70] [71], especially when elastic scattering data of the projectile-target system are available at the same energy of the SCE reaction cross sections. In this way, the SCE reaction matrix elements can be directly extracted from the experimental cross sections and connected to the nuclear response to two-body operators, as those discussed above for F and GT cases. However, other quasi-elastic mechanisms in the projectile-target collision are in principle allowed. For example, multi-nucleon transfer reactions, where the colliding partners exchange nucleons, could have a non-negligible contribution to SCE channel. In particular, the transfer of a proton/neutron from the projectile to the target (stripping process) followed by the transfer to the projectile of a neutron/proton from the target (pick-up process) is a two-step mechanism which feeds the same outgoing channel as the direct one-step SCE reaction induced by two-body nucleon-nucleon interaction. The two-step mechanism is sensitive to the nucleon-nucleus mean field potential and cannot probe the nucleon-nucleon interactions which originates the F and GT response of nuclei. This is an obstacle that should be taken into account, especially in heavy-ion induced SCE reactions, and can be possibly minimized by an appropriate choice of the experimental conditions. From the theory point of view this problem has been extensively debated in the past [72], [73], [74], [75], [76], [77], [78], [79] with major advances achieved thanks to the development of microscopic approaches for the data analysis. As a general finding, the two-step mechanisms tend to be small at incident energies far above the Coulomb barrier. This has been reported in ($^{12}$C,$^{12}$B) [80], ($^{12}$C,$^{12}$N), ($^{13}$C,$^{13}$N) [81] and in ($^{7}$Li,$^{7}$Be) reactions [77], [82], [83], [84], [85], [78], [86], [87], [88] explored at different energies from 5 to 70 MeV/u and on different targets. In references [78], [89] it was shown that quantitative information on GT matrix elements can be extracted from ($^{7}$Li,$^{7}$Be$_{gs}$(3/2$^-$)) and ($^{7}$Li,$^{7}$Be$_{0.43}$ MeV(1/2$^-$)) measured cross sections for isolated transitions. The results, obtained at about 8 MeV/u bombarding energy for light neutron rich nuclei as $^{11}$Be, $^{12}$B, $^{15}$C and $^{19}$O, indicate that a good accuracy (better than 10%) is achieved, providing that a fully consistent microscopic approach is used for the ISI, FSI and the reaction form factors.

An interesting aspect of heavy-ion induced SCE reactions is that a significant amount of linear momentum is available during the collision and it is transferred to the final asymptotic state, even at forward angles. This feature is normally considered a drawback of heavy-ion induced SCE reactions, as the typical focus is in studying the *L*=0 modes, namely the GT one. However, this property is interesting since neither *β*-decay nor light ions induced CE reactions can effectively probe the nuclear response to the higher multipoles of the isospin (F-like) and spin-isospin (GT-like) operators. Nowadays much interest is given to this aspect of nuclear response for its implications in *0νββ* decay matrix elements [90], [91] where high order multipoles are considered to give a major contribution [92]. Thus, the exploration of heavy-ion induced SCE reactions has recently regained a great interest, with the consequent need to develop suitable experimental techniques and advanced theoretical analysis for a detailed description of the data (see Section 5).

## 2.2 Heavy-Ion induced Double Charge Exchange Reactions

A Double Charge Exchange (DCE) reaction is a process induced by a projectile *a* on a target *A*, in which two protons (neutrons) of the target are converted in two neutrons (protons), $\Delta Z_A = \pm 2$, $\Delta N_A = \mp 2$, being the mass number *A* unchanged, with opposite transition simultaneously occurring in the projectile, $\Delta Z_a = \mp 2$, $\Delta N_a = \pm 2$. In the isospin representation, DCE reactions probe the double isovector excitations generated, at four-body level, by $\tau_{a\pm}\tau_{a\pm}\tau_{A\mp}\tau_{A\mp}$ combination of the isospin rising and lowering operators acting on two nucleons in the projectile *a* and the target *A*, respectively. If we limit to only the target excitations, DCE transitions can also occur as result of $(\pi^+,\pi^-)$ or $(\pi^-,\pi^+)$ reactions or $\beta\beta$-decays, the latter allowed only for positive *Q*-value.

Similarly to SCE reactions, DCE probe nuclear response to the isospin degree of freedom, despite here the second order effects are selected. It is useful to recall the main features of known nuclear processes connected to second order isospin operators. *2νββ*-decays, induced by the heavy gauge bosons of the weak interaction, are sensitive to the nuclear response to a sequence of two GT operators acting independently and probing the low momentum component of nuclear wave functions. *0νββ*-decays, which are also induced by the weak interaction, are connected to the nuclear response to two-body isospin operators in a broad range of momenta distributed around 0.5 $fm^{-1}$ and consequently in a wide range of multipolarities [92]. Pion-induced DCE reactions require the isospin components of the strong interaction acting twice. At a nucleonic level, two independent nucleons interact sequentially with the $\pi$ fields. In the first step, the charged incident pion is converted to a neutral one as follows n($\pi^+,\pi^0$)p; in the second step the neutral pion is converted to a charged one as follows n($\pi^0,\pi^-$)p. A similar sequence occurs for DCE induced by negative pions according to the following reaction chain p($\pi^-,\pi^0$)n followed by p($\pi^0,\pi^+$)n. Due to the spin-less nature of pions spin-isospin nuclear responses are not directly accessed and thus are difficult to observe. Extensive studies of $(\pi^+,\pi^-)$ were performed in the 80's [22], [23], [93] leading to the observation of second order collective excitations as the Double Isobaric Analogue State (DIAS) or the Isobaric Analogue State built on the top of the Giant Dipole Resonance (GDR-IAS). Instead, no Double Gamow-Teller (DGT) was observed, maybe due to the above mentioned weak sensitivity to spin modes for this probe.

An important feature of DCE reactions induced by nuclear collisions is that no light projectiles can be practically used. The lightest projectiles allowed are tritons or $^3$He, and even in these cases the (t,3p) or the ($^3$He,3n) reactions are very challenging from the experimental point of view and, to our knowledge, never explored. Also moving to heavier projectiles the experiments become rather demanding. First pioneering explorations of the heavy-ion induced DCE reactions are the ($^{18}$O,$^{18}$Ne), ($^{18}$O,$^{18}$C) and ($^{14}$C,$^{14}$O) reactions, which were performed at Berkeley, NSCL-MSU, IPN-Orsay, ANU-Pelletron, Los Alamos laboratories [94], [25], [95], [26], [27] at energies above the Coulomb barrier. The main purpose was to determine the mass of neutron rich isotopes by reaction Q-value measurements. However, these experiments were not conclusive for deeper spectroscopic investigations, mainly due to the poor statistical significance of the few DCE observed events; hence, no other experiments were proposed. Also the theory, which was initiated to study the DCE reaction mechanism [28], [29], soon followed the trend and the field was abandoned for a long time.

In the recent years, major interest has raised for DCE studies, especially because of their possible connection to $\beta\beta$-decays. New reactions have been considered, such as the ($^8$He,$^8$Be) [96], the ($^{11}$B,$^{11}$Li) [97] or the ($^{12}$C,$^{12}$Be) [98], explored at RIKEN and RCNP at energies between 80 and 200 MeV/u. The ($^8$He,$^8$Be) was used to search for the tetra-neutron (4n) system by the $^4$He($^8$He,$^8$Be)4n at 186 MeV/u [96]. The ($^{11}$B,$^{11}$Li) and the ($^{12}$C,$^{12}$Be) were investigated with the main goal to find the DGT resonance and provide quantitative information about the DGT sum-rule, important for modern nuclear structure theories [99]. Another new DCE reaction, ($^{20}$Ne,$^{20}$O) have been introduced by us, with the aim to probe $\beta\beta$-like nuclear response. Preliminary results of this reaction will be introduced in Section 4.5. In addition to that, important results have been recently achieved by the renewed use of the ($^{18}$O,$^{18}$Ne) reaction in upgraded experimental conditions [31], [32]. In reference [31] the $^{40}$Ca($^{18}$O,$^{18}$Ne)$^{40}$Ar was studied at 15 MeV/u at the MAGNEX facility of the INFN-LNS [100], showing that high mass, angular and energy resolution energy spectra and accurate absolute cross sections are at our reach, even at very forward angles. In addition, a schematic analysis of the reaction cross sections demonstrated that relevant quantitative information on DCE matrix elements can be extracted from the data.

In analogy to the case of heavy-ion induced SCE reactions, an important issue for the DCE is to quantify the contribution coming from multi-nucleon transfer reactions. In this case the effects start from the 4th order in the nucleon-nucleon potential since two protons (neutrons) should be stripped from the projectile and two neutrons (protons) picked-up from the target. In Ref. [31] it was shown that, under the experimental conditions set for the experiment at INFN-LNS, the contribution of multi-nucleon transfer was negligible (less than 1%). Similar results are found in the preliminary analysis of the other explored cases. Consequently, the leading DCE reaction mechanism is connected to nucleon-nucleon isovector interaction, which acts between two neutrons (protons) in the projectile and two protons (neutrons) in the target for the ($^{18}$O, $^{18}$Ne) and the ($^{20}$Ne, $^{20}$O) reactions, respectively. A useful way to consider the DCE process is by means of the exchange of two charged $\pi$ or $\rho$ mesons between the involved nucleons. An interesting question is whether the two mesons are exchanged independently of each other in analogy to *2νββ*-decays or in a correlated way, as in the *0νββ*-decays. This last question is quite interesting for the connection of DCE reactions to *0νββ*-decays. This aspect is also important from the point of view of nuclear reaction theory, since it could indicate a new way to access nucleon-nucleon short-range correlations (see Section 5).

## 2.3  DCE reactions and *0νββ* decays

The availability for the first time of valuable data on DCE reactions raises the question whether they can be used toward the experimental access to *0νββ* decay NMEs. Although the DCE and *0νββ* decay processes are mediated by different interactions, there are a number of important similarities among them:

- Parent/daughter states of the *0νββ* decay are the same as those of the target/residual nuclei in the DCE;
- Short-range Fermi, Gamow-Teller and rank-2 tensor components are present in both the transition operators, with relative weight depending on incident energy in DCE. Performing the DCE experiments at different bombarding energies could give sensitivity to the individual contribution of each component;
- A large linear momentum (~100 MeV/c) is available in the virtual intermediate channel in both processes [11]. This is a distinctive similarity since other processes such as single *β* decay, *2νββ* decay, ligh-ion induced SCE cannot probe this feature [101]. An interesting development is the recently proposed μ-capture experiments at RCNP [102];

- The two processes are non-local and are characterized by two vertices localized in a pair of valence nucleons;

- Both processes take place in the same nuclear medium. In-medium effects are expected to be present in both cases, so DCE data could give a valuable constraint on the theoretical determination of quenching phenomena on $0\nu\beta\beta$. One should mention, for example, that in single $\beta$ decay, $2\nu\beta\beta$ decay [4] and SCE reactions [47], the limited model space used in the calculations and the contribution of non-nucleonic degrees of freedom and other correlations require a renormalization of the coupling constants in the spin-isospin channel. However, an accurate description of quenching has not yet been fully established and other aspects of the problem can give important contributions [103];

- An off-shell propagation through virtual intermediate channels is present in the two cases. The virtual states do not represent the asymptotic channels of the reaction and their energies can be different from those (measurable) at stationary conditions [104]. In practice, a supplementary contribution of several MeV to the line width is present in the intermediate virtual states. This is related to the transit time of a particle (neutrino in one case and pair of nucleons in the other) along the distance of the two vertices of the $0\nu\beta\beta$ decay and DCE processes. The situation is very different in SCE reactions, where the intermediate states of $0\nu\beta\beta$ decay are populated as stationary ones and in $2\nu\beta\beta$ decay, where the neutrinos and electrons are projected out from the nucleus. No effective broadening of the line width is thus probed in SCE and $2\nu\beta\beta$ decay.

The descriptions of NMEs for DCE and $0\nu\beta\beta$ decay present the same degree of complexity, with the advantage for DCE to be "accessible" in laboratory. In Refs. [105] and [106] such analogy have been investigated and a good linear correlation between double GT transitions to the ground state of the final nucleus and $0\nu\beta\beta$ decay NMEs is reported for pf-shell nuclei. However, a simple relation between DCE cross sections and $\beta\beta$-decay half-lives is not trivial and needs to be explored.

# 3   The NUMEN Project

NUMEN proposes to access the nuclear matrix elements entering the expression of the life time of the $0\nu\beta\beta$ decay by measuring cross sections of DCE reactions in a wide range of incident energies. The project stems out as a natural evolution of the successful pioneering investigation of the $^{40}$Ca($^{18}$O,$^{18}$Ne)$^{40}$Ar DCE reaction performed at INFN–LNS [31].

A key aspect is the use of the K800 Superconducting Cyclotron for the acceleration of the required high resolution and low emittance heavy-ion beams and of the MAGNEX large acceptance magnetic spectrometer for the detection of the ejectiles. The use of the powerful techniques for particle identification and high-order trajectory reconstruction, implemented in MAGNEX (described in Section 4.5), allows to reach the challenging sensitivity and resolution required to measure DCE reactions, characterized by very low cross section over a large background coming from other reaction channels. The INFN-LNS set-up is today ideal for this research in worldwide context. For some cases, described in this article, the measured quantities can be accessible with the present facility. However, a main limitation on the beam current delivered by the accelerator and on the maximum rate accepted by the MAGNEX focal plane detector must be sensibly overcome in order to systematically provide accurate nuclear structure information to the neutrino physics community in all the studied cases. The upgrade of the INFN-LNS facilities in this view is part of this project.

## 3.1   The NUMEN goals

The experimental approach toward the determination of $0\nu\beta\beta$ decay NMEs is one of the main goals of our project. For that, we need to test if the DCE measured cross sections and in turn DCE matrix elements are connected to $0\nu\beta\beta$ decay NMEs as a smooth and thus controllable function of the projectile energy $E_p$ and of the mass of the system $A$. If the effort is successful, then the result will provide a new experimental approach to extract NMEs for $0\nu\beta\beta$ decay. This implies an accurate description of the reaction mechanism, factorized in a reaction part and a nuclear structure part, the latter factorized in a projectile and target matrix elements. The development of a consistent microscopic description of the DCE reaction and the nuclear structure part is essential to explore this opportunity. The use of the quantum approach for the Distorted Wave Born Approximation (DWBA) or Coupled Reaction Channel (CRC) cross sections with form factors including transition densities from state-of-art nuclear structure approaches is a suitable framework in which this theory can be developed. Experimentally, the achievement of this first goal requires to build up a systematic set of appropriate data, facing the relative experimental challenges connected with the low cross sections, the high sensitivity and the requirement of high resolutions.

The measurement of the DCE absolute cross sections that NUMEN wish to provide could have a major impact for tuning the nuclear structure theories of *0νββ* decay NMEs. This can be considered an important additional goal of the project, achievable in a short term. As mentioned in Section 2.3, the NMEs for DCE and *0νββ* decay probe the same initial and final wave functions by operators with similar structure. Consequently, the measured DCE absolute cross sections allows to test the validity of the assumptions done for the unavoidable truncation of the many-body wave functions. The reaction part needs to be precisely controlled to this purpose, a result that NUMEN aims to pursue within a fully quantum scattering framework. Once the nuclear wave functions have been tested by DCE cross sections, the same can be used for *0νββ* decay NMEs. Promoting the development of these kinds of DCE constrained theories for the NME of the *0νββ* decay is thus an important goal that NUMEN can achieve even with a reduced experimental dataset and without assuming cross section factorization.

Finally, another goal is to provide relative NME information on the different candidate isotopes of interest for the *0νββ* decay. The ratio of the measured cross sections can give a model independent way to compare the sensitivity of different half-life experiments. This result can be achieved even in presence of sizeable systematic errors in the measured cross sections and in the extraction of DCE matrix elements, as they are largely reduced in the ratio. Performing these comparative analyses could have strong impact in the future developments of the field, especially in a scenario were fundamental choices for the best isotope candidates for *0νββ* decay need to be made.

## *3.2 The phases of the NUMEN project*

The NUMEN project is conceived in a long-range time perspective, planning to perform a comprehensive study of many candidate systems for *0νββ* decay. Moreover, this project promotes and is strictly connected with a renewal of the INFN-LNS research infrastructure and with a specific R&D activity on detectors, materials and instrumentation, as described in the next subsections. Consequently, other research activities are likely to benefit from such upgrades. NUMEN is divided into the following four phases, each one delimited by a starting point and defined by the fulfilment of an intermediate goal, which is necessary for the development of the successive phase.

*3.2.1   Phase 1: "The pilot experiment"*

In 2013, the $^{40}$Ca($^{18}$O,$^{18}$Ne)$^{40}$Ar DCE reaction was measured at the INFN-LNS laboratory together with the competing processes: $^{40}$Ca($^{18}$O,$^{18}$F)$^{40}$K SCE, $^{40}$Ca($^{18}$O,$^{20}$Ne)$^{38}$Ar two-proton (2p) transfer and $^{40}$Ca($^{18}$O,$^{16}$O)$^{42}$Ca two-neutron (2n) transfer [31]. A beam of $^{18}$O$^{4+}$ ions, extracted by the K800 Superconducting Cyclotron accelerator, bombarded a 280±30 µg/cm$^2$ Ca target, at 15 MeV/u incident energy. A total charge of 3.6 mC was integrated by a Faraday cup, placed downstream the target. The ejectiles produced in the collisions were momentum-analysed by MAGNEX [100], [107] and detected by its focal plane detector [108], [109]. An angular range of -1.2° < $\theta_{lab}$ < +8° in the laboratory frame was explored, corresponding to scattering angles in the center of mass 0° < $\theta_{CM}$ < 12°. The ejectiles identification was achieved as described in Refs. [110], [111]. The positions and angles of the selected ions measured at the focal plane were used as input for a 10$^{th}$ order ray-reconstruction of the scattering angle $\theta_{CM}$ and of the excitation energy $E_x = Q_0 - Q$ (where $Q_0$ is the ground-to-ground state reaction *Q*-value) [112], [113], [114]. An energy resolution of ~500 keV (FWHM) was obtained similarly to Ref. [115]. The absolute cross section was extracted from measured yields according to Ref. [112]. A systematic error of ~20% was estimated from the uncertainty in the target thickness and beam collection. The measured energy spectra and angular distribution for the ground state to ground state transition are published in Ref. [31].

This work showed for the first time high resolution and statistically significant experimental data on DCE reactions in a wide range of transferred momenta. The measured cross-section angular distribution is characterized by a clear oscillating pattern, remarkably described by an *L* = 0 Bessel function, indicating that a simple mechanism is dominant in the DCE reaction. This is confirmed by the observed suppression of the multi-nucleon transfer routes.

DCE matrix elements were extracted under the hypothesis of a two-step charge exchange process. Despite the approximations used in our model, which determine an uncertainty of ±50%, the obtained results are compatible with the values known from literature, signaling that the main physics content has been kept. This makes the ($^{18}$O,$^{18}$Ne) reaction very interesting to investigate the DCE response of the nuclei involved in *0νββ* research.

*3.2.2   Phase2: From the pilot experiment toward the "hot" cases*

The results of Phase 1 indicate that suitable information from DCE reactions can be extracted. The availability of the MAGNEX spectrometer for high resolution measurements of much suppressed reaction channels was essential for such a pioneering measurement. However, with the present set-up, it is difficult to suitably extend this research to the *"hot"* cases, where $\beta\beta$ decay studies are concentrated. We consider that in the $^{40}$Ca($^{18}$O,$^{18}$Ne)$^{40}$Ar experiment we have collected about 300 counts for the angle integrated transition $^{40}$Ca$_{g.s.}$ → $^{40}$Ar$_{g.s.}$. However, about one order of magnitude more yield would have been necessary for the reaction studied, especially at backward angles where large amounts of linear momentum (1-2 fm$^{-1}$) are available. Moreover:

- In the studied reaction, the *Q*-value was particularly favorable ($Q$ = -2.9 MeV), while in the DCE reactions involving candidate isotopes of interest for $0\nu\beta\beta$ the *Q*-values are more negative. A sensible reduction of the cross-section is thus expected in these cases, especially at very forward angles.

- The isotopes of interest are heavier than $^{40}$Ca, consequently the nucleus-nucleus potential in the initial and final state (ISI and FSI) are expected to be more absorptive with consequent further reduction of the cross section for direct reactions as DCE.

- The DCE cross section is expected to decrease at higher bombarding energies (at least in the energy range explored by NUMEN, i.e. 10 to 70 MeV/u) since both $\tau$ and $\sigma\tau$ components of the nucleon-nucleon effective potential show this trend. This aspect is particularly relevant considering that direct DCE cross section is sensitive to the 4$^{th}$ power of the potential strength.

- The ($^{18}$O,$^{18}$Ne) reaction, investigated in the pilot experiment, could be particularly advantageous, due to the large value of both the B[GT;$^{18}$O$_{gs}$(0+)→ $^{18}$F$_{gs}$(1+)] and B[GT;$^{18}$F$_{gs}$(1+)→ $^{18}$Ne$_{gs}$(0+)] strengths and to the concentration of the GT strength in the $^{18}$F(1$^{+}$) ground state. However, this reaction is of $\beta^{+}\beta^{+}$ kind, while most of the research on $0\nu\beta\beta$ decay is on the $\beta^{-}\beta^{-}$ side. None of the reactions of $\beta^{-}\beta^{-}$ kind looks like as favorable as the ($^{18}$O,$^{18}$Ne). For example the ($^{18}$Ne,$^{18}$O) requires a radioactive beam, which cannot be available with enough intensity. NUMEN proposes the ($^{20}$Ne,$^{20}$O) reaction, which has smaller B(GT), so a reduction of the yield could be foreseen in these cases.

- In some cases, e.g. $^{136}$Xe or $^{130}$Xe, gas or implanted target will be necessary, which are normally much thinner than solid state films obtained by evaporation or rolling technique, with a consequent reduction of the collected yield.

- The achieved energy resolution (typically about half MeV, see Sections 4.4) is not always enough to separate the ground from the excited states in the final nucleus (see Table 4.2). In these cases the coincident detection of gamma rays from the de-excitation of the populated states is necessary, but at the price of reducing the yield (see Section 6.7).

All of these considerations suggest that the beam current of the DCE experiments must be much increased. In particular, for a systematic study of the many "hot" cases of $\beta\beta$ decays, an upgraded set-up, able to work with a two or three orders of magnitude higher current than the present, is necessary. This goal can be achieved by a substantial change in the technologies used in the beam extraction and transport, in the target and in the detection of the ejectiles. For the accelerator, the use of a stripper-induced extraction is an adequate choice as presented in Section 6.1. For the targets, the development of radiation tolerant cooled systems is under study (see Section 6.4). For the spectrometer, the main foreseen upgrades are:

- The increase of the maximum accepted magnetic rigidity (see Section 6.2);
- The replacement of the present wire-based gas tracker with a new tracker system based on Micro Pattern Gas Detector (see Section 6.5);
- The replacement of the wall of silicon pad stopping detectors with a dedicated array of smaller size detectors based on radiation hard compliant technologies (see Section 6.6);
- The development of an array of detectors around the target for measuring the coincident gamma rays generated in the DCE reactions (see Section 6.7).

During the NUMEN Phase 2, the R&D activity necessary for the above mentioned upgrades is going to be carried out still preserving the access to the present facility. In the meanwhile, experiments with integrated charge of tens of mC (about one order of magnitude higher than that collected in the pilot experiment) are going to be performed. These require several weeks of data taking for each reaction, since thin targets (a few $10^{18}$ atoms/cm$^2$) are mandatory in order to achieve enough energy and angular resolution in the measured energy spectra and angular distributions. The attention is presently focused on a few favorable candidate cases for $\beta\beta$ decay, as discussed below, with the goal to achieve conclusive results for them. In addition, during Phase 2 a deeper understanding of the main features which limit the experimental sensitivity, resolution and systematic errors is being pursued.

In this framework, we study the ($^{18}$O,$^{18}$Ne) reaction as a probe for the $\beta^+\beta^+$ transitions and the ($^{20}$Ne,$^{20}$O), or alternatively the ($^{12}$C,$^{12}$Be), for the $\beta^-\beta^-$, with the aim to explore the DCE mechanism in both directions. Since NMEs are time invariant quantities, they are common to a DCE and to its inverse, so the contextual measurements of $\beta^+\beta^+$ and $\beta^-\beta^-$ reactions represent a useful test bench of the procedure to extract NME from the measured DCE cross section.

The choice of the target isotopes is a result of a compromise between the interest of the scientific community to specific isotopes and related technical issues. In particular, the possibility to separate g.s. to g.s. transition in the DCE measured energy spectra and the availability of thin uniform target of isotopically enriched material was considered. We started by selecting two systems, the $^{116}$Cd-$^{116}$Sn and $^{76}$Ge-$^{76}$Se pairs. For these nuclei the ground states are resolved from excited states by MAGNEX (being respectively 562 keV for $^{76}$Ge, 559 keV for the $^{76}$Se, 1.29 MeV for $^{116}$Sn and 513 keV for $^{116}$Cd) for both ($^{18}$O,$^{18}$Ne) and ($^{20}$Ne,$^{20}$O) reactions. In addition, the production technologies of the thin targets are already available at INFN-LNS. We are also exploring the $^{130}$Te($^{20}$Ne,$^{20}$O)$^{130}$Xe reaction and are planning to study $^{106}$Cd by ($^{18}$O,$^{18}$Ne). For each measured system, the complete net of reactions involving the multi-step transfer processes, characterized by the same initial and final nuclei, as shown in Fig. 3.1, are studied under the same experimental conditions.

During the Phase 2, the data reduction strategy is going to be optimized and the link with the theoretical physics strengthened, especially in the view of the construction of a "universal" framework, where $\beta\beta$-decay and DCE reactions are coherently analyzed (see Section 5).

The experimental activity of NUMEN Phase 2 and the analysis of the results is the main aspect of the NURE project [116] recently awarded by the European Reseach Council. The synergy between the two projects is an added value which significantly enhance the discovery potential already achieved in NUMEN Phase 2.

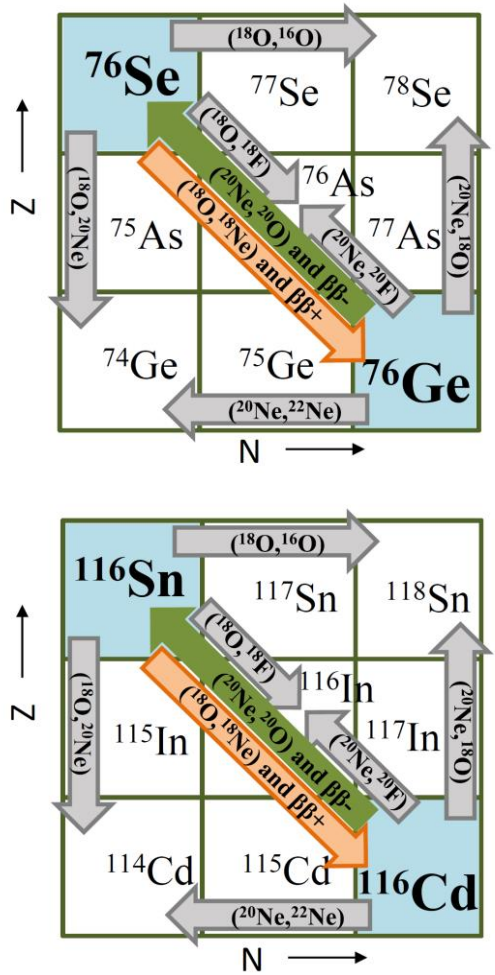

Figure 3.1 Scheme of the complete net of processes studied in the case of the $^{116}$Cd – $^{116}$Sn and $^{76}$Ge – $^{76}$Se pairs of nuclei of interest for the ββ decay. Inside the arrows, the reaction used to populate the final nuclei is indicated.

### 3.2.3 *Phase 3: The facility upgrade*

Once all the building blocks for the upgrade of the whole facility will be ready at the INFN-LNS, the NUMEN Phase 3, will incorporate the disassembling of the old set-up and re-assembling of the new will start. An estimate of about 18-24 months is evaluated. During this period, the data analysis of the NUMEN Phase 2 experiments will continue. In addition, tests of the new detectors and selected experiments will be performed with Tandem beams at INFN-LNS and in other laboratories in order to provide possible pieces of still missing information in the explored reactions network.

*3.2.4  Phase 4: The experimental campaign with upgraded facility*

The NUMEN Phase 4 will consist of a series of experimental campaigns at high beam intensities (some pµA) and integrated charge of hundreds of mC up to C, for the experiments in which γ-coincidence measurements are required, spanning all the variety of *0νββ* decay candidate isotopes of interest, like: $^{48}$Ca, $^{76}$Ge, $^{76}$Se, $^{82}$Se, $^{96}$Zr, $^{100}$Mo, $^{106}$Cd, $^{110}$Pd, $^{116}$Cd, $^{110}$Sn, $^{124}$Sn, $^{128}$Te, $^{130}$Te, $^{136}$Xe, $^{130}$Xe, $^{148}$Nd, $^{150}$Nd, $^{154}$Sm, $^{160}$Gd, $^{198}$Pt.

Based on the know-how gained during the experimental activity of Phase 2, the Phase 4 will be devoted to determine the absolute DCE cross sections and their uncertainties. Hopefully, the use of upgraded theoretical analyses will give access to the challenging NMEs *0νββ* decay that is the ambitious goal of NUMEN.

# 4  NUMEN experiments

The NUMEN experimental activity with accelerated beams consists of two main classes of experiments, corresponding to the exploration of the two directions of isospin transfer $\tau^-\tau^-$ and $\tau^+\tau^+$, characteristic of $\beta^-\beta^-$ and $\beta^+\beta^+$ decays, respectively.

In particular, the $\beta^+\beta^+$ direction in the target is investigated using an $^{18}$O beam and measuring the ($^{18}$O,$^{18}$Ne) DCE induced transitions, together with the other reaction channels involving the same beam and target. Similarly, the $\beta^-\beta^-$ direction is explored via the ($^{20}$Ne,$^{20}$O) reaction, using a $^{20}$Ne beam and detecting the reaction products of the DCE channel along with other open channels characterized by the same projectile and target.

During Phase 2, we are performing explorative investigations of the two types of experiments, highlighting the strengths and the limiting aspects of the adopted technique and establishing the best working conditions.

## *4.1  The experimental apparatus*

The experiments are performed at INFN-Laboratori Nazionali del Sud (Italy), taking advantage of the high performing experimental facilities there installed, mainly constituted by the Superconducting Cyclotron and the MAGNEX magnetic spectrometer.

The K800 Superconducting Cyclotron (CS) accelerator provides the required beams, namely $^{18}$O and $^{20}$Ne, at energies ranging from 10 MeV/u to 80 MeV/u with high energy resolution (1/1000) [115] and low emittance (~$2\pi$ mm mr) [117]. The present limit of the cyclotron beam power (about 100 W), discussed in Section 6.1.1, is not an issue for the NUMEN Phase 2, since more stringent limitations come from the present detectors (as discussed later in this Section). It represents a major obstacle for Phase 4, where beam power of few kW are foreseen.

The MAGNEX magnetic spectrometer is a large acceptance optical device with a large aperture vertically focusing quadrupole lens followed by a horizontally bending magnet. A detailed description of MAGNEX is found in Ref. [100], [107]. In the present situation, the maximum magnetic rigidity achieved is ~1.8 Tm, corresponding to a maximum accepted energy of about 46 MeV/u for ($^{18}$O,$^{18}$Ne) experiment and 24 MeV/u for ($^{20}$Ne,$^{20}$O). A slight (~20%) increase in the field of the magnetic elements, corresponding to ~40% in accepted energy, is possible without major concerns on iron saturation, providing that more powerful power supplies are used (as discussed in Section 6.2).

MAGNEX was designed to investigate processes characterized by very low yields and allows the identification of heavy ions with high mass ($\delta A/A \sim 1/160$), angle ($\delta\theta \sim 0.2°$) and energy resolutions ($\delta E/E \sim 1/1000$), within a large solid angle ($\Omega \sim 50$ msr) and momentum range ($-14\% < \delta p/p < +10\%$). It also allows to measure at zero degrees, which is the most important domain to explore in the NUMEN research, thus creating ideal laboratory conditions. High-resolution measurements for quasi-elastic processes, characterized by differential cross-sections falling down to tens of nb/sr, were already performed by this setup [100], [118], [119], [120]. The crucial issue of MAGNEX is the implementation of the powerful technique of trajectory reconstruction, which allows solving the equation of motion of each detected particle to high order ($10^{th}$ order) [113], [121], [122], [123], [124]. In this way, an effective compensation of the high order aberrations induced by the large aperture of the magnetic elements is achieved. The use of the sophisticated data reduction approaches based on the differential algebra is a unique feature of MAGNEX, that have been developed in the years (see Section 4.5). This guarantees the above mentioned performances and its relevance in the worldwide scenario of heavy-ion physics [125] [126] [127] [128] [129].

The MAGNEX Focal Plane Detector (FPD) consists of a large (active volume 1360 mm × 200 mm × 96 mm) low-pressure gas-filled tracker followed by a wall of 60 silicon pad detectors to stop the particles and provide the acquisition trigger signal [130]. A schematic view of the present FPD is shown in Figure 4.1. A set of wire-based drift chambers ($DC_i$) measures the vertical positions ($Y_i$) and angles ($\varphi_i$) of the reaction ejectiles, while the induced charge distributions on a set of segmented pads allow to extract the horizontal positions ($X_i$) and angles ($\theta_i$). The energy loss measured by the multiplication wires and the residual energy at the silicon detectors are used for atomic number identification of the ions. The mass and charge identification is performed exploiting the relation between measured position and energy in the dispersive direction, as described in Section 4.5.2. The use of silicon detectors to measure the residual energy is crucial to allow a high resolution mass discrimination, avoiding time of flight measurements and consequently without the introduction of additional start detectors [111]. The performances of the present FPD are described in Ref. [108] and listed in Table 4.1. Recently, we performed an upgrade of the detector described in Ref. [108], keeping the same configuration but passing from 4 to 6 position sensitive drift chambers, in order to reduce cross-talk phenomena, increase the signal-to-noise ratio and improve the field uniformity in the multiplication region.

Table 4.1 Main characteristics of the MAGNEX focal plane detector. Resolution (FWHM) of the main parameters measured by the FPD. See Refs. [108], [111].

|  | Achieved results |
|---|---|
| Intrinsic energy loss resolution for $^{18}$O (single wire) | 6.3 % |
| Energy loss resolution | 4 % |
| Horizontal position resolution | 0.6 mm |
| Horizontal angle resolution | 5 mr |
| Vertical position resolution | 0.6 mm |
| Vertical angle resolution | 5 to 9 mr |

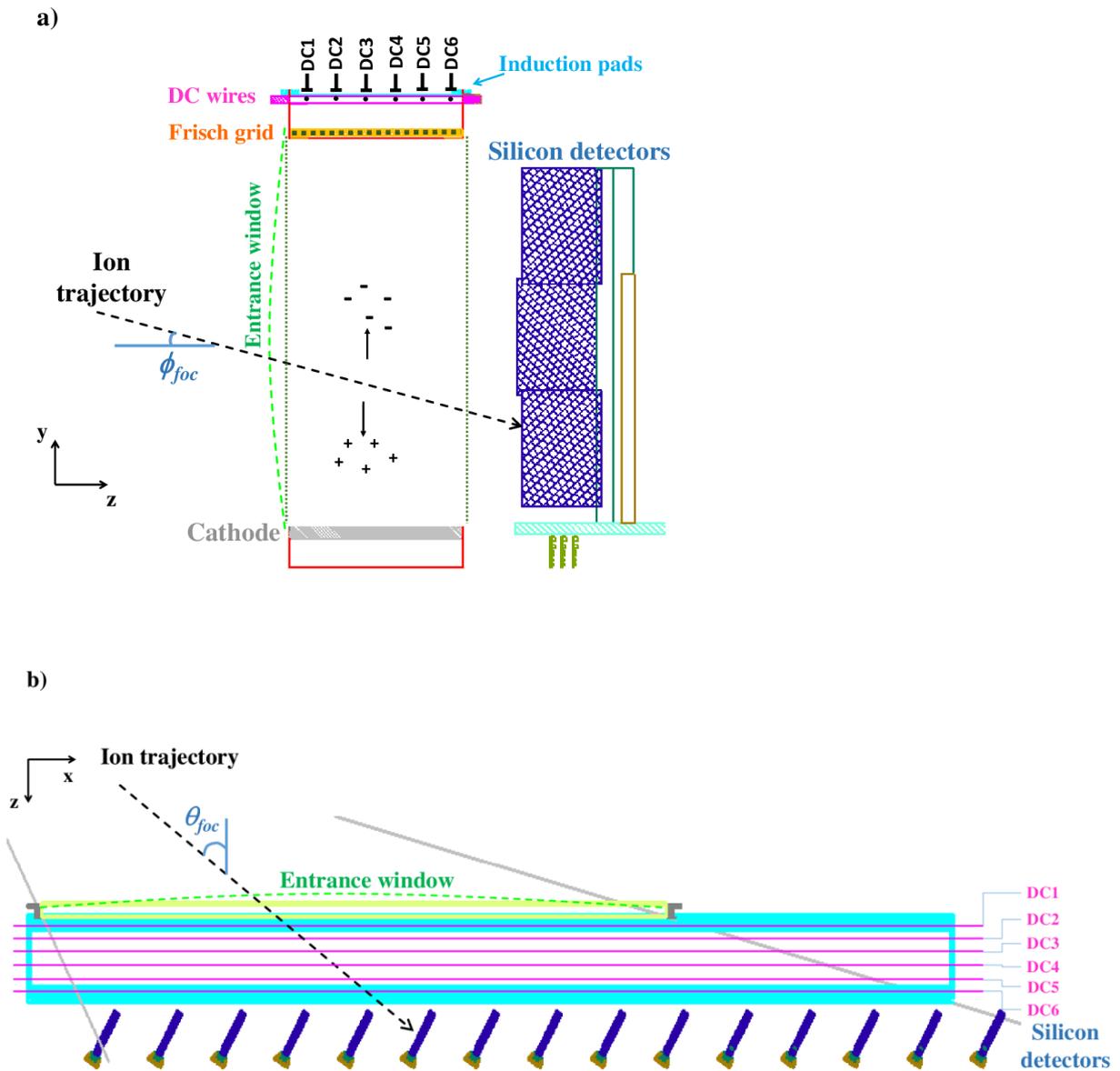

Figure 4.1 Schematic view of the Focal Plane Detector: a) side view; b) top view.

The present FPD, coupled with the MAGNEX spectrometer, is a detector able to discriminate from light to heavy ions with a 0.6% mass and charge resolution and a 2% atomic number resolution. The tracking measurement sensitivity guarantees an overall energy resolution of about 1/1000, which is close to the limit of the optics for the used beams (~1/1350 for a beam spot of 2 mm diameter [131]). However, the present FPD has an intrinsic limitation in the tolerable rate of few kHz of incident ions, due to the slow drift time of the positive ions in the drift chambers and the presence of long multiplication wires. This, in turns, reduces the acceptable beam current to few tens of enA for the NUMEN experiments. Such limitation will be overcome in the new design of the FPD planned within the NUMEN project (see Section 6.5).

A relevant issue regarding the present detection system is related to the radiation tolerance of the silicon detectors, which is of the order of $10^9$ particles/cm$^2$ in terms of fluency for the heavy ions to be detected in the NUMEN experiments. This issue has a moderate impact in the experimental activity of Phase 2, resulting in a still acceptable replacement rate of the silicon detectors, but would be not tolerable in the upgraded conditions (Phase 4). Thus a specific R&D activity is under study to get rid of this problem, as discussed in Section 6.6.

## *4.2 Experiments with $^{18}O$ beam ($β^+β^+$ direction)*

For the experiments of this class, the reaction channels where the main interests lie are listed below:

- Elastic and inelastic scattering ($^{18}O,^{18}O$)
- DCE reaction ($^{18}O,^{18}Ne$)
- SCE reaction ($^{18}O,^{18}F$)
- Two-proton pickup ($^{18}O,^{20}Ne$)
- One-proton pickup ($^{18}O,^{19}F$)
- Two-neutron stripping ($^{18}O, ^{16}O$)
- One-neutron stripping ($^{18}O,^{17}O$)

The purpose is to build a coherent set of experimentally driven information on nucleus-nucleus potentials and wave function of the projectile and the target, thus providing stringent constraints to the theoretical calculations.

One of the main challenges of such experiments is the measurement at very forward angles, including 0°. This is performed by placing the spectrometer with its optical axis at +4° with respect to the beam axis. Thanks to its large angular acceptance, an angular range $-1° < \theta_{lab} < 10°$ is covered. The MAGNEX quadrupole and dipole magnetic fields are set in order that the incident beam, after passing through the magnets, reaches a region which is beside the FPD but external to it (see simulations of the beam trajectory shown by the red line in

Figure 4.2).

In these experiments, the incident beam ($^{18}O^{8+}$) has magnetic rigidity ($B\rho$) higher than the ejectiles of interest. A full high-order simulation, including the tracking of the ion trajectories inside the magnetic field and the complete geometry of the spectrometer is performed to describe in each experiment the motion of the beam coming out from the target and of the emitted ejectiles along the spectrometer and downstream to the FPD region. A Faraday Cup designed to stop the beam and to measure the incident charge for each run is located in the high-$B\rho$ region besides the FPD as schematically shown by the red rectangle in

Figure 4.2 (high-$B\rho$ region).

## *4.3 Experiments with $^{20}$Ne beam ($\beta^-\beta^-$ direction)*

The reaction channels in the class of experiments with $^{20}$Ne$^{10+}$ beams are the following:
- Elastic and inelastic scattering ($^{20}$Ne,$^{20}$Ne)
- DCE reaction ($^{20}$Ne,$^{20}$O)
- SCE reaction ($^{20}$Ne,$^{20}$F)
- Two-proton stripping ($^{20}$Ne,$^{18}$O)
- One-proton stripping ($^{20}$Ne,$^{19}$F)
- Two-neutron pickup ($^{20}$Ne,$^{22}$Ne)
- One-neutron pickup ($^{20}$Ne,$^{21}$Ne)

For these experiments, the incident beam ($^{20}$Ne$^{10+}$) has a magnetic rigidity which is lower than the reaction ejectiles of interest. Thus, for a fixed magnetic field setting, the beam will be bent more than the ejectiles of interest.

The spectrometer optical axis is typically placed at -3° in order to cover a wide angular range including zero-degree. The quadrupole and dipole magnetic fields of MAGNEX are set in order that the $^{20}$Ne$^{10+}$ beam reaches the low-$B\rho$ region beside the FPD as shown by the green line in

Figure 4.2

Figure 4.2 (low-$B\rho$ region). A Faraday cup was designed, mounted and properly aligned to stop the beam, as schematically shown by the green rectangle in Figure 4.2.

A peculiarity of these experiments regards the treatment of the different charge states of the beam outgoing the target. The beam components characterized by charge states lower than $10^+$, namely $^{20}Ne^{9+}$ and $^{20}Ne^{8+}$, produced by the interaction of the beam with the electrons of the target material, have a magnetic rigidity which is similar to that of the ions of interest. Therefore, they enter in the FPD acceptance, generating a large background that requires the limitation of the beam intensity. Such low charge state components of the main beam have in fact an intensity of the order of $10^{-3}$ (for the $9^+$) and $10^{-5}$ (for the $8^+$), with respect to the $10^+$ beam. For example, for a typical beam of 10 enA and 15 MeV/u energy, the amount of $9^+$ and $8^+$ components at the focal plane is of the order of $10^7$ and $10^5$ pps, respectively. This is beyond the acceptable rate of the FPD. In addition, the elastic scattering on the target at forward angles by $^{20}Ne^{9+}$ and $^{20}Ne^{8+}$ beams also produces high counting rate at the focal plane.

In order to stop these unwanted $^{20}Ne$ $9^+$ and $8^+$ contaminants, two aluminum shields have been designed and mounted upstream the sensitive region of the focal plane detector. The shields act on a limited phase space region stopping mainly the $9^+$ and $8^+$ beams and the elastic scattering at very forward angles, while they do not interfere with the reaction channels of interest. Since the latter are the main contribution to the background in these experiments, a way to minimize the amount of $^{20}Ne^{9+}$ and $^{20}Ne^{8+}$ is under exploration. As it is known, the charge state distribution of a heavy-ion beam after crossing a material depends on the bombarding energy and on the chemical composition of the target [132], [133], [134]. The relevant targets for NUMEN generate an unwanted charge distribution that can be conveniently changed, minimizing the amount of $^{20}Ne^{9+}$ and $^{20}Ne^{8+}$, by adding an appropriate second target (post-stripper) downstream of the isotopic one. Recently a study of different materials to be used as post-stripper has been performed. A beam of $^{20}Ne^{10+}$ at 15 MeV/u incident energy extracted by the Superconducting Cyclotron was used to bombard different post-stripper foils positioned downstream of a $^{197}Au$ target. For each post-stripper configuration, three runs were performed with different magnetic field settings in order to accept the $^{20}Ne^{10+}$, $^{20}Ne^{9+}$ and $^{20}Ne^{8+}$ ejectiles and estimate the ratios between different charge states products.

A preliminary analysis of the acquired data show that material containing Carbon and Hydrogen atoms are the most efficient to reduce the lower charge state contributions and so the most promising in this view. In particular, we have observed about two orders of magnitude reduction of the unwanted $^{20}Ne^{8+}$ and about a factor 8 for the $^{20}Ne^{9+}$ when we add a Carbon post-stripper, compared to the Au target alone. The analysis of the data is presently on going.

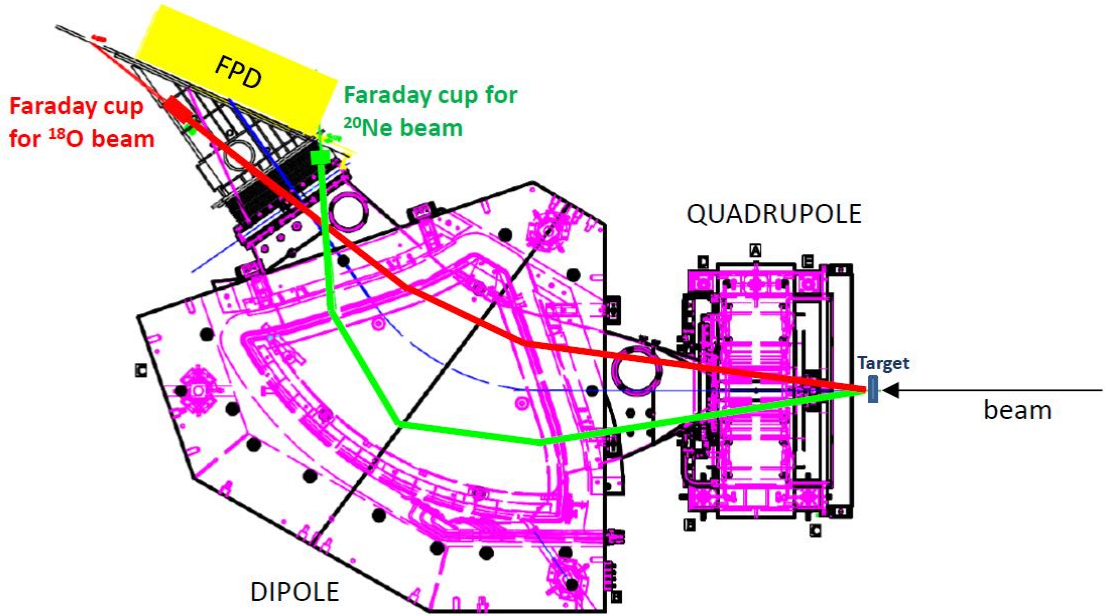

Figure 4.2 Layout of the MAGNEX spectrometer. The green and red lines represent the typical trajectories of the $^{20}Ne^{10+}$ and $^{18}O^{8+}$ beams crossing the spectrometer and reaching the Faraday cup (filled rectangles) located adjacent to the Focal Plane Detector acceptance, respectively.

## *4.4 The target systems*

The goal of the NUMEN project is to explore all systems that are candidate for *0νββ* decay in both directions of isospin transfer $\tau^-\tau^-$ and $\tau^+\tau^+$. This is achieved using the isotopic materials of interest as targets for the DCE reactions ($^{18}O$, $^{18}Ne$) and ($^{20}Ne$, $^{20}O$), respectively, as explained in Section 3. The targets used in the Phase 2 of NUMEN are thin films manufactured at the INFN-LNS chemical laboratory, typically by evaporation procedure (for thicknesses lower than 1 mg/cm$^2$) or by rolling (in case of larger values). For maximum beam intensities of about 20 enA and for present beam energies, the maximum power dissipated in the target is of the order of 0.2 W, which is small enough to avoid any needs for target cooling systems, differently from the case with upgraded facility, discussed in Section 6.4.

In Table 4.2 the most studied isotopes as *"hot"* cases for *0νββ* decay researches are listed. In each line, the isotope targets needed to explore *β⁻β⁻*-like transitions and their partners for *β⁺β⁺*-like transitions are indicated. For each nucleus, the excitation energy of the first excited state of the residual nucleus $E^*(2^+)$ is also reported.

In order to separate the ground state-to-ground state transition from that to the first excited state of the residual nucleus, sufficient energy resolution in the measured excitation energy spectra is required. This resolution mainly depends on three factors, namely the intrinsic energy resolution of the MAGNEX spectrometer ($\delta E_{MAGNEX}/E \sim 1/1000$), the energy spreading of the cyclotron accelerated beam ($\delta E_{CS}/E \sim 1/1000$) and a contribution due to the straggling and energy loss of the beam and ejectiles in the target film $\delta E_{TARGET}$. A contribution due to the kinematic effect should also be considered in principle but for quasi-elastic reactions at forward angles, including DCE, it is very small, so it will be neglected here. The total energy resolution is thus:

$$\delta E \sim \sqrt{\delta E^2_{MAGNEX} + \delta E^2_{CS} + \delta E^2_{TARGET}} \qquad (4.1)$$

$\delta E_{TARGET}$ depends, for a given beam, on the target film material and thickness and on its uniformity. Thus, the request on resolution of the measured energy spectra implies stringent requests on the target characteristics. To this purpose, the target thickness for each experiment is chosen in order to be small enough to guarantee a sufficient $\delta E_{TARGET}$ to separate the transition to the ground state from the transition to the first excited state at $E^*(2^+)$. In this respect, isotopes with corresponding high $E^*(2^+)$ in the residual nucleus allow to produce targets with larger thickness, which is advantageous in studies of rare processes to increase the count rate.

The average thickness of the target foils is determined by direct weighting the foils and by measuring the energy loss of α particles from a $^{241}$Am source traversing them.

A crucial requirement of the target construction is the uniformity. A non-uniform target causes, in fact, a spreading in the energy loss of the ion traversing the target. Specific studies on the target uniformities produced by the evaporation procedure are in progress (see Section 6.4) in order to find the best conditions for each atomic species.

Table 4.2 Isotopes marked with * are candidate for spontaneous ββ-decay.

| Isotope Target for β⁻β⁻ | E*(2⁺) [keV] | Isotope Target for β⁺β⁺, β⁺EC, ECEC decay | E*(2⁺) [keV] |
|---|---|---|---|
| $^{48}$Ca* | 3832 | $^{48}$Ti | 983 |
| $^{76}$Ge* | 563 | $^{76}$Se | 559 |
| $^{78}$Se | 614 | $^{78}$Kr * | 455 |
| $^{82}$Se* | 655 | $^{82}$Kr | 776 |
| $^{92}$Zr | 934 | $^{92}$Mo * | 1509 |
| $^{96}$Zr* | 1582 | $^{96}$Mo | 778 |
| $^{96}$Mo | 778 | $^{96}$Ru * | 833 |
| $^{100}$Mo* | 536 | $^{100}$Ru | 539 |
| $^{106}$Pd | 512 | $^{106}$Cd * | 633 |
| $^{110}$Pd* | 374 | $^{110}$Cd | 658 |
| $^{116}$Cd* | 513 | $^{116}$Sn | 1294 |
| $^{124}$Sn* | 1132 | $^{124}$Te | 603 |
| $^{124}$Te | 603 | $^{124}$Xe * | 354 |
| $^{128}$Te* | 743 | $^{128}$Xe | 443 |
| $^{130}$Te* | 840 | $^{130}$Xe | 536 |
| $^{130}$Xe | 536 | $^{130}$Ba * | 357 |
| $^{136}$Xe* | 1313 | $^{136}$Ba | 818 |
| $^{136}$Ba | 818 | $^{136}$Ce * | 552 |
| $^{148}$Nd* | 302 | $^{148}$Sm | 550 |
| $^{150}$Nd* | 130 | $^{150}$Sm | 334 |
| $^{154}$Sm* | 82 | $^{154}$Gd | 123 |
| $^{160}$Gd* | 75 | $^{160}$Dy | 87 |
| $^{198}$Pt* | 407 | $^{198}$Hg | 412 |

## *4.5 NUMEN data reduction*

The data reduction procedure resembles that described in Refs. [112] [135]. In the following we describe the steps of a typical data reduction for the reaction $^{116}$Cd($^{20}$Ne,$^{20}$O)$^{116}$Sn at 15 MeV/u. The aim is to identify the $^{20}$O$^{8+}$ ejectiles and to reconstruct the scattering angle and excitation energy spectra for the residual $^{116}$Sn nucleus.

### *4.5.1 Calibration procedures*

The first step is the calibration of the *x* and *y* parameters measured by the focal plane detector [108](see Figure 4.1 for definition), which are the basic coordinates for the application of the ray-reconstruction technique implemented in MAGNEX [113]. In order to obtain the horizontal position parameters $x_1$, $x_2$, $x_3$, $x_4$, $x_5$, $x_6$ at the focal plane, a relative calibration of the response of the induction pads for each DC detector is performed. Then the position of the avalanche of a typical event can be determined extracting the center of gravity of the discrete distribution. A proper centroid-finding algorithm was developed with this aim, which accounts for the particular geometrical configuration of the pads with respect to the multiplication wires [109].

The vertical position of the ions in the FPD is determined at six different $z$ positions by the measurements of the drift times of electrons moving towards the six multiplication wires (see Figure 4.1). The absolute calibration of the vertical position is obtained taking as a reference the shadows of the horizontal silicon-coated wires used to support the entrance Mylar window (visible in Figure 4.3), which appear as regularly spaced minima in the $y_i$ spectra (see Figure 4.3). The absolute position of such wires is determined by optical measurements in the laboratory frame.

The $x_i$ and $y_i$ parameters are used to reconstruct the ion track through the detector and deduce the horizontal ($\theta_{foc}$) and vertical angles ($\phi_{foc}$), respectively.

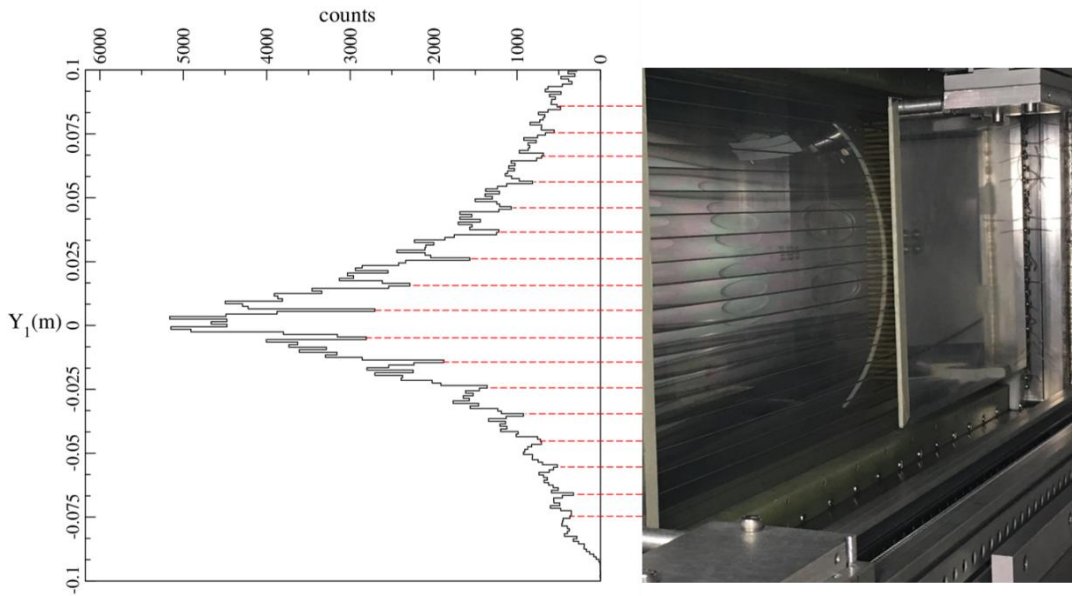

Figure 4.3 Typical $Y_i$ spectrum with no events selection. The minima indicated by the red dashed lines correspond to the horizontal silicon coated wires used to support the entrance Mylar window (visible in the photograph).

### 4.5.2 *Particle identification*

In the NUMEN experiments, the ejectiles of interest are typically in the region of mass $18 \leq A \leq 22$ and atomic number $8 \leq Z \leq 10$. After crossing the experiment target, the electron stripping of the ions is not full and therefore different charge states ($q$) are distributed at the focal plane for each ejectile isotope species (see Section 4.3), making the ion identification more challenging. For this reason, an appropriate Particle Identification (PID) technique is necessary to distinguish the ions of interest, among the whole range of *A* and *Z* produced in the collision. The PID technique adopted in MAGNEX experiments guarantees this requirement.

The atomic number of the ejectiles is identified by the standard *ΔE-E* technique. A typical *ΔE-E* two-dimensional plot is shown in Figure 4.4 (upper panel) for a single silicon detector together with a coarse graphical contour that includes the oxygen ejectiles. The plotted parameters are the residual energy measured by the silicon detectors ($E_{resid}$) in abscissa, and the total energy loss in the FPD gas section $\Delta E_{tot}$ in ordinate. The latter is obtained as the sum of the six $\Delta E_{CDi}$ and corrected for the different path lengths in the gas.

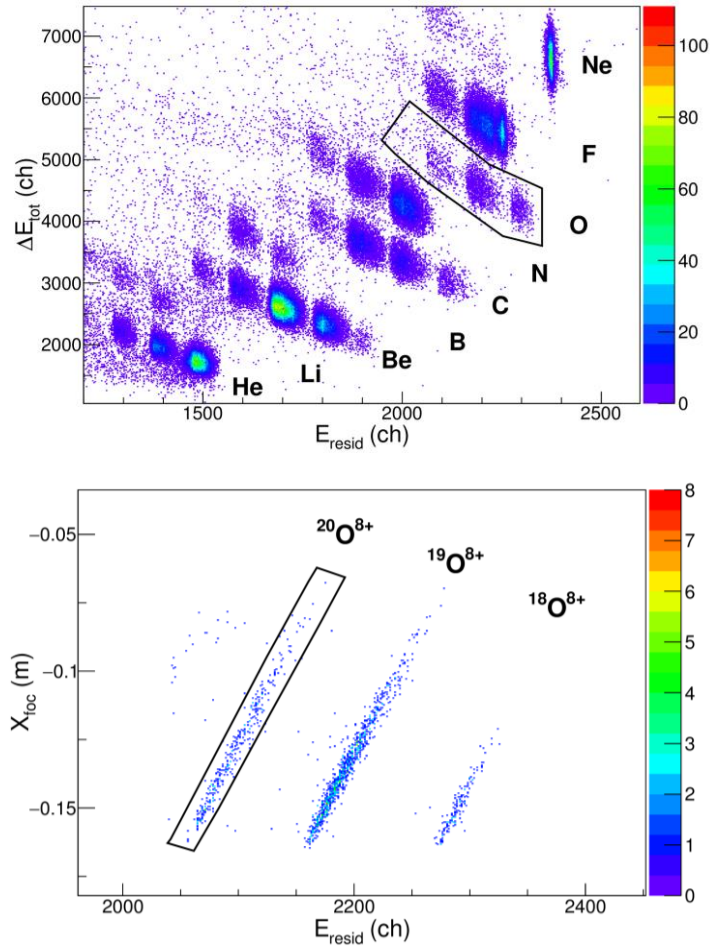

Figure 4.4 (Upper panel) Typical $\Delta E_{tot}$ vs $E_{resid}$ plot for the ejectiles detected in the reaction $^{20}$Ne + $^{116}$Cd at 15 MeV/u incident energy for a single silicon detector. The different atomic species and a coarse graphical contour on the oxygen region are also indicated. (Lower panel) Typical $X_{foc}$-$E_{resid}$ plot after applying the graphical condition on the $\Delta E_{tot}$-$E_{resid}$ for the same silicon detector. The different oxygen isotopes and a graphical contour selecting the $^{20}$O$^{8+}$ ejectiles are indicated.

For the mass identification, an innovative particle identification technique for large acceptance spectrometers, which exploits the properties of the Lorentz force, was introduced in Ref. [111]. When dealing with a large acceptance device as MAGNEX, the best resolution in the identification technique is achieved performing a precise reconstruction of the ions kinetic energy, as demonstrated in Ref. [111]. However, when a high mass resolution is not necessary, as in the experimental conditions of the NUMEN reactions, which involve oxygen, fluorine and neon ions, the identification procedure is successfully performed using the $X_{foc}$-$E_{resid}$ correlation. The relationship between the two measured quantities ($X_{foc}$ and $E_{resid}$) is approximately quadratic with a factor depending on the ratio $\sqrt{m}/q$

$$X_{foc} \propto \frac{\sqrt{m}}{q}\sqrt{E_{resid}} \qquad (4.2)$$

Therefore, in a $X_{foc}$ versus $E_{resid}$ plot the ions are distributed on different loci according to the ratio $\sqrt{m}$/q. The clear separation between the different oxygen isotopes is evident in Figure 4.4 (lower panel), where the $X_{foc}$-$E_{resid}$ matrix is shown for the data selected with the graphical condition on the- $\Delta E_{tot}$ -$E_{resid}$ one (Figure 4.4 upper panel). In this plot the selection of the $^{20}O^{8+}$ ejectiles can be safely performed, as indicated by the graphical contour in the lower panel of Figure 4.4.

*4.5.3 Ray reconstruction technique*

The ray-reconstruction procedure is then applied to the identified set of data, in order to extract the momentum vector at the reaction point of the ejectiles and the absolute cross section. In order to perform an accurate trajectory reconstruction of the measured data, a precise model of the spectrometer response in the specific magnetic setup of the experiment is necessary. The way to test the accuracy of such a model comes from a comparison between the measured phase space parameters at the focal plane and the simulated events for the selected reaction.

In the first step of the procedure, a direct transport map is generated. It describes the evolution of the phase-space parameters from the target point through the spectrometer field up to the focal plane. In the MAGNEX case, the formalism of the differential algebra [136] [137] implemented in the COSY INFINITY program [138] is used to build the transport map up to the 10$^{th}$ order. To test its goodness, a set of events corresponding to the considered reaction is generated by Monte Carlo routines [139]. Since the DCE reaction channel has a very low yield, the tuning of the transport map using the DCE data is quite difficult. Therefore, an alternative approach, quite advantageous, is to compare simulations and experimental data for reaction channels with higher yield, e.g. the elastic scattering, and then to use the same transport map to reconstruct the experimental DCE data. To this aim, the $^{20}Ne^{8+}$ ejectiles elastically scattered from $^{116}Cd$ target nuclei, which have a magnetic rigidity very close to the $^{20}O^{8+}$, were identified in the experimental data and tracked through the spectrometer by the application of the direct transport map. The comparison between the simulation and the experimental data for the elastic scattering is shown in Figure 4.5. Despite the highly non-linear aberrations, the simulated data give a rather faithful representation of the experimental ones both in the horizontal ($\theta_{foc}$-$X_{foc}$ plot) and vertical ($Y_{foc}$-$X_{foc}$ plot) phase spaces [113]. The $^{20}$O$^{8+}$ ejectiles connected to the excitation of the ground and the first excited state at 1.3 MeV of the residual $^{116}$Sn nucleus were also tracked through the spectrometer by the application of the same direct transport map. The comparison with the experimental data is found in Figure 4.6. Some states at arbitrary excitation energy were also simulated to explore the whole phase space, scanning, in particular, the excitation energy parameter. In the simulated data of Figure 4.6 the bulges visible at $0.9 < \theta_{foc} < 1.0$ rad correspond to aberrations.

Once a reliable direct transport map has been obtained, it can be inverted by the COSY INFINITY program and applied to the measured final coordinates in order to obtain the initial phase space parameters at the target point. These are directly related to the modulus of the ejectile momentum and the scattering angle. Indeed, from the initial vertical $\phi_i$ and horizontal $\theta_i$ angles, the laboratory scattering angle $\theta_{lab}$ is extracted. Then, from the reconstructed momentum, the initial kinetic energy of the ejectiles is deduced. The corresponding $Q$-values, or equivalently the excitation energy $E_x = Q - Q_0$, where $Q_0$ is the ground state to ground state $Q$-value, are finally obtained by a missing mass calculation based on relativistic energy and momentum conservation laws, assuming a binary reaction.

A plot of $\theta_{lab}$ versus $E_x$ is shown in Figure 4.7. The $^{116}$Sn ground state region is visible as vertical and straight locus around $E_x = 0$ even with the low collected yield, as expected since the $E_x$ parameter does not depend on the scattering angle for transitions to the $^{116}$Sn states. This demonstrates that the effects of the aberrations observed in Figure 4.6 have been satisfactorily compensated. The efficiency cut on the bottom of the distribution is due to the presence of the protection screen that limit the FPD acceptance (see Section 4.3).

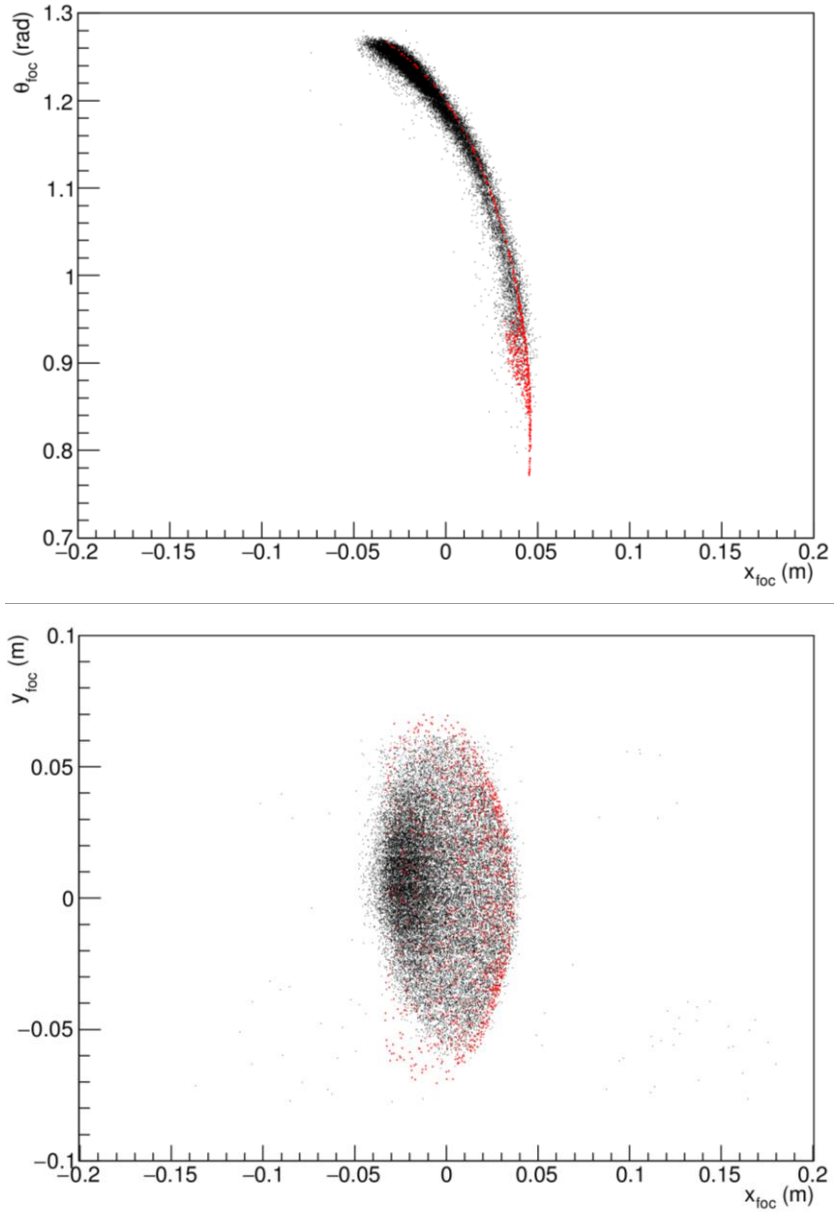

Figure 4.5 Comparison between the experimental (black points) and the simulated (red points) data in the $\theta_{foc}$-$X_{foc}$ (upper panel) limited to events at -0.01 < $Y_{foc}$ < +0.01 m and $Y_{foc}$-$X_{foc}$ (lower panel) limited to events at $\theta_{foc}$ > 1.0 rad representations for the elastic scattering $^{20}$Ne + $^{116}$Cd at 3° < $\theta_{lab}$ < 14°.

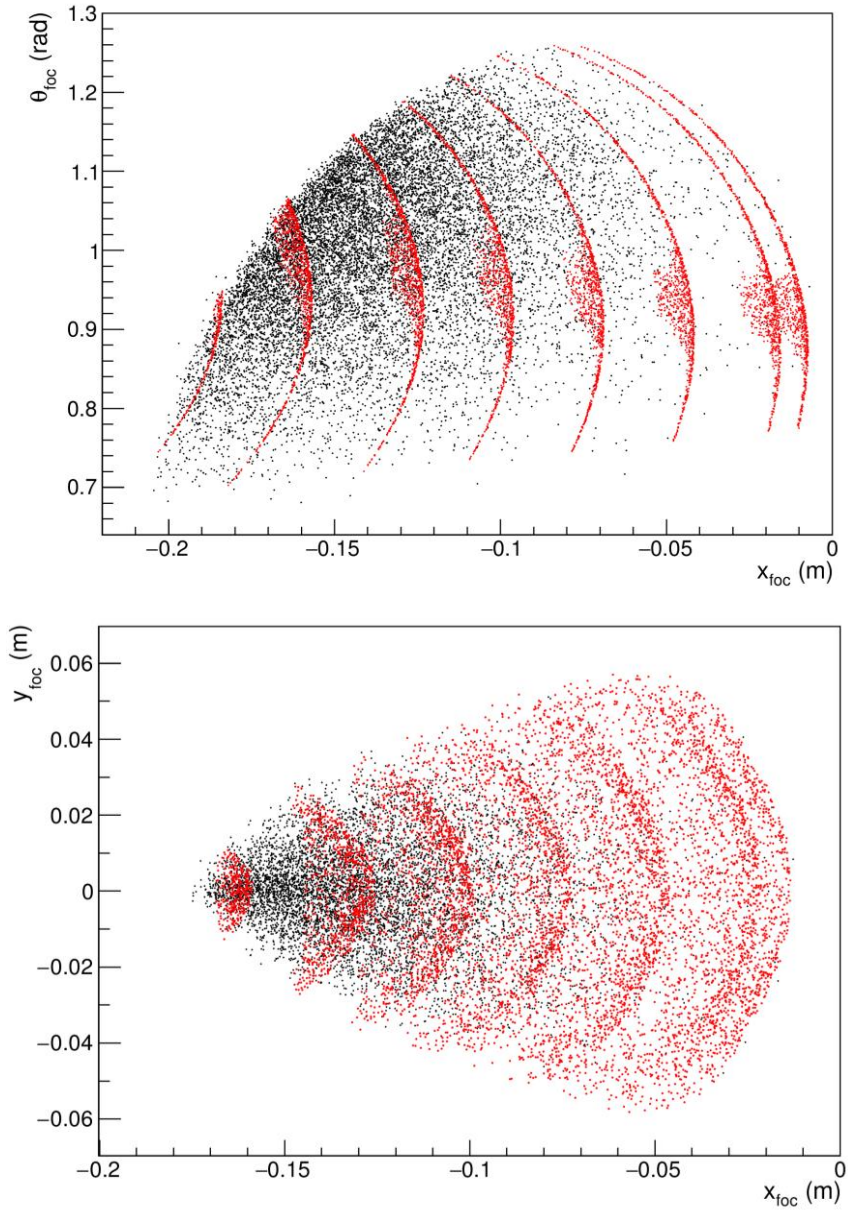

Figure 4.6 Comparison between the experimental (black points) and the simulated (red points) data in the $\theta_{foc}$-$X_{foc}$ (upper panel) limited to events at $-0.01 < Y_{foc} < +0.01$ m and $Y_{foc}$-$X_{foc}$ (lower panel) limited to events at $\theta_{foc} > 1.0$ rad representations for the $^{116}$Cd($^{20}$Ne,$^{20}$O)$^{116}$Sn reaction at $3° < \theta_{lab} < 14°$.

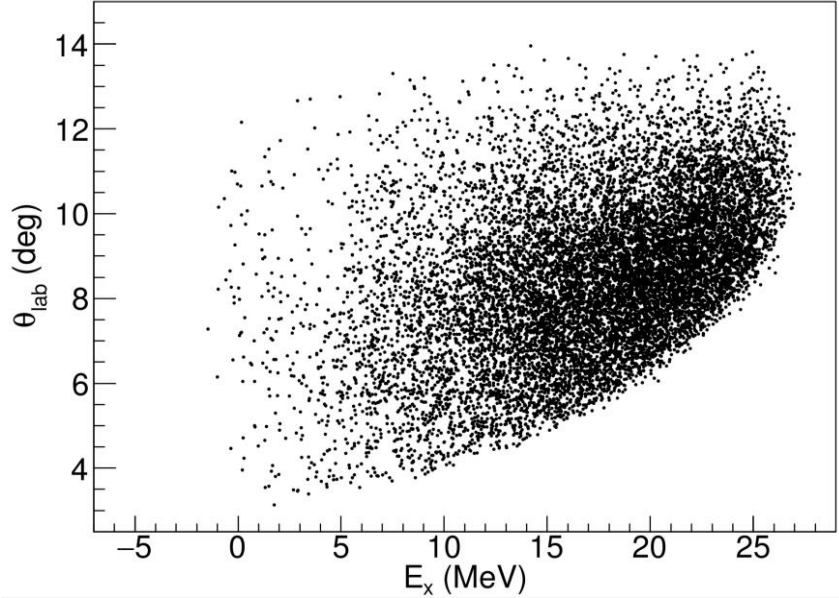

Figure 4.7 Plot of the reconstructed $\theta_{lab}$ versus the $^{116}$Sn excitation energy ($E_x$) for the $^{116}$Cd($^{20}$Ne,$^{20}$O)$^{116}$Sn reaction at 15.3 MeV/u.

After the ray reconstruction procedure, the scattering angle $\theta_{lab}$ and the excitation energy $E_x$ are obtained. A major achievement of the ray reconstruction technique is the very small systematic error obtained in the horizontal $\theta_i$ (-0.01° ± 0.04°) and vertical $\phi_i$ (-0.05° ± 0.3°) angles, as demonstrated in Ref. [113]. In addition a high resolution is also obtained in $\theta_i$ (0.2°) and $\phi_i$ (0.7°) angles. In the reconstruction of the scattering polar angle in the laboratory frame $\theta_{lab}$ both the horizontal and the vertical angles contribute according to basic geometrical relations. The reconstruction of the vertical angle has a significant contribution only at very forward angles. As an example the overall error induced by an uncertainty of $\delta\phi_i = 1°$ on the scattering angle is less than $\delta\theta_{lab} = 0.08°$ at $\theta_{lab} = 40°$ and $\delta\theta_{lab} = 0.8°$ at $\theta_{lab} = 5°$. This has to be taken into account for the NUMEN experiments, which are performed at very forward angles including $\theta_{lab} = 0°$. Regarding the reconstructed momentum modulus, a resolution of 1/1800 with an accuracy better than 1/1600 is obtained for the reaction channels of interest.

When dealing with very rare processes, as the DCE reactions, other important parameters of the experimental measurement are the cross section sensitivity and the rejection factor. In particular, looking at the reconstructed $\theta_{lab}$ versus $E_x$ plot shown in Figure 4.7, we can see that there are no spurious counts in the region between -7 and -2 MeV. This corresponds to sensitivity better than 1 count within 5 σ confidence level in an energy range of 1 MeV. To estimate the rejection factor in the region of interest for the transition $^{116}$Cd($^{20}$Ne,$^{20}$O)$^{116}$Sn$_{g.s.}$ (approximately from -600 keV to

+600 keV), we estimated the total ejectile flux emerging from the target seen by the solid angle aperture of the spectrometer according to the following formula: $N_{reac}^{tot} = N_{targ} N_{beam} \sigma_{reac} \Delta\Omega_{MAGNEX}^{\varepsilon}/4\pi$. The use of this formula requires the knowledge of: the number of ions/cm$^2$ deduced from the target thickness ($N_{targ}$), the number of incident ions measured by the Faraday Cup ($N_{beam}$), the solid angle seen by MAGNEX taking into account the transport efficiency of the specific setup (in the present case $\Delta\Omega_{MAGNEX}^{\varepsilon}$ ~ 0.041 sr) [114] and the total reaction cross section $\sigma_{reac}$ for the system $^{20}$Ne + $^{116}$Cd ($\sigma_{reac} \cong \pi R^2$~1.8 b). After the $B\rho$ selection by the dipole magnetic field, the particle identification and the ray-reconstruction technique, we are able to obtain a tiny amount of spurious counts in the DCE region of interest (< 0.25) that corresponds to a rejection factor of better than $4 \times 10^{-9}$ in the region of interest.

# 5 Theoretical aspects

Within the NUMEN project, theoretical developments aim at reaching a full description of the DCE reaction cross section, including also competing channels that may lead to the same final outcome, and at investigating the possible analogies with double beta decay. The main lines of investigation are discussed below.

## 5.1 DCE reactions as a two-step process and analogies with 2νββ decay

The most conventional description of the DCE mechanism would be to consider it as a two-step process [140]. The latter is given by two uncorrelated single charge exchange events where, after the first event, the system propagates before a second charge exchange occurs. Each of the SCE processes is induced by the action of one-body operators on the projectile and the target nucleus. Thus, this process can be considered as a two-step one-body DCE reaction. The two-step DCE process is depicted diagrammatically in Figure 5.1.

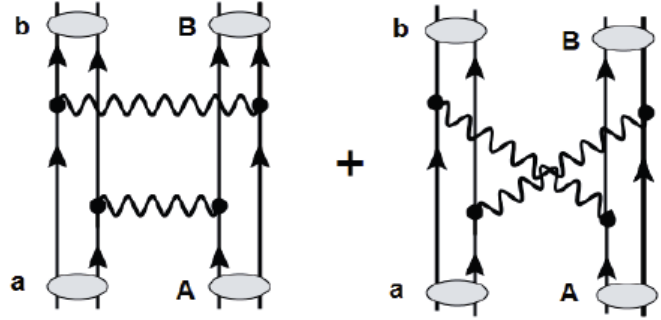

Figure 5.1 Graphical representation of the one-body two-step DCE reaction by hadronic interactions. The reaction $a(N_a, Z_a) + A(N_A, Z_A) \rightarrow b(N_a \pm 2, Z_a \mp 2) + B(N_A \mp 2, Z_A \pm 2)$ proceeds by the sequential, but independent, action of two charge changing strong interaction events, i.e. the exchange of charged mesons. Each of the interaction events acts like a one-body operator on the target and projectile, respectively.

The reaction matrix element, connecting the incident channel $\alpha = a + A$ and the final channel $\beta = b + B$, is written down easily, in DWBA, as a quantum mechanical second order amplitude:

$$M_{\alpha\beta}^{(DCE)}(\mathbf{k}_\beta, \mathbf{k}_\alpha) = \langle \chi_\beta^{(-)}, bB | T_{NN} \mathcal{G}^{(+)}(\omega) T_{NN} | aA, \chi_\alpha^{(+)} \rangle \quad (5.1)$$

Initial and final state interactions are taken into account by the distorted waves, $\chi(r)$, obeying incoming and outgoing spherical wave boundary conditions, respectively. The intermediate nuclear propagator may be represented in terms of the complete system of the nuclear eigenstates $|\gamma\rangle = |C_\gamma, c_\gamma\rangle$ which are reached by a SCE reaction,

$$\mathcal{G}(\omega) = \sum_{\gamma=c,C} |\gamma\rangle G_\gamma(\omega_\gamma, \omega_\alpha)\langle\gamma| \quad (5.2)$$

where the reduced Green function $G_\gamma = 1/(\omega_\alpha - \omega_\gamma + i\eta)$ describes the propagation of the intermediate scattering states. Above, $\omega_{\alpha,\gamma}$ is the total center-of-mass energy as defined by the rest masses and the kinetic energy in the incident ($\alpha$) or intermediate ($\gamma$) channel, and $\boldsymbol{k}_{\alpha,\beta}$ defines the relative momentum in the entrance ($\alpha$) or exit ($\beta$) channel.

Either of the first and second reaction step is described by the charge-changing part of the nucleon-nucleon T-matrix, which in non-relativistic notation is given by:

$$T_{NN} = \left(t_{01} + t_{11}\boldsymbol{\sigma}_a \cdot \boldsymbol{\sigma}_A + t_{T1}S_{12}(\sigma_a, \sigma_A)\right)\left(\tau_+^{(a)}\tau_-^{(A)} + \tau_-^{(a)}\tau_+^{(A)}\right) \quad (5.3)$$

including spin-scalar ($t_{01}$) and spin-vector ($t_{11}$) central interactions and rank-2 tensor interactions given by the usual rank-2 tensor operator $S_{12}$ and with a form factor $t_{T1}$.

A suitable representation is obtained by expressing the 2-step amplitude in momentum space:

$$M_{\alpha\beta}^{(DCE)}(\mathbf{k}_\beta, \mathbf{k}_\alpha) = \sum_{c,C} \int \frac{d^3 k_\gamma}{(2\pi)^3} M_{\beta\gamma}^{(SCE)}(\mathbf{k}_\beta, \mathbf{k}_\gamma) G_\gamma(\omega_\gamma, \omega_\alpha) \widetilde{M}_{\gamma\alpha}^{(SCE)}(\mathbf{k}_\gamma, \mathbf{k}_\alpha) \quad (5.4)$$

showing that the DCE amplitude is obtained as a superposition of one-step SCE reaction amplitudes.

The fully quantum mechanical DCE differential cross section is then given by:

$$d\sigma_{\alpha\beta}^{(DCE)} = \frac{m_\alpha m_\beta}{(2\pi\hbar^2)^2} \frac{k_\beta}{k_\alpha} \frac{1}{(2J_a+1)(2J_A+1)} \sum_{\substack{M_a, M_A \in \alpha \\ M_b, M_B \in \beta}} \left|M_{\alpha\beta}^{(DCE)}(\mathbf{k}_\beta, \mathbf{k}_\alpha)\right|^2 d\Omega \quad (5.5)$$

averaged over the initial nuclear spin states and summed over the final nuclear spin states, respectively. Reduced masses in the incident and exit channel are denoted by $m_{\alpha,\beta}$, respectively.

As it appears from Eq. (5.4), the description of DCE reactions in terms of the convolution of two uncorrelated SCE processes exhibits close analogies with 2νββ decay [92]. Thus one could possibly extract information on *2νββ* decay nuclear matrix elements, from the study of DCE reactions, and make a comparison to the observations related to double-beta decay events.

## 5.2 Single charge exchange cross section

As it is shown in the previous Section, the DCE cross section can be estimated, within the DWBA theory, considering a suitable folding of two SCE reaction amplitudes. Thus it is particularly important to improve also the description of the SCE process, with the special purpose of factorizing the corresponding cross section into reaction and structure parts. Indeed this separation allows one to isolate the nuclear matrix elements relevant to β-decay processes.

We consider a charge changing reaction between projectile, *a*, and target, *A*, nuclei:

$$^a_z a + ^A_Z A \rightarrow {}^a_{z\pm 1} b + {}^A_{Z\mp 1} B$$

The charge changing process induced by two-body interactions acting on the projectile-target nucleons is depicted in Figure 5.2.

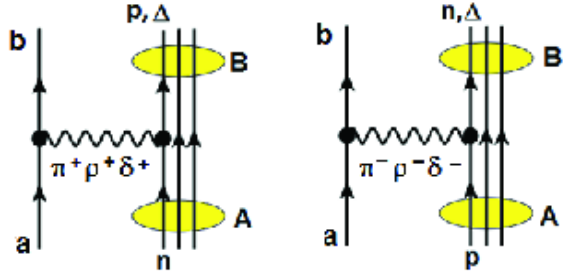

Figure 5.2 Graphical representation of a HI-SCE reaction by hadronic interactions, corresponding to νβ processes. Both (n,p)-type (left) and (p,n)-type (right) reactions, as seen in the A →B transition in the target system, are displayed, indicating also the exchanged meson.

In the momentum representation, the full reaction amplitude $M_{\alpha\beta}$ is given by folding the reaction kernels with the distortion coefficient $N_{\alpha\beta}$:

$$M_{\alpha\beta}(\mathbf{k}_\beta, \mathbf{k}_\alpha) = \sum_{ST} \int d^3 p \, K^{(ST)}_{\alpha\beta}(\boldsymbol{p}) N_{\alpha\beta}(\mathbf{k}_\beta, \mathbf{k}_\alpha, \boldsymbol{p}) \quad (5.6)$$

where the kernels $K_{\alpha\beta}$ are expressed by suitable products of the nuclear interaction form factors in the spin-isospin (ST) channels, $V_{ST}$, and projectile and target transition form factors $F_{ST}$:

$$K^{(ST)}_{\alpha\beta}(\boldsymbol{p}) = (4\pi)^2 V^{(C)}_{ST}(p^2) F^{ab\dagger}_{ST}(\boldsymbol{p}) \cdot F^{AB}_{ST}(\boldsymbol{p})$$

$$+ \delta_{S1}(4\pi)^2 \sqrt{\frac{24\pi}{5}} V^{(T)}_{ST}(p^2) Y^*_2(\hat{\boldsymbol{p}}) \cdot \left[F^{ab\dagger}_{ST}(\boldsymbol{p}) \otimes F^{AB}_{ST}(\boldsymbol{p})\right]_2$$

(5.7)

with $F_{ST}^{(ab)}(\mathbf{p}) = \frac{1}{4\pi} \langle J_b M_b | e^{+i\mathbf{p}\cdot\mathbf{r}_a} \mathcal{O}_{ST} | J_a M_a \rangle$.

In the relation above, $\mathcal{O}_{ST}$ indicates the spin-isopin transition operator and $J_{a,b}$ denotes the spin of the projectile (target), with its projection $M_{a,b}$.

Owing to the same spin-isospin structure of the transition operators, one can show that the transition form factors are directly connected to beta decay strengths (in particular, for transitions with multipolarity $L = 0$) [141].

The distortion coefficient is given by:

$$N_{\alpha\beta}(\mathbf{k}_\beta, \mathbf{k}_\alpha, \mathbf{p}) = \frac{1}{(2\pi)^3} \langle \chi_\beta^{(-)} | e^{-i\mathbf{p}\cdot\mathbf{r}} | \chi_\alpha^{(+)} \rangle \qquad (5.8)$$

where the distorted waves obey the Schroedinger equation, according to the optical potential in the entrance and exit channels.

Under suitable conditions of low momentum transfer $q_{\alpha\beta}$, the reaction amplitude expression given by Eq. (5.6) can be factorized as:

$$M_{\alpha\beta}(\mathbf{k}_\beta, \mathbf{k}_\alpha) = M_{\alpha\beta}^{(B)}(\mathbf{q}_{\alpha\beta})(1 - n_{\alpha\beta}) \qquad (5.9)$$

obtaining in leading order the Born-amplitude $M^{(B)}{}_{\alpha\beta}$ scaled by a distortion coefficient. In particular, the distortion coefficient can be easily derived assuming a Gaussian shape for the transition form factors, in the strong absorption limit.

This factorization is particularly important because, as discussed above, the Born amplitude is linked to projectile and target beta decay strengths [48].

Explicit calculations, based on the code HIDEX [142], [78], have been performed for the reaction $^{18}O + {}^{40}Ca$ at $T_{lab} = 270$ MeV. We stress again that these studies are particularly useful for the extension to the treatment of DCE reactions (see Section 2.2).

Transition form factors were evaluated on the basis of QRPA calculations. We note that very recent developments will allow one to compute the spectroscopic amplitudes (and in perspective the radial transition densities) also in the scheme of IBFFM2 (IBM for odd-odd nuclei) [143]. This will open the possibility to give predictions with IBM approaches [143] [144] [145] in the field of charge exchange reactions, where an experimental feedback is available.

Among the most important conclusions reached so far, we emphasize the role played by the imaginary part of the optical potential (i.e. by absorption effects) in determining the amplitude of the distortion coefficient, see Eq. (5.8), and then of the full reaction cross section. This is illustrated in Figure 5.3, where one can see that the results of the full DWBA calculations practically coincide with calculations performed by taking into account only the imaginary part of the optical potential.

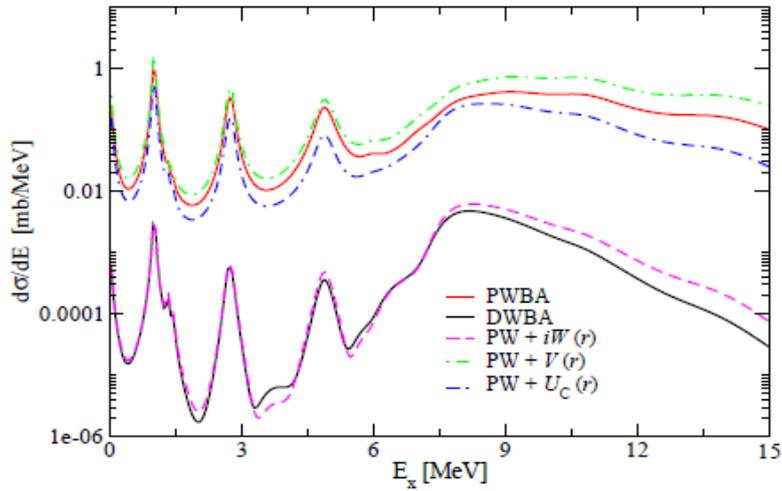

Figure 5.3 SCE cross sections as a function of the target excitation energy, $E_x$, integrated over the full angular range, for Gamow-Teller like transitions, with $J = 1^+$ in both projectile and target. The different curves show the effect of Coulomb potential ($U_C(r)$), of real ($V(r)$) and imaginary ($W(r)$) components of the optical potential and of the full potential (DWBA), with respect to Plane Wave Born Approximation (PWBA) calculations.

These results support the black disk approximation as a suitable tool to depict the reaction dynamics. This assumption is particularly convenient because one can easily make predictions for the distortion coefficient of Eq. (5.9). Results obtained in the limit of small momentum transfer are represented in Figure 5.4, for reactions with three projectile masses ($A_P=12, 18, 28$) and a $^{40}$Ca target, as a function of the beam energy.

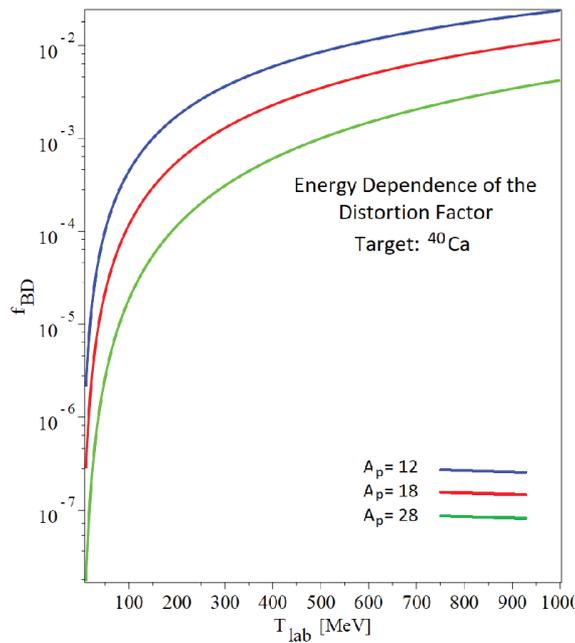

Figure 5.4 The distortion coefficient, as a function of the beam energy, for reactions with a $^{40}$Ca target and three projectile mass choices: $A_P$=12 (blue line), 18 (red line), 28 (green line).

## 5.3  *Analogy between DCE reactions and 0νββ decay*

In this Section, we enter into a deeper discussion of the analogies motivating the investigation of DCE reactions as a tool to probe the nuclear structure input to the matrix elements of nuclear double beta decay.

Collisional (or direct) charge exchange by isovector nucleon-nucleon interactions is the most probable cause for transferring charge between colliding ions at energies well above the Coulomb barrier, as discussed e.g. in [146], [147], [80], [148].

The SCE process can, of course, occur in higher order. As discussed above, a second order charge exchange reaction may be given by two independent, sequential SCE reactions, i.e. the ions need to interact twice by two-body interactions between a target and a projectile nucleon. As far as the reaction mechanism is concerned those double-SCE reactions resemble closely nuclear *2νββ* decay (see Section 2).

However, it is possible to envisage a nuclear Double-Charge Changing (DCC) scenario of a close structural similarity to *0νββ* decay. In the presence of a second nucleus, a correlated DCE transition can be initiated by a particular type of hadronic two-body operator and proceed as a one-step reaction. A correlated pair of nucleons (through the exchange of a neutral meson) changes its total charge by two units under emission of a virtual pair of charged mesons captured by the second ion and there inducing a complementary DCC transition. In an isolated nucleus the two emitted mesons would be reabsorbed immediately thus restoring the nucleonic charge, and no net effect would be observed. The process would be of no particular interest except for contributing to the nuclear short range correlations, causing nuclear momentum distributions to deviate by 10 to 15% from those expected for independent quasiparticles [149], [150], [151], [152]. If, however, the intermediate charged mesons are absorbed by a second nucleus, both ions will be found to have changed their charge by two units in a complementary manner. Different from a conventional two-step heavy ion charge exchange reaction the whole reaction proceeds as one-step reaction with respect to counting the ion-ion interaction.

That physical process is depicted in Figure 5.5 for a nn → pp transition. The figure shows the involved nuclear configurations that are realized by the coherent action of charge-changing meson-nucleon interactions. Overall, the process scrutinized here corresponds on the target side to a $nn \to pp + \pi^-\pi^-$ reaction. In free space, the corresponding charge-conjugated reaction $pp \to nn + \pi^+\pi^+$ reaction and other double-pion production channels were in fact already investigated experimentally at CELSIUS and COSY [153], [154], [155], [156], [157], [158], [159], [160], [161], [162], [163] and later also at HADES [164]. Thus these correlations are expected to be present also in more involved nuclear reactions. Theoretical studies combining meson exchange and resonance excitation have been performed by the Valencia group [165] and in somewhat extended form by Xu Cao et al. [166].

We conclude by noting that, owing to the intrinsically different nature of the two-step and one-step DCE processes, one expects different cross sections and angular distributions for the two mechanisms, which could be finally disentangled from the total experimental DCE cross section. Work is in progress in this direction.

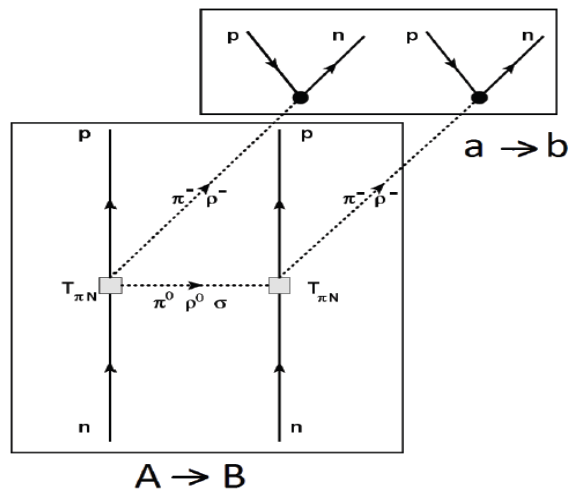

Figure 5.5 Generic diagram illustrating the hadronic surrogate process for $0\nu\beta\beta$ decay. A virtual $nn \to pp\pi^-\pi^-$ scattering process, causing the $\Delta Z = +2$ target transition $A \to B$, is accompanied by $nnp^{-1}p^{-1}$ double-CC excitation in the projectile. As indicated, other isovector mesons as e.g. the $\rho$-meson isotriplet will contribute, too.

## *5.4 Competing channels*

To interpret the experimental cross sections and properly isolate the DCE contribution, the description of competing processes leading to the same exit channel is mandatory. The latter are essentially multi-nucleon transfer reactions. An example is shown in Figure 5.6, for the reaction $^{20}Ne + ^{116}Cd \to ^{20}O + ^{116}Sn$ explored by NUMEN.

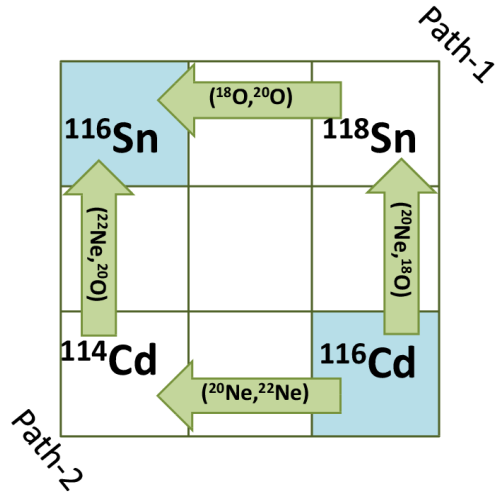

Figure 5.6: Schematic illustration of (2n-2p) or (2p-2n) transfer paths for the reaction $^{20}$Ne + $^{116}$Cd $\rightarrow$ $^{20}$O + $^{116}$Sn.

The theoretical description of these reactions is generally afforded still within the DWBA approach, though coupled channel calculations (based on the FRESCO code [167]) are also feasible. An important point to consider is the improvement of the spectroscopic information contained in such a theoretical description.

Recently, the formalism for two nucleon transfer reactions [168] [169] has been elaborated within the microscopic IBM-2 [170], [171] [172], [173], [174], [175]. This development is particularly appealing; indeed microscopic IBM allows one to calculate in a realistic way the transition matrix elements for heavy, medium nuclei and it has been exploited for the evaluation of $0\nu\beta\beta$ decay NMEs [176] [4].

Spectroscopic amplitudes of two-proton and two-neutron transfer reactions, which are competing with the double charge exchange, have been computed for the combinations represented in Figure 5.6, both with microscopic IBM-2 and with Shell Model (SM) calculations, and inserted as input in the FRESCO code. The resulting multi-nucleon transfer reaction cross sections, obtained within the two model schemes, differ by less than a factor 2.

We also found that these competing processes are far from saturating the total detected experimental cross section, thus pointing to a dominant role of the meson exchange double charge exchange contribution. Indeed, the calculated transfer cross section is of the order of $10^{-4} - 10^{-3}$ nb, to be compared with the preliminary extracted experimental cross section of the order of few nb.

On the other hand, the calculated two-proton transfer cross section $^{20}$Ne + $^{116}$Cd $\rightarrow$ $^{18}$O$_{gs}$ + $^{118}$Sn$_{gs}$, as obtained employing SM spectroscopic amplitudes, is of the order of 10 nb, which is close to the experimental value. The latter result can be considered as a further check of the good performances of the FRESCO calculations.

The role of nucleon transfer with respect to direct charge exchange can be investigated also for the SCE reaction $^{20}$Ne + $^{116}$Cd → $^{20}$F + $^{116}$In. In this case, one has to evaluate the cross section corresponding to a sequential two nucleon transfer, to be compared with the one step process, i.e. with DWBA calculations of the direct SCE process.

We conclude by stressing that the most ambitious goal of the theoretical analysis supporting the NUMEN project will be to formulate a theory of heavy ion DCE reactions, taking into account coherently the interplay of the competing direct and multistep-transfer channels. In combination with microscopic nuclear structure methods, a description of DCE reactions as a tool for spectroscopic investigations is achieved.

# 6   Upgrade of the experimental setup

## 6.1   *The particle accelerator for NUMEN: the INFN-LNS Superconducting Cyclotron*

### *6.1.1   The present accelerator*

The INFN-LNS Superconducting Cyclotron (CS) is a three-sector compact accelerator with a pole radius of 90 cm and an overall yoke diameter of 380.6 cm. Two pairs of superconducting coils allow the production of a maximum magnetic field of 5 T at the center. Using 20 trim coils wound on each of the 6 sectors (120 in total), an isochronous magnetic field is achieved which allows to accelerate and extract all ions, from molecular hydrogen, $H_2^+$, up to uranium in a wide range of energies, between 10 and 80 MeV/u [117], [177].

A disadvantage of the CS compactness is the lack of orbit separation at the last turns. The fact that the extraction is performed through two electrostatic deflectors, ED1 and ED2 (see Figure 6.1), limits the extraction efficiency to 50-60%. Furthermore, in spite of the water cooling, thermal issues arise on the first electrostatic deflector, ED1, when the extracted power exceeds 100 W.

The NUMEN experiment plans to use mainly oxygen and neon beams with intensity up to $10^{14}$ pps. The required energies are in the range 15-70 MeV/u, which corresponds to a beam power of 1-10 kW. The extraction of 1-10 kW beams is not feasible using the ED nor through the existing extraction channel. Moreover, the existing extraction channel has no thermal shields to dissipate the beam power coming from the beam halos, such that only beams with a transversal size not larger than 8 mm can be extracted.

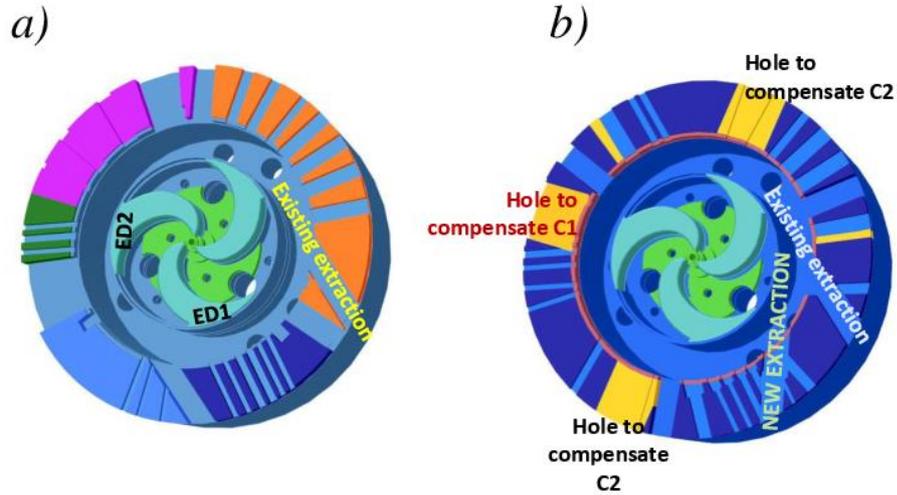

Figure 6.1 Views of the changes foreseen for the central iron ring of the CS. Part *a)* is a top view of the present CS. The existing extraction channel is in the orange sector. In part *b)* the results of the optimization process are shown. The highlighted areas show where the iron is removed to eliminate unwanted field harmonics.

*6.1.2  The CS upgrade*

For the reasons mentioned above, the CS will be upgraded with a new extraction channel designed for extraction by stripping [178], [179], [180].

In this case, ions are accelerated with a charge state $Z-1 \leq q \leq Z-3$ and they become fully ionized after crossing a stripper foil. The use of a suitable stripper foil allows the beam trajectory to escape from the region of the cyclotron pole. All ions with mass number $A < 40$ and energy of interest for NUMEN are fully stripped with efficiency higher than 99% [181]. Beam losses inside the cyclotron are below 100-200 W, which is a reasonable value as far as activation is concerned.

Table 6.1 List of the ions to be extracted by stripping and their expected power.

| Ion | Energy | Isource | Iaccelerated | Iextracted | Pextracted |
|---|---|---|---|---|---|
| | MeV/u | eμA | eμA | pps | Watt |
| $^{12}C^{4+}$ | 45 | 400 | 60(4+) | $9.4\times10^{13}$ | 8100 |
| $^{12}C^{4+}$ | 60 | 400 | 60(4+) | $9.4\times10^{13}$ | 10800 |
| $^{18}O^{6+}$ | 29 | 400 | 60(6+) | $6.2\times10^{13}$ | 5220 |
| $^{18}O^{6+}$ | 45 | 400 | 60(6+) | $6.2\times10^{13}$ | 8100 |
| $^{18}O^{6+}$ | 60 | 400 | 60(6+) | $6.2\times10^{13}$ | 10800 |
| $^{18}O^{7+}$ | 70 | 200 | 30(7+) | $2.7\times10^{13}$ | 5400 |
| $^{20}Ne^{7+}$ | 28 | 400 | 60(7+) | $5.3\times10^{13}$ | 4800 |
| $^{20}Ne^{7+}$ | 60 | 400 | 60(7+) | $5.3\times10^{13}$ | 10280 |

Table 6.1 summarizes the expected results for the beam power delivered at the exit of the cyclotron for some of the studied cases. Conservative values of the beam currents delivered by the ion source and accelerated by the cyclotron are given.

Beam dynamics is a crucial feature of extraction by stripping, since the focusing properties of the magnetic field are not provided like in the electrostatic case. The axial and radial envelopes have to be maintained as small as possible along the entire extraction channel, by using the minimum possible number of correcting elements. The design of the extraction channel has to be done carefully and many parameters have to be considered since each extraction trajectory is different from the others.

A detailed beam dynamics optimization [178], [179], [180] and an appropriate design of the stripper foil system [182] allows to make the trajectories of the beams listed in Table 6.1 cross the same exit point, as shown in Figure 6.2 (right panel).

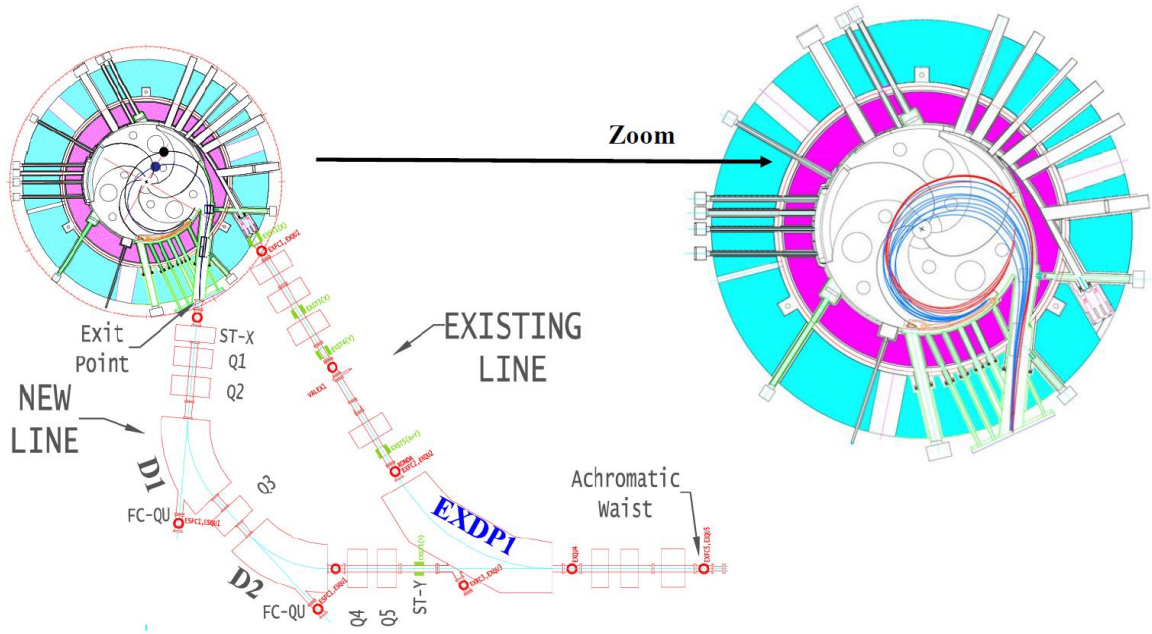

Figure 6.2 The left part shows the new beam line for extraction by stripping. The right part shows a zoomed view with a few trajectories studied along the new extraction channel for different ions and energies

Since extraction by stripping is a multi-turn extraction, it is mandatory to consider in the calculations the energy spread introduced after the crossing of the stripping foil. According to [183], this value has been fixed at ± 0.3% (for 90% of the flux) for all ions and energies. However, a beam with a lower energy spread can be delivered to the MAGNEX hall, by a dedicated design of the transport line. Figure 6.2 shows the new extraction channel and the elements of the new extraction line, which joins to the existing transport line at the magnet ED1. This line is designed to handle beams extracted by stripping with an energy spread up to ± 0.3% and to compensate the chromaticity of the extraction path, such to produce an achromatic beam waist at the position "achromatic waist" marked in Figure 6.2 [184]. The new beam transport line to MAGNEX is shown in Figure 6.3 and corresponds with the FRAISE (FRAgment Ion SEparator) line. FRAISE is designed for the production and separation of radioactive ion beams, but, when used with stable isotopes, it can guarantee the transport of high intensity beams and can be used to limit the energy spread to ± 0.1%.

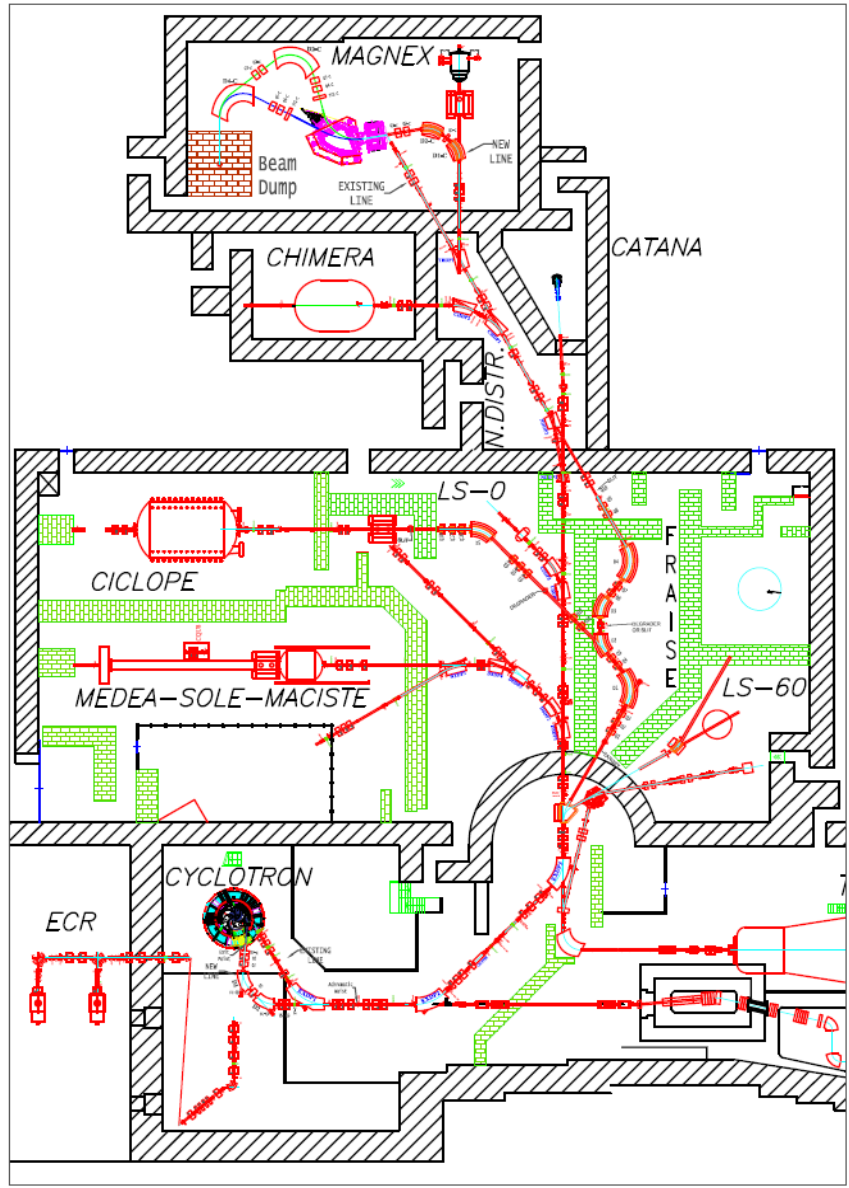

Figure 6.3 INFN-LNS experimental-hall layout. From the CS the beams pass through FRAISE and reach MAGNEX. The beam bump is also shown.

### 6.1.3 *What needs to be modified in the CS*

The extraction by stripping of beams with power of about 10 kW requires a large acceleration chamber inside the cyclotron in order to minimize beam losses and to have a better vacuum conductance. Therefore, we plan to increase the vertical gap in the pole region from 24 mm to 30 mm. This will be achieved by installing two new liners with a reduced thickness with respect to the present ones.

The beam dynamics study indicates that a new extraction channel into the cryostat of the cyclotron is required, having different direction and larger cross section with respect to the existing one. Then, the present cryostat has to be replaced by a new one [185]. The design of the present superconducting coils dates back to 35 years ago: new technologies are available nowadays to achieve the same performance using smaller coils with higher current density. This simplifies the design of the cryostat and reduces the consumption of cryogenic liquids. The smaller size of the new coils allows also to increase the vertical gap in the extraction channel from the current 30.5 mm up to 60.5 mm. This wider clearance makes easier the insertion of the magnetic channels, namely additional iron elements placed after the pole radius that change locally the magnetic field, aiding the radial focusing and slightly steering the beam when necessary.

According to our simulations, two magnetic channels are enough for all ions to be extracted by stripping, even if positions have to be changed according to the particle and energy to be extracted. Two compensating bars have to be placed inside the cryostat to minimize field perturbations introduced by the magnetic channels. Also the position of compensating bars changes according to the extracted beam, see [178], [180].

Since the extraction by deflector is maintained for the other ions of the CS operating diagram [117], [177] the new penetrations and subsystems have to be designed to avoid interferences with the existing ones.

Figure 6.1 shows the present and the final shape of the central ring of the yoke. In Figure 6.1 *a*, the existing deflector penetrations are in the dark blue and green sectors. The beam exits the cyclotron through the hole in the orange sector (the shape of the hill and the inner wall of the cryostat are drawn for reference). In Figure 6.1 *b*, the optimized central ring profile is shown in dark blue. The iron in the central ring (± 12.5 cm above and below the median plane) has been redistributed considering the new penetrations necessary for the extraction by stripping and filling the unused holes. The yellow areas highlight the extra iron to be removed from the central ring to correct the field perturbations introduced by the new non-symmetric penetrations. With this iron configuration, the first and second harmonics of the magnetic field versus radius are kept under control along the entire acceleration path. In particular, the first harmonic is kept below 5 Gauss and the second harmonic is even smaller. Details on the used methodology, the current sheet approximation, as well as further considerations on the studies to be done are published in [186].

## 6.2 *The upgraded magnetic system for MAGNEX*

Exploring the nuclear response to DCE reactions at different incident energies could reveal important details on the relative weight of isospin, spin-isospin and tensor components in the overall nuclear matrix elements. For this reason, it is important to overcome the present limit of 1.8 Tm as maximum magnetic rigidity for the reference trajectory in MAGNEX, corresponding to about 46 MeV/u for the ($^{18}$O,$^{18}$Ne) and 24 MeV/u for the ($^{20}$Ne,$^{20}$O) reactions. In collaboration with Danfysik A/S, NUMEN has considered different options, ranging from the complete replacement of the present room-temperature large size magnets to superconducting ones to lower impact solutions. The best trade-off was to keep the present magnetic configuration and to increase the supplied current, by upgrading the power supply units and consequently the water cooling system. An increase up to ~20% of the magnetic field is still achievable with the present magnets with a tolerable change of magnetic field shape due to iron saturation. The corresponding gain in maximum detectable ion energy is about 40%.

### 6.2.1 *The quadrupole magnet*

The MAGNEX quadrupole system consists of a 20 cm bore radius, 60 cm long water-cooled quadrupole magnet coupled to a large mirror plate to isolate the spectrometer from the feed beamline. At present it is supplied with a maximum current of 1000 A with 90 turns per coil. This gives a pole tip field of 0.947 T and a field gradient in the median plane of 4.96 T/m at an effective magnetic length of 607 mm along the optical axis. If the current is increased from 1000 A to about 1630 A then the pole tip field is increased about 20% to 1.139 T with a gradient of 6.0 T/m in the median plane. With a water pressure drop of 10 bar over the magnet, the temperature increase is estimated to about 24 °C. The needed excitation current can be reduced from 1600 A to for example 1500 A by adding 70 mm thick iron plated on the outside of the yoke. It is expected that this modification is possible with the existing magnet support structure. With a water pressure drop of 10 bar over the magnet the temperature increase is estimated to about 20 °C. Figure 6.4 shows the Opera field strength mappings including the field clamps.

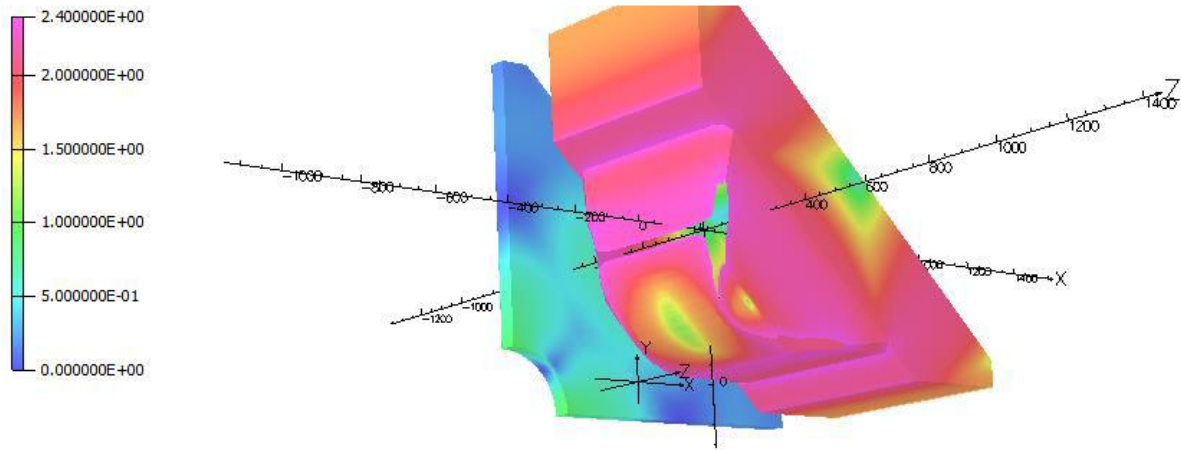

Figure 6.4 Magnetic field up to 2.4 T on the iron surface as calculated in an Opera-3D model of the as-build magnet at an excitation current of 1600 A.

### 6.2.2 *The dipole magnet*

The dipole system consists of a 1.6 m curvature radius bending magnet with flat quadrupole and sextupole corrector coils mounted on the dipole poles and field clamps at each end of the dipole. The magnet is presently supplied with a maximum current of about 920 A with 120 turns per coil, to give a center field of about 1.15 T. The model calculation shows that a 20% increase from a center field of 1.15 to 1.38 T can be obtained with a current of about 1150A. With a pressure drop of 8 bar over the magnet the temperature increase is estimated to about 20°C. Figure 6.5 shows the Opera field strength mappings including the field clamps.

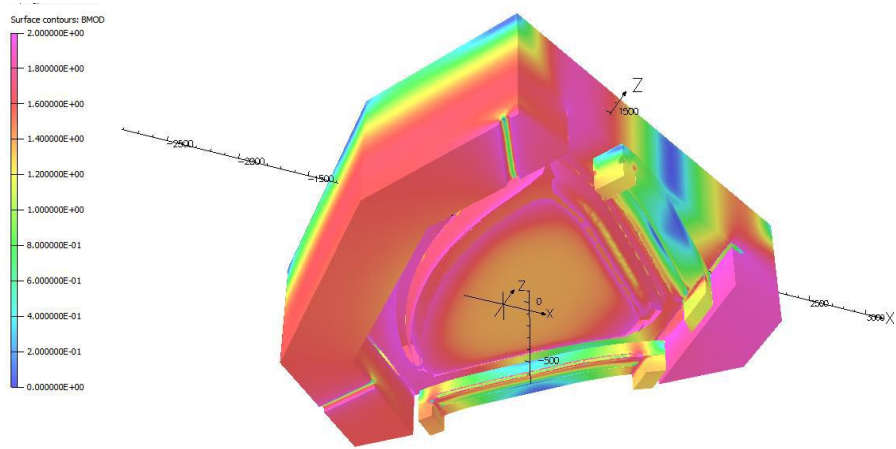

Figure 6.5 Magnetic field up to 2 T on the iron surface of the dipole magnet at center field of 1.38 T.

## 6.3 The new beam dump

A key requirement for the NUMEN experiment is the beam transport, which passes through the MAGNEX spectrometer to the beam dump (see Figure 6.3). The beam dump must substain a power up to 10 kW and have enough shielding to reduce significantly the background to the MAGNEX focal plane detector due to the neutron and gamma radiation produced by the interaction of the beam on the dump.

A borated concrete cube with a side of about 5 m surrounds and shields the beam dump. The installation of this beam dump in the MAGNEX experimental room is challenging. The beam can get out from the MAGNEX spectrometer through two possible exits, one on the high-$B\rho$ side of the focal plane detector and another on the low-$B\rho$ side (see Figure 4.2). The exit side depends on the specific reaction under investigation. To switch from one exit to the other, it will be necessary to dismount a beam pipe line of about 1.5 m and a steering magnet. No other components of the two exit beamlines need to be removed. Different positions of the beam dump were evaluated: the final solution is shown in Figure 6.3 and consists of a beam dump placed on the floor plane, which allows to mitigate the radiation at the focal plane detector and to minimize the changes needed in the MAGNEX experimental room. In particular, to host such a large beam dump the MAGNEX room must be enlarged by about 2 m and the spectrometer must be rotated of 60° with respect its original position. As a consequence, the installation of a new beam line at the entrance of the spectrometer is required (see Figure 6.3). This solution matches the constraints of the existing experimental hall.

## 6.4 Design of the targets

The particular features of the NUMEN project pose three severe constraints on the design of the targets: (1) they will be illuminated by very intense ion beams; (2) they will be very thin, in order to minimize the energy spread of both the beam and the reaction products; (3) the target nuclides are heavy isotopes of several atomic species.

Concerning the use of intense beams at INFN-LNS, it must be recalled that they are made of fully stripped $^{18}$O and $^{20}$Ne ions. Their intensity ranges from ~ 10 enA (NUMEN phase 2) to ~ 60 eμA (NUMEN phase 4). The beam profile has a radial Gaussian distribution and the FWHM of the beam spot is about 2 mm. Therefore, a disk-shaped target is the most suitable choice. A diameter of ~ 1 cm is sufficient to keep the rate of the spurious reactions between the ions in the beam tail and the target frame to a negligible level. The target thickness will be in the range 250 – 1200 nm, depending on the isotope, in order to guarantee the required energy resolution of the detected products.

The main drawback of the use of intense ion beams is the large production of heat during their passage through the target. Therefore, the design of the targets must face with the heat dissipation as the primary problem.

The heat production is mainly due to the energy loss by ionization, while other nuclear processes, like scattering and fragmentation, contribute less than 1%. The heat dissipation technique explored in NUMEN will take into account only the heat from ionization.

The rate of heat produced by an ion beam linearly depends on the beam intensity $I$, on the ratio $Z/A$ of the target, on the density and, for thin targets, on the thickness. The dependence on the beam energy is approximately inverse, while the projectile atomic number $z$ contributes as $z^2$. This leads to a heat production rate for the various combinations of beam, targets and energies, which ranges from 0.03 to 0.14 J/s·μm·μA [187]. Therefore, the beam creates a nearly uniform source of heat inside the target, which must be dissipated in order to avoid severe damages in the target.

*6.4.1 Techniques for the heat dissipation*

Thin targets in nuclear experiments with ion beams are usually layers surrounded by a rigid frame. The heat generated by the beam is dissipated by maintaining the frame at very low temperature. In general, this technique is efficient under two conditions: a good thermal conductivity of the target and low beam intensity. Both requirements will be missed in NUMEN Phase 4, because all target nuclides have conductivity within the range 0.5 – 97 W/(m·K) and the beam intensity will reach 60 eμA. A preliminary study [188] demonstrated that a target-frame system (sketched in Figure 6.6) requires beam intensities of less than ~ 5 eμA to avoid the melting. In particular, at low energy this limit reduces to ~ 1 eμA. In addition, a difficulty in building and handling a self-sustained layer of heavy isotopes must be taken into account. Therefore, the option of a self-sustained target with cooled frame was discarded.

Since the low thermal conductivity is the main responsible of the slow flow of the heat from the beam spot to the cold frame, the considered strategy is based on the deposition of the target nuclide on a thin substrate of highly conducting material, as sketched in Figure 6.7. The cold frame clamps the substrate all around the target, in order to maintain it at low temperature. The aim is to allow for a large amount of heat to flow from the target to the highly conductive substrate and then to quickly reach the cold region. Three tasks have to be accomplished: i) finding a suitable thermally conductive material, ii) exploring the feasibility of a controlled uniform deposition of the target nuclides on this material, iii) checking whether the fast heat transfer is possible. The first nuclide under study was the Sn target and the ion beam was $^{18}$O at 15 MeV/u.

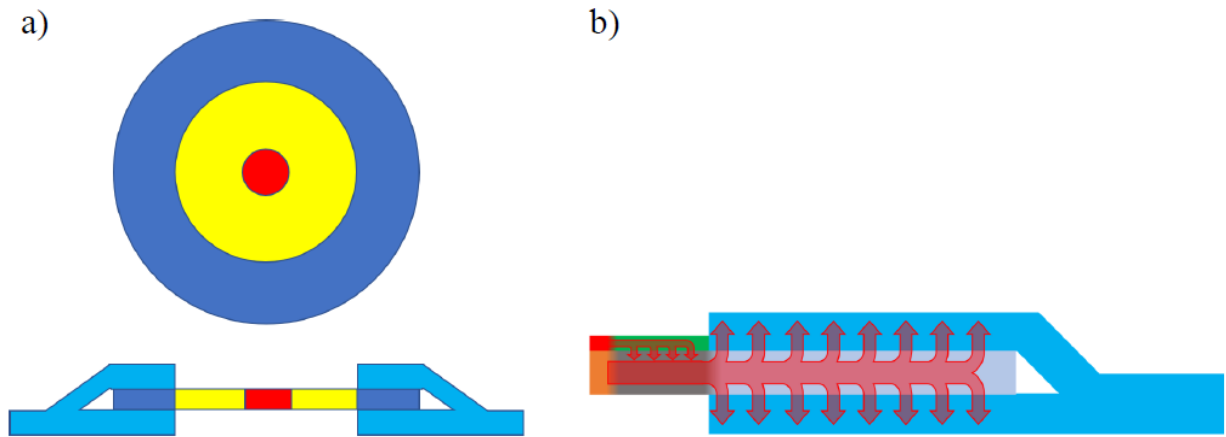

Figure 6.6 (a): scheme (not in scale) of the top view (upper part) and side view (lower part) of the target. The red circle represents the region illuminated by the beam spot, the blue crown is the cooled clamped region, while the yellow part represents the region through which the heat passes from the center to the cold frame; (b) schematic side picture of the flow of heat (magenta/pink arrows) from the center of the target to the frame at cold temperature.

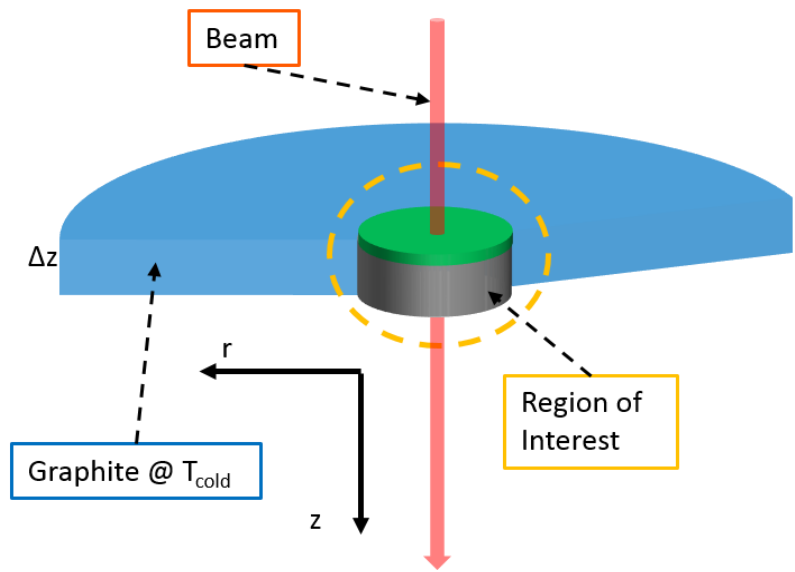

Figure 6.7 Sketch (not in scale) of the target-substrate system. In red the beam, in green the target deposition (Tin), in grey the graphite substrate under the deposition, in blue the graphite substrate clamped by the cold frame.

### 6.4.2 *The substrate*

The properties of the commercial pyrolytic graphite largely meet the requirements of the substrate. It is a stack of graphene layers, whose thickness ranges from 10 to 100 μm and more. The thermal conductivity along the surface of the stack is ~ 1950 W/m·K (nearly two orders of magnitude larger than that of the nuclides of interest), with density 2.2 g/cm$^3$. It has a good mechanical resistance and high melting temperature (~ 3900 K). The thermal conductivity along the thickness is very low, ~ 3.5 W/m·K, but this does not worsen the speed of the heat flow, as we shall see below. In addition, keeping in mind that in this kind of experiments a thin layer, usually carbon, is placed downstream the target, to strip completely the produced ions from the residual electrons, the pyrolytic graphite substrate can fulfil this task.

A disadvantage in using the substrate is the energy spread of the reaction products due to the energy loss. A thickness of 10 μm is a suitable compromise as a stripper and affects the energy resolution within the acceptable tolerance [187]. The interaction of the beam with the pyrolytic graphite can produce spurious product; however, the reaction $Q$-value for DCE in $^{12}$C target is typically much more negative than that on the target nuclides. Therefore, the spurious products would affect only the high excitation energy region of the measured DCE spectra. Supplementary runs with pyrolytic graphite only will be recorded in order to subtract this background.

*6.4.3 Target deposition on pyrolytic graphite*

Test targets were produced by using the Electron Beam Deposition technique, which is quite common in the R&D of electronic devices, where the substrate is preferably Si. Although in nuclear physics research it is common to deposit the target onto a standard carbon substrate, the deposition on pyrolytic graphite is very unusual and challenging and requires a dedicated study to optimize the process.

This exploration is divided in two phases; the first ongoing phase corresponds to search for the best parameters of the deposition and characterization of the natural isotopes of the deposited targets. The second will be the deposition and characterization of the isotopically enriched targets of interest.

Two different samples of pyrolytic graphite were studied, having different physical properties (density, thermal conductivity etc.). One of them was a standalone graphite sheet whereas the second one featured an adhesive layer on one side. For convenience, tests started on the standalone graphite.

In the first target prototype, a nominal layer of 500 nm of tin was deposited on the pyrolytic graphite substrate, which was maintained at room temperature. A Field Emission Scanning Electron Microscopy (FESEM) analysis was performed on this sample. Figure 6.8a shows the top view of the film surface, which appears significantly non-uniform. In order to estimate the dimension of the tin structures, the sample was cut to obtain a side view (Figure 6.8b). It shows bumps and clusters, whose height ranges from a few nm to about 1 μm. This thickness non uniformity affects the energy resolution in the measurement of the reaction products. For this reason, this trial was considered unsuccessful and calling for an improvement of the deposition technique.

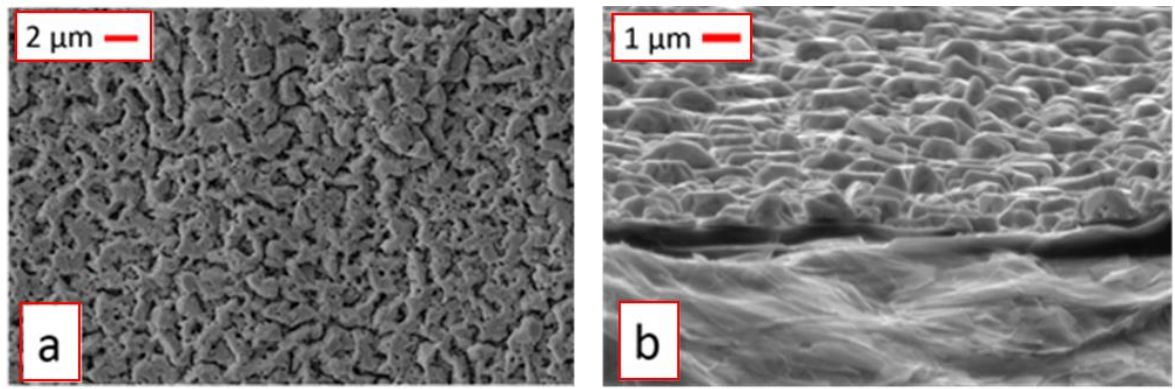

Figure 6.8 a) FESEM image of the Sn film on pyrolytic graphite substrate (top view); b) side view of the same sample; in the upper part the Sn film, the lower part shows the graphite substrate. The Sn has been deposited at room temperature by Electron Beam Deposition.

The main parameters that affect the uniformity of the deposited layer of a given material on a given substrate are: a) the temperature of the substrate during the deposition, b) the rugosity of the substrate surface, c) the presence of a suitable buffer layer (in general few atomic layers of a metal) on the substrate surface, d) the annealing of the sample at suitable temperature.

As a first step, the samples obtained with substrate at room temperature (e.g. Figure 6.8) underwent several annealing processes at different temperatures and durations. No appreciable effects concerning the thickness uniformity were obtained. Then several 500-nm-thick films were deposited on substrates warmed at different temperatures. Some improvements were observed, as reported in Figure 6.9, where the FESEM microscopy of a deposition performed at 150° is shown. The channels between the grains look narrower than Figure 6.8a, indicating a better uniformity of the coverage. The structures visible in fig 6.8b are no longer present and the thickness of the tin film appears more homogeneous.

While the last result looks satisfying, further tests are necessary to confirm the reproducibility of the deposition process. Several systematic studies are planned for different thicknesses of the deposited films and for pyrolytic graphite substrates with different physical characteristics.

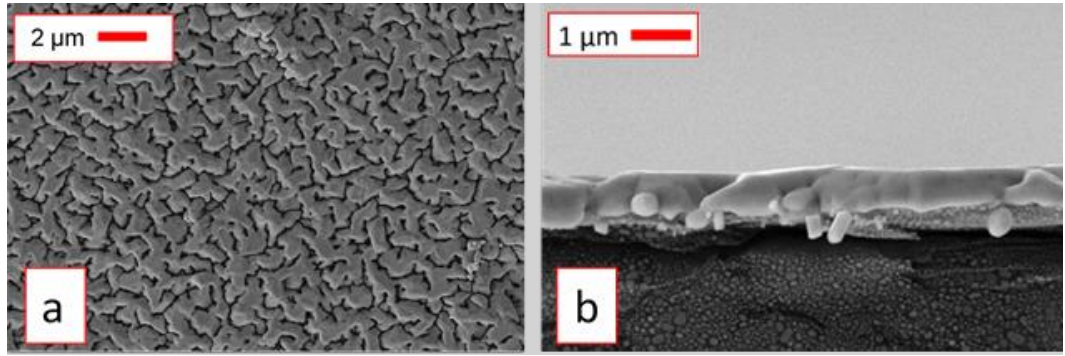

Figure 6.9 a) FESEM image of the Sn film deposited on pyrolytic graphite substrate at 150° (top view); b) side view of the same sample; in the upper part the Sn film (light grey), the lower part shows the graphite substrate (dark grey). The side view was captured on the edge of the sample. In addition to the deposition on the surface, some drops of tin were formed on the side of the graphite.

### *6.4.4 Heat transfer*

The fast transfer of the heat is the most crucial point in the design of the targets, because of the low melting temperature of each nuclide. As mentioned above, the direct transfer through the target is slow, due to the low thermal conductivity of the target material. The heat hence cumulates in the region under the beam spot, where the temperature increases above the melting point.

In the scheme of Figure 6.7, the heat is expected to flow mainly through the pyrolytic graphite thanks to its high conductivity. The spatial distribution and the time evolution of the temperature is governed by the heat equation, whose solution is reported in [187]. Thanks to the cylindrical symmetry of the system, the temperature depends only on the radial distance $r$ from the beam $z$-axis, on the depth $z$ along the beam and on the time $t$. The target-substrate system has a discontinuity of the conductivity between tin and graphite and between the radial and $z$ direction inside the graphite. Moreover, the heat source in the region under the beam spot has different values in tin and in graphite. These discontinuities suggested to solve the heat equation by using a numerical technique. The dimensions of the system were similar to those of the deposition sample: 5 mm for the radius of tin and graphite, 500 nm and 10 μm for their thickness. The initial temperature (before the beam starts) was 100 K everywhere and the cold region (clamped by the frame) maintained the same temperature at all times. This temperature is easily achievable in the cold frame, as boundary condition, by using commercial cooling systems.

The results of the calculation demonstrate that:
a) temperature distribution in both tin and graphite reaches the steady state within 5 s
b) in the steady state the temperature along $z$ is nearly uniform

c) the temperature increases everywhere vs. time
d) the hottest point is always the centre of the beam spot, on the surface of the tin
e) the maximum temperature does not overcome 430 K, below the melting point (~ 505 K) of Sn and below the points where Sn starts to deteriorate.

These results show that the technique of the "conducting substrate" is successful in cooling the very thin targets of NUMEN. The plan for the future includes the investigation of the other target nuclides, following similar steps to those of the tin.

An experimental activity to explore the response of target prototypes under comparable beam power dissipation from low energy heavy-ion beams is going to start soon at UNAM facilities.

## *6.5 NUMEN Focal Plane Detector tracker*

A new 3D tracker for the MAGNEX focal plane detector, designed to work with the upgraded facility is under development within the NUMEN Phase 2 project.

The present FPD gas tracker [108], based on a series of drift chambers and on the use of long multiplication wires, is intrinsically limited to a few kHz rate, due to the slow drift of positive ions from the multiplication wires to the Frish grid (see Section 4.1)

The new tracker should allow high resolution measurements of the phase space parameters at the focal plane ($X_{foc}$, $Y_{foc}$, $\theta_{foc}$, $\varphi_{foc}$) needed for accurate ray-reconstruction also at the high rate conditions foreseen after the facility upgrade. The identification of the reaction ejectiles in charge, atomic number and mass, which is a crucial aspect for heavy-ion detectors, will be accomplished by a dedicated particle identification wall, as discussed in Section 6.6.

The new tracker will be a large volume gas-filled detector covering the present FPD size (1360mm × 200mm × 100mm) and will basically consist of:
- A gas drift region delimited by a cathode and an electron multiplication plane,
- An electron multiplication element, namely a Micro-Pattern Gas Detectors (MPGD),
- A segmented readout board

The incident charged particles coming from the dipole cross a thin Mylar window (1.5 to 6 μm thickness, depending on the particular case) and leave a track of ionized atoms and primary electrons in the low-pressure gas (typically from 10 to 100 mbar) between the cathode and the electron multiplication element. Under a uniform electric field, the electrons drift with constant velocity, whose actual value depends on the voltage and gas pressure. Thus, the drift time of electrons and consequently the vertical position and angle are measured. Reaching the multiplication element, electrons are accelerated in the strong electric field in correspondence of the holes. The resulting electron jets are then directed towards the segmented readout board where the horizontal position and angles are measured.

The choice of the gas mixture is one of the aspects under investigation. The use of a gas quencher such as pure isobutane at low pressure seems promising for operations with heavy ions at low pressure, even if its response to high rates needs to be studied.

A prototype of reduced size (100 mm × 100 mm × 100 mm), conceived to guarantee a direct scalability to the full scale final detector, is under construction in order to identify the most performing and reliable solutions. In particular, the geometry guaranteeing a uniform electric field in the drift region, the applied voltages, the gas mixtures and working pressure, the gas flowing system, the multiplication technology and the read-out and front-end electronics will be the main features to be tested with the small size prototype.

*6.5.1   The drift region*

The drift region delimited by a cathode and a MPGD plane is designed to set a uniform electric field of about 50 V/cm. The uniformity of the electric field in the drift region is guaranteed by a field cage made by a printed circuit in the lateral planes and by a partition grid consisting of gold-plated tungsten wires arranged at 5 mm one from the other in the front and back planes. It provides a smooth distribution of the voltage and a safe value of the current (about 10 µA) flowing in the circuit. A double series of wires is mounted in the front and back sides of the gas chamber to reduce the disturbances of external potentials as those generated by the high-voltage-supplied PID detectors or by the Mylar entrance window.

Electrostatic simulations based on the Poisson-Superfish code [189] have been performed in order to model the drift region. The code calculates the static electric field in the detector geometry by generating a triangular mesh and solving the field equations by a procedure of successive over relaxations for each mesh point. An example of output plot is shown in Figure 6.10. The cathode is modeled as a conductive element at -2000 V. The shaping wires at increasing voltages from -2000 V to -1000 V are shown in section and generate a fairly uniform field in the internal drift region, as displayed by the almost-parallel equipotential lines. A much stronger electric field is simulated in the electron multiplication region. On the right-hand part of the figure, a conductive element simulates the presence of the PID wall, with an applied voltage of -2000 V. This high value of voltage is set in order to test the field uniformity inside the cage even in the worst cases of high bias voltage needed for the PID detectors. Thanks to the double series of shaping wires, the perturbation generated by the high negative voltage applied to PID wall does not affect appreciably the internal region, where the field is maintained uniform.

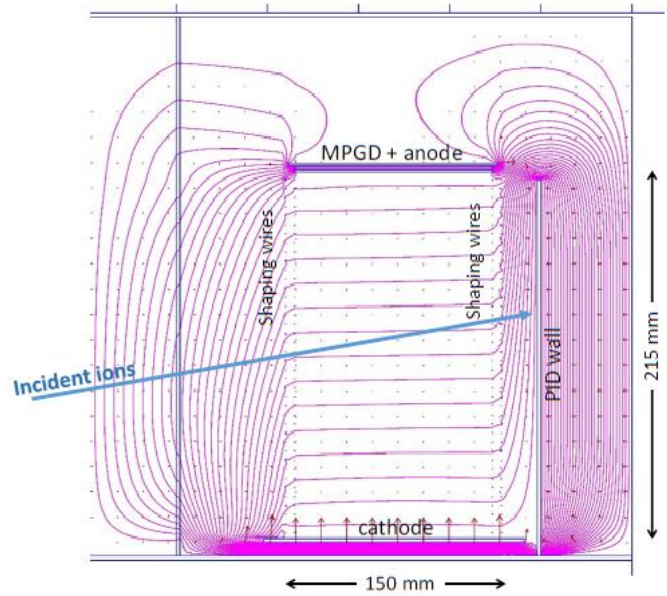

Figure 6.10 Output plot of the Poisson-Superfish calculations for the FPD electric field. The magenta lines are the equipotential lines. The arrows represent the calculated electric field.

*6.5.2 The electron multiplication region*

The above-mentioned rate limitation of the present tracker can be overcome by replacing the multiplication wires with MPGD.

Promising examples of MPGD are the GEM (Gas Electron Multipliers) [190] or Thick-GEM (THGEM) foils [191], [192]. Common feature among these structures is a narrow amplification gap of typically 50–100 μm for GEM and 400-600 μm for THGEM, compared to many millimeters for wire-based structures. The short drift path for ions overcomes the space charge effect present in wire chambers, where the slowly drifting ions may remain in the gas volume for milliseconds, and affect the electric field.

In recent developments [193], detectors based on this technology have been proven to work up to several MHz/mm$^2$, i.e much beyond the expected rates in NUMEN.

The high rate capability of the MPGDs, makes them a very attractive technology for the MAGNEX FPD. However most of the GEM-based detectors operate at atmospheric pressure and beyond: this is not feasible for NUMEN, where the ideal working pressure for the spectrometer energy resolution is about 10-50 mbar. In addition, GEMs are often used at energies where all particles behave as Minimum Ionizing Particles (MIP). In the cases foreseen for NUMEN, ions with much larger ionizing power than MIP will be detected. In addition, different ions reach the detectors during the same experiments, thus requiring a broad dynamic range for the tracker detector (typically larger than 30:1). Low-pressure THGEM have been already used to detect not-MIP particles [194], thus making the use of THGEM very appealing for NUMEN.

The development of suitable technologies for the construction of a MPGD-based tracker, working at low pressure and wide dynamic range, will be a key issue of the R&D activity during NUMEN Phase 2.

*6.5.3 The segmented readout board*

One of the main objectives is to design a modular, scalable, radiation-hard architecture for the readout, which, in addition, meets the demanding requirements in terms of high event rate, easy maintenance and precise synchronization. The design takes advantage of the possibility to use a brand new full-custom Front-End (FE) and Read-Out (RO) electronics, described in details in Section 6.8.

The electron jets emerging from the MPGD foil are directed towards a first layer of 750 μm pitch strips, corresponding to a capacitance of 22 pF. Each strip of this layer is capacitively coupled to a twin strip in a second layer. The charge pulse induced in the twin strip is then integrated by the

FE and shaped. The shaped signal is compared to a suitable threshold and the logic high output of the comparator identifies the hit strip. In this scenario, the position is extracted by one strip only, without the need for the calculation of the center of mass.

The capacitance of each channel of the tracker is optimized to match with the performance of the selected FE Application Specific Integrated Circuit (ASIC). The drift time is also measured by the FE at sub-nanosecond resolution.

An innovative scheme for the connection of FE electronics to the anode board was developed. The main objective was to place the FE in air, such to simplify the heat dissipation, the maintenance and, above all, the interconnections to RO. An advantage of this strategy is also the possibility to adopt countermeasures with respect to high level of radiation during the experiment. The working principle of the new FPD segmented readout board is displayed in Figure 6.11.

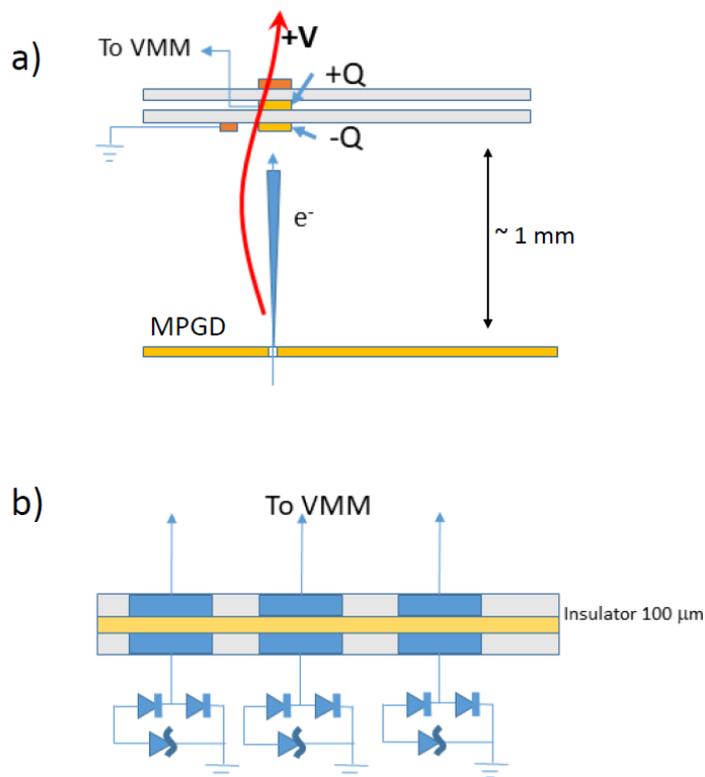

Figure 6.11 (a) Working principle of the new FPD segmented readout board, (b) detailed view of the Electrostatic Discharge Strategy (EDS)

## *6.6 Particle Identification*

As discussed in Section 6.5 the present gas tracker of the MAGNEX spectrometer must be upgraded in order to cope with the challenging high rate of heavy ions expected in NUMEN Phase 4 experiments. The use of micro-patterned electron multipliers, keeping the present geometry of the drift sections of the detector, is well-suited for tracking purposes. However, this technology is unable to provide accurate information on ion energy loss, reducing the overall particle identification capabilities of the set-up. In addition, the large area (50 × 70 mm$^2$) Silicon detectors used by MAGNEX as active ion stopping devices are not suitable neither for high fluencies of heavy-ion tracks nor for high-rate applications. Deterioration of the detector response is indeed observed starting from $10^9$ implanted ions (far from the overall $10^{13}$ ions expected by the NUMEN Phase 4 experimental campaigns), while the expected signal pile-up probability in a single detector is too large for the NUMEN Phase 4 expected rates. As a consequence, PID must be demanded to a dedicated wall of telescope detectors downstream the tracker, not available in the present MAGNEX configuration.

Many aspects must be considered to design a suitable detection system for NUMEN, matching the fundamental PID requirement to identify ions unambiguously in the region of O, F and Ne atomic species. The most relevant are related to:

  i. the radiation hardness, since the expected overall heavy-ion fluency will be of the order of $10^{12}$ ions/(cm$^2 \cdot$ yr);
 ii. the energy resolution $\delta E/E$, which must be better than ∼ 2%, in order to maintain the present performance in terms of atomic number and mass identification ($\delta Z/Z$ ∼ 1/48 and $\delta A/A$ ∼ 1/160 [111]) or at least to allow a clear identification of the ejectiles of interest for NUMEN, characterized by an atomic number $Z$ ∼ 10 and mass number $A$ ∼ 20. The energy resolution should also be good enough to guarantee the same sensitivity in the cross section measurements, which is limited by the spurious events inside the identification graphical cuts, as discussed in Section 4.5;
iii. the time resolution, such to guarantee an accurate Time Of Flight (TOF) measurement of the ejectiles from the target to the focal plane and the drift time of primary electrons in the gas tracker. The TOF measurement with resolution better than 2-3 ns [128] is necessary to effectively suppress the background in the coincident events between MAGNEX and the γ-ray calorimeter (see Section 6.7). The drift time is also used to reconstruct the vertical track of the ejectiles, for which a time resolution better than 5 ns would be acceptable [108];

iv. the degree of segmentation, in order to keep the double-hit event probability below 3% in the whole FPD, modules of about 1 cm$^2$ area are proposed;

v. the geometrical efficiency, which should be high enough to obtain accurate measurement of the absolute cross section and to reduce the background coming from events with partial charge collections, which could reduce the overall sensitivity of NUMEN to rare DCE events;

vi. the detectors thickness, which must be chosen in order to stop the ejectiles of interest in a wide dynamical range of incident energies (15 to 70 MeV/u);

vii. the scalability, which should guarantee that several thousand detectors can be easily built, assembled and managed at reasonable price, also in terms of time required for the calibration procedures;

viii. the coupling with the FPD tracker, which requires that the PID wall should work in a low-pressure gas environment (typically $C_4H_{10}$ at 10-50 mbar), where the presence of high voltages could be an issue.

Several nuclear physics experiments [195] [196] [197] [198] have adopted the telescope solution to study and identify reaction products. This consists of at least two detectors assembled such that the particles of interest cross the first and stop into the second. The correlation between the energy loss signal in the thin detector ($\Delta E$ stage) and the residual energy ($E_r$) deposited in the stopping one is connected to the atomic number $Z$ of the detected ion through the Bethe-Bloch formula [199]. Due to the good energy resolution and linearity, thin Si detectors are typically used as $\Delta E$ stage, followed by a thick Si or scintillator detector (CsI, NaI, etc.) or even a gas detector. This configuration easily provides a good $Z$ identification, acceptable energy resolution and a high stopping efficiency. In some activities [196] [197] this solution has been further improved also by pulse shape analysis [200] for the identification of ions stopping in the first stage of telescope. Nevertheless, all these solutions are limited by the radiation hardness of silicon.

Another possibility is the use of an array of plastic + inorganic phoswich scintillators [201] readout by means of Silicon Photo Multipliers (SiPM). A phoswich detector is the combination of two different scintillators, chosen to have different decay times, optically coupled to a single photodetector [199]. In this way, the shape of the output pulse from the SiPM is dependent on the relative contribution of scintillation light from the two scintillators.

During the R&D of NUMEN Phase 2, two main approaches are under investigations for the PID wall, namely Silicon Carbide (SiC) telescope and the phoswich array. Telescopes based on thin SiC detectors and inorganic scintillators are also under consideration.

*6.6.1 Silicon Carbide Detectors*

Among the "robust" radiation-hard materials, SiC has recently received special attentions also thanks to technological improvements. SiC is a wide-band semiconductor and due to its composition it is the only stable compound in the binary phase diagram of the two group IV elements, silicon and carbon. It is thermally stable up to about 2000 °C, even in oxidizing and aggressive environments. Among all the wide band-gap semiconductors, silicon carbide is presently the most intensively studied and it has the highest potential to reach market maturity in a wide field of device applications [202].

The first requirement for the new PID wall is the radiation hardness, i.e. the inertness of the detectors to high doses of particle irradiation. This is strictly related to the damage of the lattice created by traversing particles. SiC, due to its wide gap and strength of its chemical bonds, is a very valid alternative to Si for the production of radiation-hard detectors.

The usual design of a solid-state detector includes a diode structure operating under reverse bias, where a space charge region is formed. Ionizing particles produce ionization in a semiconductor when they are slowed down or absorbed. Thus, electron-hole pairs are formed and are then separated by the electric field and collected at the electrodes, yielding a current pulse in the detection circuit. The current generated is directly correlated with the deposited energy. A detector should have a low concentration of impurities and defects, as they cause a decrease in the current pulse amplitude due to recombination of electron-hole pairs and scattering of charge carriers. Moreover, a low concentration of dopant impurities extends the thickness of the space charge region, i.e. the detection active region. The wide band-gap of SiC (3.28 eV) is useful, as it reduces significantly the rate of thermal noise. On the other hand, it also represents a disadvantage: a particle with a certain energy, ideally converting all its energy for the generation of electron-hole pairs, generates about 3 times more charge carriers in Si (band-gap 1.12 eV) than in SiC. Detectors based on SiC, therefore, have lower pulse amplitudes. However, the heavy ions to be detected in NUMEN generate a large number of primary charge carriers, whose statistical fluctuations are not an issue. Furthermore, SiC-based detectors still have a high Signal to Noise Ratio (SNR) at temperatures which are unattainable for Si-based devices, which instead need external cooling to keep the intrinsic carrier level sufficiently low [203].

Traversing particles not only ionize the lattice but also interact with the atomic bodies via the electromagnetic and strong forces. The result is that atoms are displaced and create interstitials, vacancies and more complex structures. In addition, diffusing Si atoms or vacancies often form combinations with impurity atoms, like oxygen, phosphorus or carbon. All these lattice displacements or defects dislocation populate new levels changing the initial semiconductor properties. The resulting macroscopic changes are: i) enhancement of the leakage current; ii) change of the depletion voltage, mainly due to the creation of additional acceptor levels; iii) decrease of the charge collection efficiency, due to new defects acting as traps for the generated carriers.

The result of Monte Carlo simulations of defects dislocation generated by $^{18}$O and protons in a ΔE-E SiC telescope is shown in Figure 6.12 (left panel). The number of dislocations created on the ΔE or E stage of telescope by protons is about two orders of magnitude smaller than for the heavy oxygen ions.

Further interesting considerations arise from the leakage currents and their dependences on the ions fluency [204]. The leakage current of a p-n junction consists of a diffusion term coming from the quasi-neutral areas and of a generation term coming from the depletion region [205]. The diffusion term depends essentially on temperature and band gap. Increasing the band-gap from 1.1 eV of Silicon up to 3.2 eV of 4H-SiC corresponds to a reduction of this component by several orders of magnitude at room temperature. The generation term is instead quite sensitive to the damages created by particles passing through. In Figure 6.12 (right panel) the calculation of the relative increase of leakage current of a SiC detector as a function of the ions fluency for oxygen and proton beams is shown. As expected, the leakage current increases by several orders of magnitude after high doses irradiation. Such an enhancement can be still tolerable by a SiC detector, due to the lower absolute values of leakage current (about five order of magnitude less than Si).

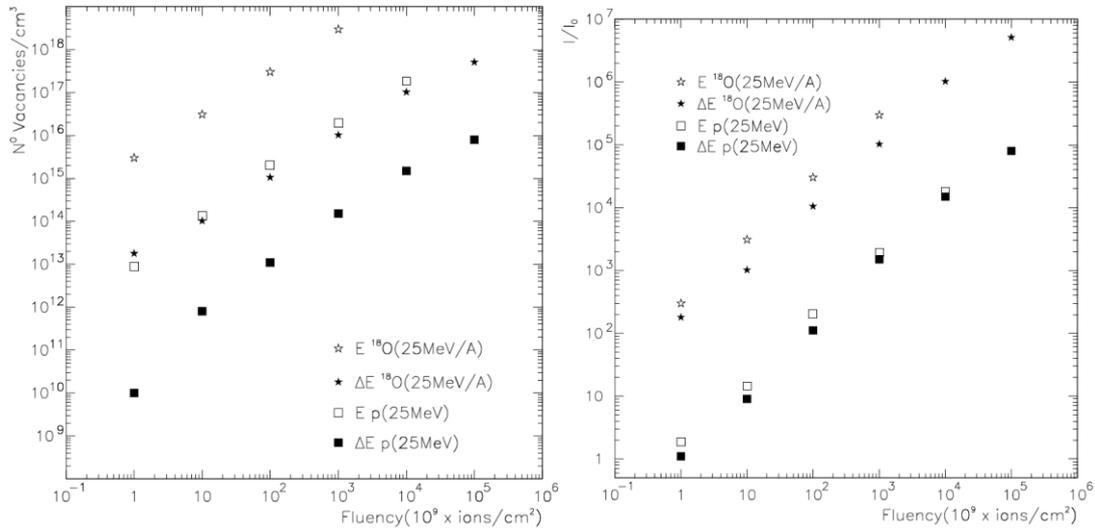

Figure 6.12 (left panel) Monte Carlo simulations of the number of defects generated in a $\Delta E$-$E$ SiC telescope as a function of ion fluency; solid symbols indicate the first stage of the detector, assuming a thickness of 100 μm, while the empty symbols the second stage assuming a thickness of 1000 μm. The simulated data are related to the oxygen and protons beams at 25 MeV/u. In (right panel) the predicted reverse current increments are shown, for the same conditions of panel (a).

Defect analysis and even defect engineering was long investigated by several R&D collaborations, e.g. Rose, RD48 and RD50 at CERN [206].

Several studies on radiation hardness of SiC to the MIPs are available in literature [206]. Non Ionizing Energy Loss (NIEL) effects in case of a MIP are small and the cooling of silicon detectors at about -20 °C is still one of the best solutions to the radiation damage of detectors. The situation will be completely different for NUMEN, because it works with heavy ions up to tens of MeV/u, where a huge number of defects is created and the NIEL is maximum.

Radiation hardness of SiC devices irradiated with heavy ions stopping in small SiC detectors ($2 \times 2$ mm$^2$, 30 μm thick) was investigated in Ref. [207]. The results proved that the detectors are able to accept fluencies as large as $10^{14}$ heavy ions/cm$^2$, thus matching with main requirements of NUMEN in terms of radiation hardness.

For time resolution, SiC detectors can profit from the high saturation velocities of the charge carriers ($2 \times 10^7$ cm/s) in the semiconductor – two times higher than in silicon - and to the possibility to effectively operate the devices at or close to the carrier velocity saturation condition.

This is because the breakdown field in SiC is 2 MV/cm, seven times higher than in Si or GaAs: the junctions on SiC can hence reach extremely high internal electric field in the depleted region. Electric field as high as $10^5$ V/cm has been reached without suffering junction breakdown or significantly increasing the reverse current [208]. A timing resolution of hundreds of ps has been measured for SiC pixel detector [209].

All these considerations and preliminary results actually support the INFN R&D-activities in the field of SiC technology in order to build the first ΔE-E SiC telescope [210]. The thickness of the telescope must be chosen in order to permit the detection of the ejectiles in the wide dynamical range of incident energies, i.e. 10 to 40 MeV/u for the ($^{20}$Ne,$^{20}$O) DCE reactions and 10 to 70 MeV/u for the ($^{18}$O,$^{18}$Ne) DCE reactions (see Section 3.2). In both kinds of reactions, the range of the ejectiles in silicon carbide varies from ~150 μm to ~ 2700 μm. An appropriate thickness for the ΔE stage would be ~100 μm, which correspond to an energy loss $\Delta E$ ~25 MeV for $^{20}$O at 40 MeV/u and ΔE ~180 MeV for $^{18}$Ne at 10 MeV/u. For the second stage ($E$) a thicker SiC detector able to stop the ejectiles for the whole energy range for the NUMEN experimental campaign is required.

The R&D work, which is part of the SiCILIA collaboration [210], aims at developing innovative processes targeted at the massive production of both the thin (~100 μm) and thick (~500-1000 μm) SiC detectors with large area (about 1 cm$^2$) and unprecedented low level of defects.

The thin detector will be based on deposition of epitaxial layers. In the past few years steep improvements in the density of defects of the substrates and of the epitaxial layers have been achieved, with a consequent large reduction of micropipes and stacking faults. As a consequence, a new opportunity of constructing bipolar devices (p/n junctions, transistors, …), that greatly benefit by the reduction of this kind of defects is open. It is now possible to build detectors characterized by lower leakage current and a better SNR. For the thick stage of the telescope the use of an intrinsic wafer instead of the usual n$^+$ substrate is under investigation. This choice would give the opportunity to have very thick undoped layer (500-1000 μm). First prototypes have been constructed and preliminary characterization in terms of resolution, timing and radiation hardness is encouraging [210].

*6.6.2 Phoswich array*

An alternative strategy is also being pursued for the NUMEN PID wall. Such a solution follows somehow more traditional guidelines even though it is innovative from some perspectives.

The proposed detector is based on the well-known phoswich technique, where a fast and a slow scintillator are coupled to form a telescope. The selected scintillators composing each cell are a 200 μm PILOT-U (fast, ~1.8 ns decay time) and a 5000μm CsI(Tl) (slow, ~3μs decay time), with active area 1 cm × 1 cm. The overall detector would consist of about 2500 of such modules. The thicknesses of the two stages of the proposed phoswich are suitable for the whole dynamical range of incident energies mentioned before (from 10 to 70 MeV/u). The range of the ejectiles in the CsI(Tl) scintillator varies from ~200 μm for the lower energies to ~3000 μm for the highest. The energy loss ΔE in the 200 μm Pilot-U stage is large enough to have a good signal ($\Delta E$ ~20 MeV for $^{20}$O at 40 MeV/u and $\Delta E$ ~120 MeV for $^{18}$Ne at 10 MeV/u).

The scintillation light readout is performed by means of a 6mm × 6mm SiPM produced by SensL [211]. Two single channel prototypes were initially built and tested with the products of an $^{16}$O beam at 320 MeV impinging on a $^{27}$Al target and with a $^{7}$Li beam at 46 MeV on a ($^{7}$LiF + C) target [212]. The shape of the output signal from the phoswich detector depends on the relative contribution of scintillation light from the fast plastic scintillator (Pilot-U) called *fast component* and the *slow component* from the second stage of CsI scintillator. A BafPro filter & amplifier module [213] [214] was used to replicate the detector signal in two copies, filtering out from one copy the fast light and amplifying both. Indeed, the particle discrimination power results better in this way than in the case of the more traditional fast signal selection. The scatter plot of the total light (X-axis) versus the slow light (Y-axis) is shown in Figure 6.13 (upper panel). A clear separation of the different ion species detected is clearly observed, for a fixed value of the slow component (energy released in the CsI scintillator), the higher Z ions have a larger total light because they lose more energy in the Pilot-U stage. The energy resolution in the reference case with $^{16}$O beam at 320MeV, measured on the slow signal, which represents the residual energy in the thick CsI(Tl) layer, is of ~1.6% and meets very well the required 2% [215]. The encouraging results led to the second step, consisting in the construction of a 10 × 10 array of identical phoswich detectors. The array, shown in Figure 6.14, was coupled to a corresponding array of SiPMs by means of a special frame built by a 3D printer.

As for the radiation hardness, direct irradiation tests have to be performed. Literature data indicate that Cs(Tl) could survive the foreseen data taking life of NUMEN. The fast Pilot-U layer is not expected to show the same hardness, and therefore it will likely be replaced by a thin layer of pure CsI, which has similar features in terms of light yield and decay time but a much better radiation hardness [216]. However, due to the rather low cost of this solution, one could also plan to replace parts of the detector during its operational life. Tests of this prototype array with gamma-ray sources have shown a good response uniformity, and in-beam tests will start soon.

Regarding time resolution of the proposed detector, the response of the thin ΔE stage is very fast, since its decay time is a few ns, thus ensuring a satisfactory discrimination in Z as shown in Refs. [212], [215]. The time resolution of this detector is potentially better than 1 ns, which is consistent with the NUMEN requirements.

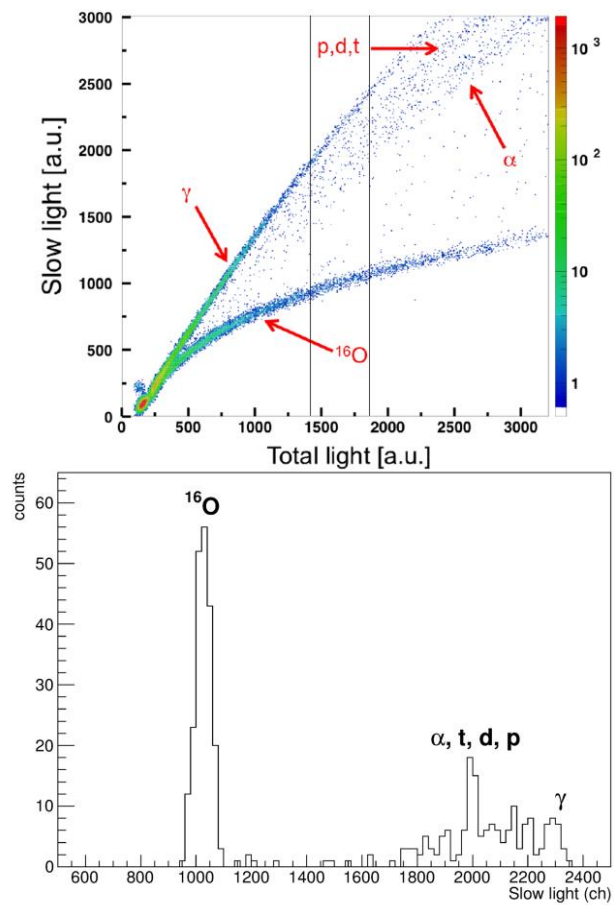

Figure 6.13 Upper panel: slow light versus total light for test with $^{16}$O at 320 MeV incident energy on $^{27}$Al target. The separated loci correspond to γ, (p, d, t), α and $^{16}$O. Lower panel: Projection on the slow axis of the upper panel with a selection on the total light axis as indicated by the black lines in the upper panel.

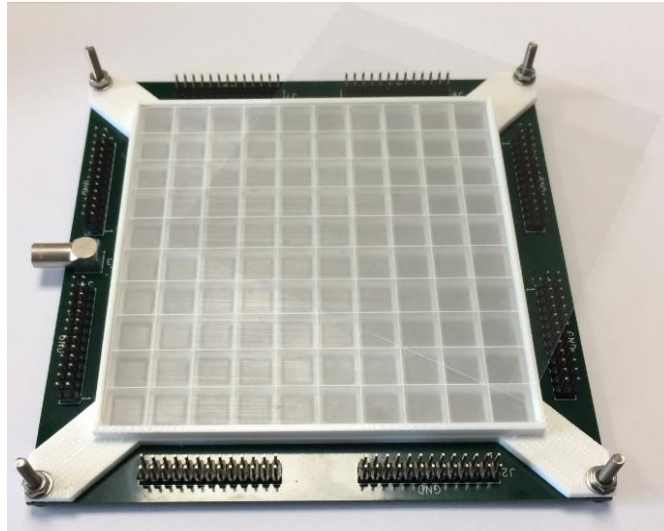

Figure 6.14 The 10x10 array of phoswich detectors, assembled to the board containing a corresponding 10x10 array of SiPMs by means of an adapter frame built by a 3D printer. The foil of PILOT-U fast scintillator, to be coupled to the matrix, is shown on top.

## *6.7 The Gamma Calorimeter for NUMEN*

The nuclear cases targeted in the NUMEN project consist of deformed and non-deformed nuclei that will be populated by means of the DCE or competing reactions. For non-deformed target nuclei and at low reaction energies, the typical energy resolution of MAGNEX with beams provided by the CS (about 0.2%) is sufficient to discriminate between the ground state ($I^\pi = 0^+$) and first excited states ($I^\pi = 2^+$) of both projectile-like and target-like species. However, for nuclei of interest in moderately and strongly deformed mass regions, such as $^{110}$Pd, $^{150}$Nd and $^{160}$Gd, and in nearly all cases at high reaction energies (40-50 MeV/u), this energy resolution is not sufficient. In such cases, a gamma detector array has been considered as an ancillary device to the magnetic spectrometer providing the necessary discrimination between nearby energy sates. Such an array should have characteristics similar to that of a calorimeter, with a large angular coverage. Inorganic scintillators, e.g. LaBr$_3$(Ce), CeBr$_3$, LYSO(Ce), LuAG(Ce), and GAGG(Ce) are considered as possible candidates for the gamma detectors.

The main requirements of the gamma calorimeter array are: sufficient energy resolution, large solid angle coverage and high photopeak efficiency, high count-rate capability and high granularity, good timing resolution and high radiation tolerance. They will be discussed below.

    i. Energy resolution. The first requirement of the NUMEN calorimeter is a sufficient energy resolution to separate the gamma transitions of the cascade from excited states populated in the DCE reaction. The most difficult case which can be anticipated is the one of $^{160}$Gd, with the first excited state ($I^\pi = 2^+$) at an energy of 75.3 keV and second excited state at 248.5 keV. A gamma-ray energy resolution of 25-30% should be sufficient to clearly resolve those states, and therefore even moderate resolution detectors such as inorganic scintillators can be employed.

    ii. Large solid angle coverage and high intrinsic detection efficiency. The efficiency of the array should be as high as possible in order to allow for the measurement of very low cross section processes with sufficient statistics. When the ground state of a DCE reaction is directly populated there is no emission of a gamma-ray, and, to determine its cross section, the calorimeter has to be used to veto the events correlated with the gamma transitions from excited states. Both the g.s. and excited state cross section measurements are affected by the statistical significance of the gamma spectrum, and therefore, by the calorimeter efficiency. For some cases, the solid angle coverage of the calorimeter is particularly important. The $2^+ \rightarrow 0^+$ transition of $^{160}$Gd, for example, has a large electronic conversion coefficient ($\alpha_{TOT}$ = 7.31), as is typical for low energy E2 gamma rays and high-Z elements. As a consequence, only 12% of the decay goes through the emission of the 75.3 keV gamma-ray, which strongly reduces its effective detection efficiency. Considering also that such low-energy gamma rays may be strongly absorbed by material interposed between the gamma emission point at the target position and the gamma scintillator crystals, it is clearly necessary to have a solid angle coverage as close as possible to 4π. For such low energy gamma rays the typical intrinsic photopeak

        detection efficiencies of the detector sensitive material is close to 100%. For higher energy gamma rays (up to about 1200 keV), high effective Z materials are required to enhance the photopeak detection efficiency.

iii.    The DCE cross sections could be very small, in the nb range or even below, as it has been also verified by the first experimental tests already performed to date. Contrary to that, typical total reaction cross sections are very large, in the few barns range. The average gamma-ray multiplicity in a reaction event is also expected to be high, ranging from 15-30 units at low reaction energies to a few units at the highest energies. If the pyrolytic graphite foil of 10μm is used as a backing to the target, in order to allow for sufficient heat dissipation (see Section 6.4), there will be also a significant production of gamma rays and neutrons from in-beam interactions with this foil. For a typical beam intensity such as of $5\times10^{12}$ beam particles/s (achievable after the CS cyclotron upgrade), a total gamma emission rate of the order of 1 GHz is expected. Therefore, a high detector granularity of the array will be required to cope with such a gamma-ray rate, in the calorimeter in a typical DCE experiment. The count rate in each detector should be kept low enough to avoid excessive pulse pile up as well as counting rate overload in the respective electronic channel. Pile up is an important concern, since the integration time of the scintillation pulse necessary for an adequate gamma energy resolution measurement (typically 100-200 ns) covers a few beam bunch periods. If a large detection solid angle coverage has to be designed, a couple of thousand detectors will be necessary to share the total count-rate. The large granularity is also useful to limit the Doppler broadening of DCE beam-like ejectile transitions, and opens the possibility for gamma-ray tracking.

iv.    Good time resolution. The singles gamma spectrum expected for the DCE experiments is an almost continuum of energies due to the extremely fragmented cross sections to the many exit channels of the nuclear reaction, as has already been confirmed in preliminary measurements. The separation of the rare events of production of DCE states from intense, nearly-continuous background will be only achievable if the time resolution of the system is sufficient to discriminate between subsequent beam pulses from the cyclotron accelerator, which have a typical period of 25 ns. Given the typical time resolution in the sub-nanosecond to 2 ns range, this should not be a problem for inorganic scintillators.

    v.    High radiation tolerance. Due to the high beam intensities to be used in the DCE experiments, the target region will become a very large source of radiation, including gamma rays, fast neutrons, electrons, light and heavy ions. The charged particles as well as the low energy X rays can be absorbed by a sufficient amount of solid material between the target and the calorimeter. Inorganic scintillators are normally quite tolerant to gamma and fast neutron radiation and are chosen to be the best candidates for the detector sensitive materials [217], [218]. Detailed simulations and tests are ongoing in order to properly quantify the radiation effects on detectors and electronics.

### *6.7.1 The observational limit*

The observational limit is defined as the ratio of the lowest cross section that can be measured to the total reaction cross section: $\alpha_{lim} = \sigma_{min}/\sigma_{tot}$, in the presence of both, correlated and uncorrelated background in a given experiment. This figure of merit is the most important one for the spectrometer design. Usually the observational limit is a compromise between a statistical one (related to a required number of counts), and, a background one (related to a required peak-to-background ratio), important for typical gamma arrays which are dedicated to the measurement of high multiplicity gamma cascades and associated to the "resolving power" of the spectrometer [219].

In order to precisely quantify the observational limit, a statistical relative uncertainty of the measured cross section must be fixed as a goal for an acceptable measurement. This relative variance can be some standard value such as *u* = 20%, comprising both the variances coming from limited statistics and background contributions. This requirement replaces the separate ones of final number of counts and final peak-to-background ratio [219]. The correlated background (mostly the Compton continuum) coming from the DCE gamma cascade itself is anticipated to be small since its multiplicity is expected to be low, particularly after the excitation energy gate applied in the reconstructed DCE energy spectra of the ejectiles measured by the MAGNEX FPD. In addition, the photopeak efficiencies are designed to be high which also mitigates such effect.

The uncorrelated gamma count-rate in a DCE experiment, however, will be extremely high. The probability of an additional reaction occurring together with the DCE one, within the same coincidence time window (of a few ns), can be quite significant. If this spurious reaction happens to produce a gamma-ray signal in the array with a similar energy as the transition of interest (*e.g.* the $2^+ \rightarrow 0^+$ transition), it may not be distinguishable from the true DCE gamma signal. This is the main origin of background expected in a DCE experiment. If the reaction rate becomes very large, the background originated from these accidental signals reduces the sensitivity of the system.

From statistical considerations, it is possible to obtain an analytical formula for the observational limit as a function of the relevant parameters listed in Table 6.2:

$$\alpha_{\text{lim}} = \frac{B(p_{\text{bg}}(E_\gamma), \varepsilon, \bar{k}, f)}{T f_C u^2} \quad (6.1)$$

where the function $B(p_{\text{bg}}(E_\gamma), \varepsilon, \bar{k}, f)$ is a "reduced" observational limit which characterizes the gamma spectrometer, but depends also on the $p_{\text{bg}}(E_\gamma)$ function, which is specific of the particular experiment.

Table 6.2 Calorimeter and reaction parameters involved in the observational limit formula, their meaning, and typical values expected in a DCE experiment.

| Parameter | Meaning | Typical Values |
|---|---|---|
| $\varepsilon$ | Total photopeak efficiency | 8-80% |
| $p_{\text{bg}}(E_\gamma)$ | Background probability density function | 0.1-1%/keV |
| $f$ | Ratio of cross section between $2^+$ and gs states | 0.1-10 |
| $\bar{k}$ | Average number of reactions per bunch | 0.3-6 |
| $T$ | Time duration of experiment | 1-3 weeks |
| $u$ | Required relative uncertainty of measurement | 10-20% |
| $f_C$ | Cyclotron bunching frequency | 20-40Hz |

The background probability parameter, i.e. the probability that a signal from a spurious reaction generates a signal within the energy gate around the DCE gamma transition of interest, is the most difficult to evaluate. This is because it depends not only on the energy resolution of the array, but also on characteristics of the nuclear reactions occurring in a particular experiment and on the material surrounding the target and the detectors of the array.

For background parameter values that might be typical of a LYSO(Ce) array with 60 keV resolution at 500 keV gamma-ray energy and for 50% photopeak efficiency, the absolute observational limit is obtained at a beam current corresponding to an average number of reactions per beam bunch $\bar{k}$ between 2 and 3 (the "optimum" condition). Above this beam intensity, the limit is raised due to the increasing contribution of the accidental background. As a reference example, for an $A$=150, 1 mg/cm² target, with a surface density of n = 4 × $10^{18}$ particles per cm² and a total reaction cross section of $\sigma_R$ = 3 b with a cyclotron pulse frequency of $f_c$ = 20 MHz, $\bar{k}$ = 1 corresponds to a beam intensity of about $I$ = 1.8 × $10^{12}$ particles per second. The observational limit cross section of about $\sigma_{lim}$ = 20 pb is estimated, for a measurement of one week duration and final relative uncertainty of $u$ = 20%. These values are obtained without the effect of the pyrolythic carbon foil. The reaction cross section of typical DCE experiment beams on $^{12}$C are around 1.6 b [4], with an average gamma multiplicity of about 1, and preliminary estimates indicate that the observational limit could raise by a factor of 20 by introducing the 10μm foil as a backing to the target. The precise value is difficult to estimate because, presently, incomplete information is available with regards to the gamma spectra from these reactions. Detailed simulation and experimental tests will be necessary for more reliable estimates.

*6.7.2 A preliminary design*

A possible geometrical configuration for the scintillator crystals of the calorimeter array contains about 2000 scintillator crystal units, populating a sphere of about 25 cm radius centered on the target, to which the axis of each crystal is directed. The total coverage of such a system could be about 60%-70% of the solid angle. The detectors can be packed in modules of 3 × 3 crystal units or "pixels" (Figure 6.15) of about 15 mm × 15 mm × 50 mm size. With this granularity, in a typical high rate, high multiplicity experiment (such as 1 GHz emission rate) the pile-up probability can be kept well below 10%. In order to increase the photopeak efficiency while keeping the Compton background to a minimum, signals from neighboring detectors of the 3 × 3 modules (and of neighboring modules) will have to be added. In fact, Geant4 simulations have shown that when radiation hits a given detector, there is a high chance that the Compton scattered gamma rays are captured in the neighboring crystals; in this way, the peak-to-total ratio can be increased to about 75% for 500 keV gamma rays.

The scintillation signals could be converted by SiPM devices. Other possible configurations, including standard photomultiplier tubes and detector crystal materials and shapes are under study and detailed tests and Geant4 simulations will be performed to help in the decision of the final design.

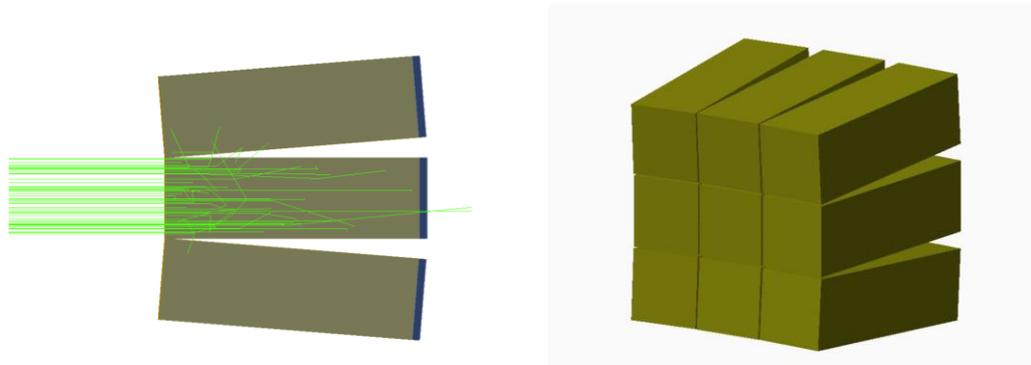

Figure 6.15 Scheme of detection modules with 3X3 pixel LYSO crystals. Left) perspective view. Right) section view illustrating Geant4 simulations of the scattering of 500 keV gamma rays hitting the central pixel.

### 6.7.3 *Simulation of the detector response to target activation*

A simulation of target activation in a specific case has been performed for LaBr$_3$(Ce) scintillator material by means of FLUKA [220]. In Table 6.3 the main characteristics of LaBr$_3$(Ce) are listed. In the simulation the chosen detector element shape is cylindrical, which should lower the cost and ease the canning process, as the material is highly hygroscopic. The diameter was chosen to be 5 cm, as a result of a simple geometrical optimization, of the reduction of lateral Compton losses, and of the availability on a manufacturer's catalog. Several thickness configurations were simulated, from 1 cm to 5 cm, and their basic features were compared in terms of the resulting performances.

It is assumed here that the released gamma-ray energy will only be considered within one single detection element: no sum of neighboring elements will be performed at this stage. However, such a kind of data analysis is not excluded a priori if needed, as it could in principle improve the overall data quality even though it can be hindered by the huge background.

Table 6.3 Main features of the LaBr$_3$(Ce) scintillator material.

| | |
|---|---|
| Density (g/cc) | 5.1 |
| < Z > | 40.5 |
| Hygroscopic | YES |
| Light yield (photons/MeV) | ~70000 |
| Emission Peak (nm) | 375 |
| Decay time (ns) | 30 |
| Energy resolution @662 keV | < 3% |

In order to produce realistic data, similar to what one can expect from the LaBr$_3$(Ce) in operational conditions, a performance of the setup similar to the usual ones reported in the literature was assumed. Therefore a 3% FWHM resolution at 662 keV was assumed, which is not the best resolution achieved by several authors with LaBr$_3$(Ce), but is a quite reasonable value obtained with standard Photomultipliers. This value was rescaled using a Poisson assumption (i.e. by the square root of the deposited energy) thus calculating the expected energy resolution in the full energy range of interest. The best tradeoff was seemingly found with 3 cm thick detectors.

For the further simulations, the $^{18}$O + $^{116}$Sn reaction at 20 MeV/u was assumed as a reference, with about 250 detectors placed on a spherical surface of 20 cm radius (≈ 97% solid angle). A quick simulation of this reaction allowed to estimate the overall shape of the inclusive gamma-ray spectrum that, as expected, is roughly exponentially decreasing. Such a distribution was normalized to a $10^{12}$ pps oxygen beam on a 1 mg/cm$^2$ Sn target by assuming as reaction cross section the geometrical one. The uncorrelated background from target activation was also simulated, and the total expected counting rate on each detector is ~ 30000 cps, easily sustainable by LaBr3(Ce) that has a scintillation decay time of about 30 ns. The overall background in a 5 ns coincidence window with the nucleus of interest detected in MAGNEX was considered and reported in Figure 6.16, along with an example of 0.8 MeV gamma-ray spectrum. The background level was also artificially scaled up by a factor ten, to account for worst-case additional unknown sources. The integral of the full energy peak divided by the corresponding integral of the background contribution represents the signal-to-background (S/B) ratio.

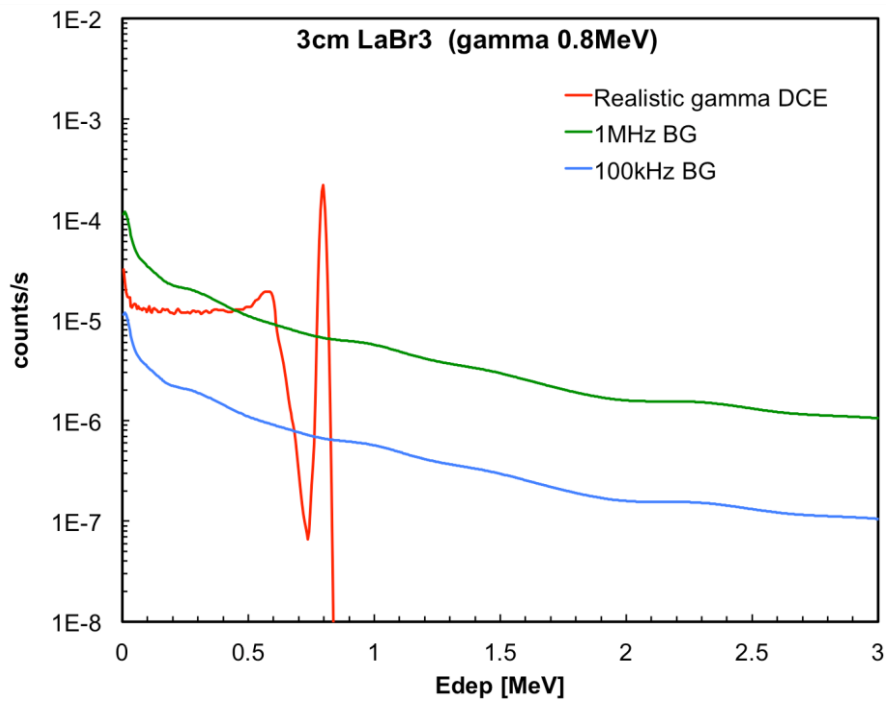

Figure 6.16 Simulated spectrum from hypothetical DCE reaction 0.8 MeV gamma rays in a 5ns coincidence window with the nucleus of interest detected in MAGNEX. Also shown is the expected contribution of the uncorrelated counts from activation of the target and other reactions (100 kHz), which was also scaled up by a factor ten (1MHz) to account for worst case additional unknown sources.

The same simulation was done with useful gamma rays ranging from 0.2 MeV to 2 MeV, the full energy efficiency (area of the full energy peak) and the signal-to-background ratio are reported in Table 6.4.

Table 6.4 The full energy efficiency and the S/B ratio in two background scenarios, for several gamma-ray energies.

| Egamma [MeV] | full energy efficiency | S/B 100 kHz | S/B 1 MHz |
|---|---|---|---|
| 0.2 | 79% | 155 | 15 |
| 0.4 | 43% | 144 | 14 |
| 0.6 | 27% | 157 | 16 |
| 0.8 | 20% | 134 | 13 |
| 1 | 16% | 120 | 12 |
| 1.2 | 13% | 135 | 14 |
| 1.4 | 11% | 140 | 14 |
| 1.6 | 10% | 133 | 13 |
| 1.8 | 9% | 177 | 18 |
| 2 | 8% | 188 | 19 |

The proposed ~ 4π detector made of about 250 LaBr$_3$(Ce) scintillators 3 cm thick, arranged on a spherical shell of 20 cm inner radius, promises to have outstanding performance in terms of gamma-ray detection efficiency and energy resolution, for the DCE reactions to be detected in the NUMEN experiments with the boosted MAGNEX spectrometer. The simulated background from other reactions can be easily sustained by the scintillators and rejected by means of the coincidence with the DCE quasi-projectile and a 5 ns wide time window. The remaining background contribution comes from uncorrelated gamma rays from the target activation. The simulations showed that even in worse (100 kHz) and worst (1 MHz) cases one can efficiently get rid of such a background with a S/B ratio respectively around 150-200 and 15-20. As a final remark it has to be stressed that a special care is needed to avoid any other possible source of additional activation, as for instance having the beam grazing or hitting thick material like the beam pipe, the target holder, etc. In such a case the background would immediately become prohibitive thus preventing any measurement.

The results achieved so far indicate that the coupling of an ancillary system such as that outlined in the previous sections should be adequate for the task of selecting and measuring the DCE cross sections to the g.s. and low-lying excited states. The expected background is tolerable, after DCE reaction selection with the MAGNEX system, except at very large beam intensities. The quality of the array, indicated by its predicted observational limit, is sufficient to allow for the measurement of very low cross sections, as required. Additional tests and simulations are still necessary for the design of the fine details of the system.

## *6.8 Front-End and Read-out Electronics*

The design of front-end and read-out electronics has been conducted in parallel with the design of the new FPD. In particular, one of the main objectives was to design a modular, scalable, radiation hard architecture, which, in addition, fulfils the strong requirements in terms of high event rate, easy maintenance and precise synchronization.

### *6.8.1 Front-end*

The front-end (FE) of the FPD tracker is based on the VMM chip, developed for the ATLAS experiment at CERN [221]. The architecture of the FE electronics is conceived to be modular and scalable to the final dimensions of the detector. The segmented anode board is designed in order to take advantage of the unique capabilities of the VMM chip, which is able to perform a digital reconstruction of the track at high event rate.

Figure 6.17 shows the architecture of VMM3, i.e. version number 3 of the VMM project, featuring higher complexity and functionality. Each channel provides the peak amplitude and time with respect to the bunch crossing clock or other trigger signal in a data-driven mode. This is accomplished as follows. Each channel is equipped with a fast comparator with an individually adjustable threshold. When a signal crosses a set threshold, a peak detection circuit is enabled. Neighbour-enable logic allows to set a relatively high threshold and yet to record very small amplitudes. At the peak, a time-to-amplitude converter is started and stopped by the trigger signal. The two amplitudes are digitized and stored in a de-randomizing buffer and readout serially with a smart token passing scheme that reads out the amplitude, timing, and addresses of the channels with relevant information only, thus dramatically reducing the data bandwidth required and resulting in a very simple readout architecture. The ASIC has 64 channels, thus easing the problem of a large number of channels. It also provides prompt information that can be used to form trigger primitives.

In the selected data transfer mode – i.e. continuous (digital) - a total of 38 bits are generated for each event in the VMM3. The first bit is used as a readout flag, the second is the threshold crossing indicator (allows discrimination between above-threshold and neighbour events). The next 6 bits define the channel address, followed by 10 bits associated with the peak amplitude, and 20 bits associated with the timing.

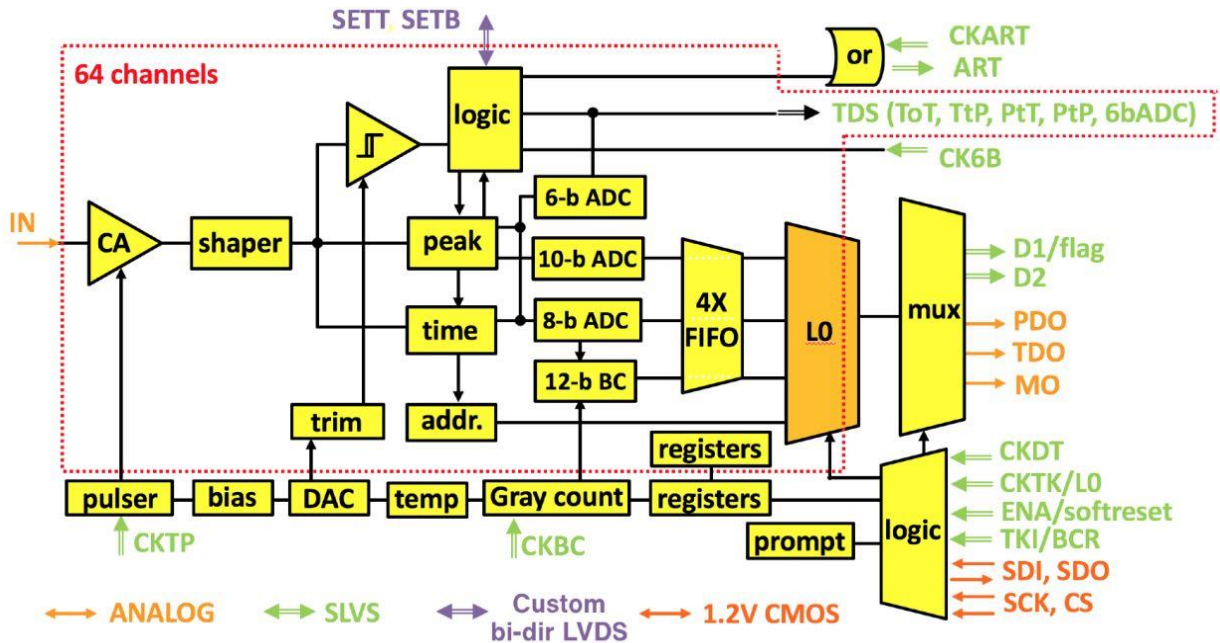

Figure 6.17 Architecture of one of the 64 channel of the VMM3.

The 38-bit word is stored in a 4-events deep de-randomizing First-In-First-Out (FIFO) (there is one such FIFO per channel) and it is read out using a token-passing scheme where the token is passed first-come first-serve only among those FIFOs that contain valid events. The first token is internally generated as needed and advanced with the token clock. The data in the FIFOs is thus sequentially multiplexed to the two digital outputs data0 and data1. The first output data0 is also used as a flag, indicating that events need to be read out from the chip. The external electronics releases a sync signal using the token clock as well (i.e. the token clock provides both advancement and data output synchronization), after which the 38-bit data is shifted out in parallel to the data0 and data1 outputs using 19 clock edges of the external data clock.

### *6.8.2 Read-Out and Slow Control*

The main tasks of the read-out (RO) electronics for the new NUMEN FPD are: (1) the real-time data collection from the FE boards and the high bandwidth data transmission towards data acquisition; (2) the remote configuration and the slow control of the FE electronics; and (3) the synchronization of the whole detector. The RO electronics architecture, designed as modular and scalable to the final size of the detectors, is based on the System On Module (SOM) manufactured by National Instruments [222]. The SOM is a board-level circuit that integrates a system function in a single module. These very versatile devices couple high performance Field Programmable Gate Array (FPGA) to powerful processor architecture and allow a graphical approach to the programming and interfacing. The tasks of SOM are the fast serial read-out of the VMM chips, the slow control of the FE and the precise synchronization of all the FE and RO boards.

### *6.8.3 Architecture of Front-End and Read-Out electronics*

The architecture of FE and RO electronics is shown in Figure 6.18. It is designed to be modular. Each module is composed by 8 VMM Asics and one SOM. The SOM configures and reads-out the VMM ASICs, and transfers data by means of a Gb Ethernet connection.

Each recorded event is constituted by the id of the hit strip, the id of the VMM chip connected to the strip, the id of the SOM module, the charge and the time. In this way it is possible to realign and reconstruct offline all pieces of data referring to one event.

A dedicated SOM module, the Synchro SOM, is devoted to the synchronization of all FE-RO modules. This task is accomplished by sending a 10 MHz clock to each module. The time stamp of each event, in this way, can be realigned and expressed in global time unit.

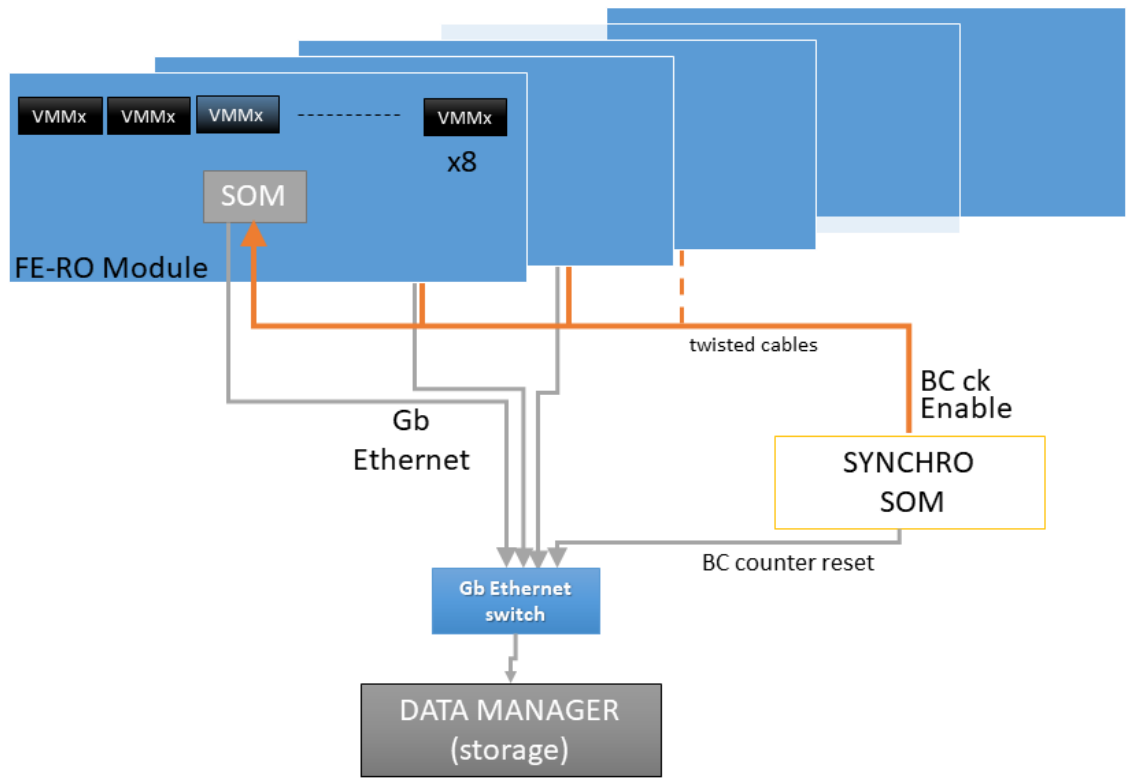

Figure 6.18 Architecture of the front-end and read-out (FE-RO) and scheme of the connection to Data manager

### 6.8.4 *FE-RO demonstrator*

A complete FE-RO chain, employing the VMM2 and the SOM, was designed and successfully tested, despite the VMM2 version still suffers from some problems correlated to the analog-to-digital conversion of charge and time. The new version VMM3, available from May 2017, introduces many improvements, especially regarding the trigger mechanisms and data communication.

A reduced scale prototype of the new FPD tracker was built and is presently being tested in connection to a FE-RO chain with the new version of VMM ASIC, the VMM3.

The work is still in progress concerning particle identification and the gamma array. The plan is to adopt the same solution for the FE-RO electronics.

### 6.8.5 *Evaluation of neutron production rates*

The significant increase in the beam current which will come from the upgrade of the CS will also cause a significant increase of the neutron flux and of the total dose in the MAGNEX hall. An early evaluation of those quantities is mandatory, both to identify possible radiation protection issues and to ensure that the significant radiation field does not interfere with the detectors and with the electronics. While all NUMEN detectors are designed to be radiation-hard, the electronic modules could be more sensitive to neutron background and will require a dedicated shielding.

In order to address the issue quantitatively, a simulation was implemented based on the FLUKA [220] Monte Carlo code. The simulation features a simplified description of the setup in the MAGNEX hall (magnets, scattering chamber, beam lines, electronics). The interaction of a reference 60 MeV/u $^{20}$Ne$^{7+}$ beam of 60 eµA current on a realistic Ge + C target produces a neutron flux of $7 \cdot 10^4$ n/(cm$^2$ s) at the focal plane detector position, where the electronic modules will be installed.

*6.8.6 Radiation Tolerance and Single Event Upsets*

The devices to be accurately characterized from the point of view of radiation tolerance are the VMM chip and the SOM.

The ATLAS collaboration provided the results of radiation tolerance test on VMM chip [223]. Deep sub-micron technologies are known to be immune to much higher Total Integrated Dose (TID) because of the increasingly thinner oxide layers which can trap smaller amounts of charge. Although not expected to be a problem, the VMM3 will be tested for TID tolerance. However, Single Event Upsets (SEU) become increasingly more serious as the technology feature size decreases: due to the smaller capacitance in the storage elements, a smaller energy depositions is sufficient to flip their state. In the VMM there are two types of storage elements that require SEU protection: the configuration register, and the state machine control logic.

While the protection of the 12-bit Bunching Cross Identification (BCID) register is being considered, no specific action is required for the FIFOs, as an occasional data corruption is not an issue. To mitigate the SEU effects in the VMM storage elements two different techniques are used:

  1. Dual Interlocked CElls (DICE) for the protection of the configuration register,
  2. The Triple Modular Redundancy (TMR) for the state machines.

The DICE uses redundancy to significantly reduce susceptibility to an upset. D-type flip flop based on the dual interlocked cell latches have redundant storage nodes and restore the cell original state when an SEU error is introduced in a single node [224]. The scheme fails if multiple nodes are upset but this is far less likely.

The TMR technique is used to protect the small number (less than 20) of storage elements of the state machines. The first version of the VMM was tested in the NSCR Demokritos Tandem accelerator. The measured cross-section is $(4.1 \pm 0.7) \times 10^{-24}$ cm$^2$/bit. Taking into account that the total number of bits for the VMM register is 3264, a total of about 300 SEU/yr/VMM is estimated for the maximum neutron flux foreseen in NUMEN ($7\times10^4$ neutrons/(cm$^2$ s), see Section 6.8.5). These SEUs can be partially recovered by register resetting.

No data exist regarding the radiation tolerance of the SOM, despite different tests with neutrons and gamma rays have been conducted on similar devices. The main criticalities are the same described for VMM, related to SEU and integrity of registers.

The NUMEN collaboration has started a dedicated test campaign of the overall electronics chain, FE and RO, in order to precisely determine the radiation tolerance performances and, in parallel, a possible mitigation strategy. First encouraging results have been recently achieved for the SOM at the nuclear reactor facility of the IPEN laboratory in Sao Paulo. In particular, the response of the SOM under a monochromatic thermal neutron collimated ($5 \times 5$ cm$^2$) beam of $10^4$ neutrons/(cm²·s) and energy lower than 1 eV show no SEU events in about 1 day irradiation. The tests performed with thermal and epithermal neutron rates gradually increasing up to $10^8$ neutrons/(cm²·s) confirm that the SOM architecture is fully compliant with the radiation hardness requirement of NUMEN. The data analysis for quantitative information is in progress and further tests are scheduled in the near future.

## *6.9 Data handling and data processing*

### *6.9.1 Data transmission and storage*

The NUMEN electronic modules (described in Section 6.8) provide a data stream which is already formatted according to the TCP/IP standard protocol and is transmitted over a standard Ethernet cable. The data rate expected to be written on disk is estimated to be between 20 MB/s and 200 MB/s, depending on the beam configuration and on the trigger settings.

Such a data flow can be handled and written on disk using commercially-available components. The conceptual layout is the following: Ethernet cables coming from the electronics are collected through a 10 Gbit/s network switch and the data flow is routed to a one- or two-CPU 32-core server (*main server*), equipped with 10 Gbit/s Ethernet cards. Only a small fraction of the cores will be busy with the disk writing. Therefore, the remaining free cores can be used for the event building (i.e. match the information of the same events coming from the different detector systems through different electronic modules) and/or for other online processing. The online processing (e.g. compression) could potentially reduce the amount of data written on disk, thus saving on the storage costs. The system will be complemented by an additional small server (*control server*) and by a *backup server*, which is ideally a clone of the main one. The control server will handle the run control and the slow control in a transparent and redundant way, being also in charge with the interaction with the electronics modules and the detector. The backup server is meant to be a quick replacement of the main server, in case of a failure during the data taking. During the normal operations, when the main server is working, the backup machine can be used for offline data processing.

The interface to the storage component is a RAID6 Fibre Channel controller: it can write on disk up to 16 Gbit/s (= 2 GB/s), which is safely above the data rate expected in NUMEN. The targeted dimension for the global RAID6 disk storage is about 500 TB, which is readily available on the market. Since NUMEN will be intrinsically made by many independent runs, with different target nuclei, an alternative layout under consideration is to have a partitioned storage (e.g. blocks of 48 × 4 TB disks, totaling 160 net TB each).

*6.9.2 Offline analysis*

The offline reconstruction of NUMEN will be performed using the MXSoft code, which is already available. MXSoft is the re-engineering in C++ of the software suite already used and validated in MAGNEX; it depends upon ROOT [225] for the storage of the final high-level information. In particular, all calibration, selection and reconstruction algorithms are kept exactly the same in MAGNEX. However, MXSoft was specifically designed to be modular and flexible, allowing to easily accommodate for possible extensions or new/alternative algorithms, and to improve the CPU performance and the memory footprint. The performance of MXSoft has been tested and validated with the recent NUMEN Phase 2 runs: based on this, the new code is expected to be appropriate and scalable up to the anticipated data rate of NUMEN Phase 4. Furthermore, the build procedure of MXSoft was designed to be cross-platform compatible, such to allow for non-problematic installation and use on Linux systems in long-term operation.

# 7 Conclusion and perspectives

Pioneering experiments on ($^{18}$O,$^{18}$Ne) and ($^{20}$Ne,$^{20}$O) DCE performed at INFN-LNS Laboratory have shown that accurate cross sections measurements at very forward angles can be done for the ground to ground state transitions. Important information on nuclear structure and specifically for NMEs connected with second order isospin nuclear response can be extracted, even within a schematic nuclear reaction model. A significant improvement of nuclear reaction theory in the view of fully microscopic quantum approach will be beneficial to get more accurate information from the measured data.

The measurement of DCE absolute cross sections and the extraction of relevant NMEs is the main activity characterizing the NUMEN project. The most ambitious goal of NUMEN is to find a connection between the NMEs extracted from DCE reactions and those characterizing $0\nu\beta\beta$ decay, at least in the nuclei in which this process is energetically allowed. In this perspective, NUMEN is exploring an original experimental approach to $0\nu\beta\beta$ decay NMEs that could have an impact in the possible evaluation of the absolute value of neutrino average mass from the hopefully future observation of this rare decay.

For neutrino physics, systematic exploration, spanning all the variety of $0\nu\beta\beta$ decay candidate isotopes, is demanded and NUMEN is fully committed to pursue this ambitious goal.

However, despite the promising results achieved to date, much remains to be done toward the determination of NME for $0\nu\beta\beta$ decay, with enough accuracy as needed by neutrino community.

As described in the paper, the project promotes a major upgrade of the INFN–LNS research facility in the direction of a significant increase of the beam intensity. This in turn demands challenging R&D in several aspects of the technology involved in heavy ion collision experiments.

The acceleration of heavy ion beams in the regime of kW power and at energies from 15 to 70 MeV/u requires a substantial change in the extraction technologies of the beam of the INFN-LNS Superconducting Cyclotron. The transport of such a beam poses serious issues of radioprotection, calling for a careful evaluation of radiation levels also involving the effects on detectors, electronics and various equipment. A critical issue is the design of thin targets for DCE experiments, considering the deterioration due to the dissipation of the enormous amount of heat deposited by the ion beam. Due to the high beam intensity, the present detectors of the MAGNEX spectrometer cannot be used. A dedicated study of new detection technologies, coping with the expected high rate and high fluency and still guarantying the same resolution and sensitivity of the present ones is mandatory. These include the search of new materials, the study of new electronics and DAQ systems, which best match the stringent experimental requirements.

Moreover, the development of the different theoretical aspects connected with the nuclear structure and reaction mechanisms involved in heavy ions induced in DCE reactions is a key issue for the achievement of the ambitious goals of the project.

Such R&D and theoretical development is a fundamental aspect of the NUMEN project, already supported by INFN.

In perspective, NUMEN aims at giving an innovative contribution in one of the most promising fields of fundamental physics. It indicates also a possible growth prospect of heavy ion physics in synergy with neutrino physics.

Within the INFN–LNS context, NUMEN promotes an important upgrade of the experimental facilities, which will be likely beneficial for other nuclear physics projects. Last but not least, an important fallout in technological and scientific developments is foreseen in a broader context of different physics fields.

## Acknowledgements

The authors wish to thank Prof. F. Iachello for the encouragement and fruitful discussions during the development of the project. The authors thank also the INFN-LNS accelerator, technical and research divisions for the support in all the phases of the project, Mr. Mauro Raimondo (DISAT – Politecnico di Torino) for the precious work in producing the FESEM images. The authors are also thankful to Trustech Company for the valuable expertise in performing the depositions.

This project has received funding from the European Research Council (ERC) under the European Union's Horizon 2020 research and innovation programme (grant agreement No 714625) and from the European Union's Horizon 2020 research and innovation programme (grant agreement No 654002).

The authors acknowledge the CNPq, FAPERJ and FAPESP institutions for the partial financial support.

## List of acronyms

ASIC = Application Specific Integrated Circuit

BCID = Bunching Cross Identification

CM = Centre of mass

CE = Charge Exchange

CCC = Coupled Channel Calculations

CS = Superconducting Cyclotron

DWBA = Distorted Wave Born Approximation

DCE = Double Charge Exchange

DGT = Double Gamow-Teller

DIAS = Double Isobaric Analogue State

DC = Drift Chambers

DICE = Dual Interlocked CElls

EC = Electron Capture

ED = Electrostatic deflectors

EDF = Energy Density Functional

F = Fermi

FESEM = Field Emission Scanning Electron Microscopy

FPGA = Field Programmable Gate Array

FSI = Final State Interaction

FIFO = First-In-First-Out

FPD = Focal Plane Detector

FE = Front-End

FWHM = Full Width at Half Maxima

GT = Gamow-Teller

GTR = Gamow-Teller Resonance

GEM = Gas Electron Multipliers

GDP = Giant Dipole Resonances

g.s.= ground state

ISI = Initial State Interaction

IBM = Interacting Boson Model

IAS = Isobaric Analogue State

INFN-LNS = Istituto Nazionale di Fisica Nucleare - Laboratori Nazionali del Sud

B$\rho$ = magnetic rigidity

MPGD = Micro-Pattern Gas Detectors

mC = milli Coulomb

MIP = Minimum Ionizing Particles

$0\nu\beta\beta$ = Neutrinoless Double Beta Decay

NIEL = Non Ionizing Energy Loss

NME/s = Nuclear Matrix Element/s

1n = one-neutron

1p = one-proton

PID = Particle Identification

p$\mu$A = pico micro Ampere

pnA = pico nano Ampere

QPRA = Quasi-particle Random Phase Approximation

RO = Read-Out

SM = Shell Model

SNR = Signal to Noise Ratio

SiC = Silicon Carbide

SCE = Single Charge Exchange

SEU = Single Event Upsets

SOM = System On Module

T = Tesla

THGEM = Thick Gas Electron Multipliers

TOF = Time of Flight

TID = Total Integrated Dose

TMR = Triple Modular Redundancy

2νββ = Two-neutrino Double Beta Decay

2n = Two-neutron

2p = Two-proton